\documentclass[11pt,twoside]{article}
\input{epsf.sty}
\usepackage{mathrsfs}
\usepackage{amsmath}
\usepackage{amssymb}
\usepackage{graphicx}
\usepackage{subfigure}
\usepackage{appendix}
\newtheorem{theorem}{Theorem}
\newtheorem{definition}{Definition}
\newtheorem{proposition}{Proposition}
\newtheorem{lemma}{Lemma}

\numberwithin{equation}{section}
\newcommand*{\QEDB}{\hfill\ensuremath{\square}}

\title{Initial-boundary value problems associated with the Ablowitz-Ladik system}
\author{Baoqiang Xia$^{1,2}$ and A.S. Fokas$^{2}$
\\
$^{1}$School of Mathematics and Statistics, Jiangsu Normal
University,\\
 Xuzhou, Jiangsu 221116, P. R. China,
 \\
 E-mail address:
xiabaoqiang@126.com
\\
$^2$Department of Applied Mathematics and Theoretical Physics,
\\University of Cambridge, Cambridge CB3 0WA, United Kingdom,
\\
 E-mail address:
t.fokas@damtp.cam.ac.uk }

\date{}
\begin{document}
\maketitle
\begin{abstract}
We employ the Ablowitz-Ladik system as an illustrative example in order to demonstrate how to analyze initial-boundary value problems for integrable nonlinear differential-difference equations via the unified transform (Fokas method).
In particular, we express the solutions of the integrable discrete nonlinear Schr\"{o}dinger and integrable discrete modified Korteweg-de Vries equations in terms of the solutions of appropriate matrix Riemann-Hilbert problems.
We also discuss in detail, for both the above discrete integrable equations, the associated global relations and the process of eliminating of the unknown boundary values.


\end{abstract}
\newpage

\section{ Introduction}
For integrable nonlinear equations, the so-called unified transform, which is also referred to as the Fokas method, provides an important generalization of the inverse scattering transform formalism from initial value problems to initial-boundary value problems (IBVPs) \cite{F1}-\cite{F6}.
This method involves the following three steps: (1) perform the
simultaneous spectral analysis of the two linear eigenvalue equations (called Lax pair) associated with a PDE; (2) express the solution of the given PDE via the solution of a Riemann-Hilbert (RH) problem; (3) by analyzing the so called global relation, eliminate the unknown boundary values in terms of given data.
The Fokas method has been successfully implemented to integrable soliton equations with $2\times2$ Lax pairs, such as the nonlinear Schr\"{o}dinger (NLS),
the Korteweg-de Vries (KdV), the modified Korteweg-de Vries (MKDV) and the sine-Gordon equations \cite{F6}, as well as to several integrable equations with $3\times3$ Lax pairs \cite{L1}-\cite{GLZ}.

A related and important question is the study of IBVPs for integrable differential-difference equations.
Recently, Biondini and collaborators \cite{BH1}-\cite{BH3} have initiated the study of implementing the Fokas method to such equations by considering
an IBVP for the integrable discrete nonlinear Schr\"{o}dinger (DNLS) equation  \cite{AL1}-\cite{AL3}:
\begin{eqnarray}
iq_t(n,t)+q(n+1,t)-2q(n,t)+q(n-1,t)-\nu |q(n,t)|^2\left(q(n+1,t)+q(n-1,t)\right)=0,
\label{nls}
\end{eqnarray}
where $\nu=\pm 1$. This equation is the reduction $p(n,t)=\nu q^\ast(n,t)$ of the Ablowitz-Ladik (AL) lattice system \cite{AL1}-\cite{AL3}:
\begin{eqnarray}
\begin{array}{l}
iq_t(n,t)+q(n+1,t)-2q(n,t)+q(n-1,t)-p(n,t)q(n,t)\left( q(n+1,t)+ q(n-1,t)\right)=0,
\\
ip_t(n,t)-p(n+1,t)+2p(n,t)-q(n-1,t)+p(n,t)q(n,t)\left( p(n+1,t)+ p(n-1,t)\right)=0.
\end{array}
 \label{aleq1}
\end{eqnarray}
Biondini and collaborators have performed the spectral analysis for the following Lax pair \cite{AL1}-\cite{AL3}:
\begin{subequations}
\begin{eqnarray}
\Phi(n+1,t,z)-Z\Phi(n,t,z)&=&
Q(n,t)\Phi(n,t,z),
\label{LPS0}\\
\Phi_t(n,t,z)-\frac{i}{2}(z-1/z)^2\sigma_3\Phi(n,t,z)&=&
H(n,t,z)\Phi(n,t,z),
\label{LPT0}
\end{eqnarray}
\label{LPST0}
\end{subequations}
where
\begin{eqnarray}
\begin{split}
&Z=\left( \begin{array}{cc} z & 0 \\
 0 &  z^{-1}\\ \end{array} \right), \quad
Q(n,t)=\left( \begin{array}{cc} 0 & q(n,t) \\
 p(n,t) &  0 \\ \end{array} \right),
 \quad
\sigma_3=\left( \begin{array}{cc} 1 & 0 \\
 0 &  -1 \\ \end{array} \right),
 \\
&H(n,t,z)=i\left( \begin{array}{cc} -p(n-1,t) q(n,t) & zq(n,t)-z^{-1}q(n-1,t) \\ zp(n-1,t)- z^{-1}p(n,t) & p(n,t)q(n-1,t) \\ \end{array} \right).
\end{split}
\label{QH0}
\end{eqnarray}
The analysis presented in \cite{BH1} differs from the analogous analysis of the corresponding PDEs in two important ways:
first, the determinants of the eigenfunctions of the $n$-part of the Lax pair depend on the potential and the independent variables;
second, the $t$-part of the Lax pair is not traceless, i.e., the matrix $H(n,t,z)$ in (\ref{QH0}) is not traceless
(it appears that this important fact was overlooked in \cite{BH1}).
As a result of these two problems, the implementation
of the Fokas method to the DNLS equation is not as effective as the corresponding PDEs.
For example, the relevant RH problem and the resulting integral representation of the solution
depend on unknown potentials; see section 3.2 in \cite{BH1}.

Here, we overcome these difficulties as follows: we first reformulate the Lax pair (\ref{LPST0})
so that the determinant of the matrix in the $n$-part of the reformulated Lax pair equals to $1$, and
the $t$-part of the Lax pair is traceless.
We then use this reformulated Lax pair in order to present an efficient way of implementing the Fokas method to differential-difference equations.
In particular, by performing the spectral analysis to the reformulated Lax pair, we solve IBVPs for the DNLS and discrete MKDV (DMKDV) equations.
The main advantage of our results in comparison with \cite{BH1} is that the jump matrices of the associated RH problems, and thus
the resulting reconstruction formulae for solutions of the IBVPs, involve only the spectral functions.
We also discuss, for both the DNLS equation and the DMKDV equation, the global relations and
show how to eliminate the unknown boundary values.

The paper is organized as follows: in section 2 we present a Lax pair which is convenient for performing the spectral analysis of the AL lattice.
In section 3, initial value problems for the AL lattice are studied, and a vanishing lemma for the associated RH problem is established.
In section 4 we implement the unified transform to analyze IBVPs for the AL lattice.
In section 5 we characterise the unknown boundary values by using the global relation.
Our results are discussed further in section 6.

\section{A Lax pair}

We normalize the determinant of the matrix appearing in the $n$-part of the Lax pair (\ref{LPS0}) to $1$,
i.e., we consider the following $n$-part \cite{Geng1}:
\begin{eqnarray}
\Phi(n+1,t,z)-\frac{1}{f(n,t)}Z\Phi(n,t,z)=
\frac{1}{f(n,t)}Q(n,t)\Phi(n,t,z),
\label{LPS}
\end{eqnarray}
where $\Phi(n,t,z)$ is a $2\times 2$ matrix, $f(n,t)=\sqrt{1-q(n,t)p(n,t)}$, and $Z$, $Q(n,t)$ are defined as in (\ref{QH0}).
Then, following \cite{Tu}, we find the following $t$-part:
\begin{eqnarray}
\Phi_t(n,t,z)-i\omega(z)\sigma_3\Phi(n,t,z)=
H(n,t,z)\Phi(n,t,z),
\label{LPT}
\end{eqnarray}
where
\begin{eqnarray}
\begin{split}
\omega(z)&=\alpha z^2+\beta z^{-2}+c,
\\
H(n,t,z)&=i\left( \begin{array}{cc} -\alpha q(n,t)p(n-1,t)-\beta p(n,t)q(n-1,t)& 2\alpha zq(n,t)-2\beta z^{-1}q(n-1,t) \\ 2\alpha zp(n-1,t)-2\beta z^{-1}p(n,t) & \alpha q(n,t)p(n-1,t)+\beta p(n,t)q(n-1,t) \\ \end{array} \right),
\end{split}
\label{H}
\end{eqnarray}
and $\alpha$, $\beta$, $c$ are three arbitrary constants.

The compatibility condition of (\ref{LPS}) and (\ref{LPT}), namely, $\frac{d \Phi(n+1,t,z)}{dt}=\left(\frac{d \Phi(m,t,z)}{dt}\right)_{m=n+1}$,
yields the following AL-type lattice equation:
\begin{eqnarray}
\begin{split}
&q_t(n,t)=2i\left(\alpha q(n+1,t)+\beta q(n-1,t)+cq(n,t)-q(n,t)p(n,t)\left(\alpha q(n+1,t)+\beta q(n-1,t)\right)\right),
\\
&p_t(n,t)=-2i\left(\alpha p(n-1,t)+\beta p(n+1,t)+cp(n,t)-q(n,t)p(n,t)\left(\alpha p(n-1,t)+\beta p(n+1,t)\right)\right).
\end{split}
 \label{aleq}
\end{eqnarray}
For $\alpha=\beta=\frac{1}{2}$ and $c=-1$, equation (\ref{aleq}) is nothing but the AL lattice (\ref{aleq1}).
For $\alpha=\beta=\frac{1}{2}$, $c=-1$ and $p(n,t)=\nu q^\ast(n,t)$,  $\nu=\pm 1$, equation (\ref{aleq}) is reduced to the DNLS equation (\ref{nls}).
Thus, the Lax pair for the DNLS equation (\ref{nls}) is given by (\ref{LPS}) and (\ref{LPT}) with $\alpha=\beta=\frac{1}{2}$, $c=-1$, $p(n,t)=\nu q^\ast(n,t)$, and
\begin{eqnarray}
\omega(z)=\frac{1}{2}\left(z-z^{-1}\right)^2.
\label{nlsw}
\end{eqnarray}
For $\alpha=-\beta=\frac{i}{2}$, $c=0$ and $p(n,t)=\nu q(n,t)$, $\nu=\pm 1$, with $q(n,t)$ real, the system (\ref{aleq}) is reduced to the integrable discrete modified KdV (DMKDV) equation,
\begin{eqnarray}
q_t(n,t)+q(n+1,t)-q(n-1,t)-\nu q^2(n,t)\left(q(n+1,t)-q(n-1,t)\right)=0.
\label{mkdv}
\end{eqnarray}
This equation possesses the Lax pair (\ref{LPS}) and (\ref{LPT}) with $\alpha=-\beta=\frac{i}{2}$, $c=0$, $p(n,t)=\nu q(n,t)$, and
\begin{eqnarray}
\omega(z)=\frac{i}{2}\left(z^{2}-z^{-2}\right).
\label{mkdvw}
\end{eqnarray}

We note that the determinant of the matrix $U(n)=\frac{1}{f(n,t)}(Z+Q(n,t))$ in the $n$-part of the Lax pair equals to $1$,
and furthermore the matrix $V(n)=i\omega(z)\sigma_3+H(n,t,z)$ in the $t$-part of the Lax pair is traceless.
It turns out that these two properties are convenient for studying the spectral functions  and
for formulating the relevant RH problems; this fact will become clear in sections 3.2 and 4.1.2.

\section{Initial value problems}

In preparation for the study of IBVPs, we first study the initial value problems
for the DNLS and DMKDV  equations: given $q(n,0)$, we will study equations (\ref{nls}) and (\ref{mkdv}) with $n\in\mathbb{Z}$ and $t> 0$, where $\mathbb{Z}$ denotes the set of integers. We require that the initial datum $q(n,0)$ decays rapidly enough as $n\rightarrow \pm \infty$ to belong to $l^1(\mathbb{Z})$, where $l^1(\mathbb{Z})$ denotes the space of the sequences $\left\{s(n)\right\}_{n\in \mathbb{Z}}$ such that its $L^1$ norm is finite, namely, $\|s(n)\|_1=\sum_{n=-\infty}^{n=+\infty}|s(n)|<\infty$ \cite{AL3}.

We note that the arguments in \cite{AL3} can be easily adapted to the modified Lax pair (\ref{LPS}) and (\ref{LPT}),
thus most of the arguments in this section are essentially the same as those in \cite{AL3}.
An improvement we have made here is that the properties of the spectral functions are similar with those of integrable PDEs; see section 3.2.1.
These properties enable us to establish a vanishing lemma which ensures the unique solvability of the associated RH problem for the DNLS equation; see section 3.2.5.
We also obtain analogous results for the DMKDV equation (in fact, we start from the general AL system (\ref{aleq})).

\subsection{The direct problem}
We define the eigenfunction $\mu(n,t,z)$ via
\begin{eqnarray}
\Phi(n,t,z)=\mu(n,t,z)Z^ne^{iw(z)t\sigma_3}.
\label{mud}
\end{eqnarray}
Then the Lax pair (\ref{LPS}) and (\ref{LPT}) becomes
\begin{subequations}
\begin{eqnarray}
&&f(n,t)\mu(n+1,t,z)-\hat{Z}\mu(n,t,z)=Q(n,t)\mu(n,t,z)Z^{-1},
\label{LPS3}\\
&&\mu_t(n,t,z)-i\omega(z)[\sigma_3,\mu(n,t,z)]=
H(n,t,z)\mu(n,t,z).
\label{LPT3}
\end{eqnarray}
\label{LPST3}
\end{subequations}
In order to derive the particular solutions of  (\ref{LPST3}), it is convenient to introduce the modified eigenfunction
\begin{eqnarray}
\Psi(n,t,z)=\hat{Z}^{-n}e^{-iw(z)t\hat{\sigma_3}}\mu(n,t,z),
\label{mud}
\end{eqnarray}
where $\hat{Z}$ and $e^{\hat{\sigma_3}}$ act on a $2 \times 2$ matrix $A$ as follows:
\begin{eqnarray}
\hat{Z}A=ZAZ^{-1}, \quad e^{\hat{\sigma_3}}A=e^{\sigma_3}Ae^{-\sigma_3}.
\label{zhatd}
\end{eqnarray}
Then (\ref{LPST3}) become
\begin{subequations}
\begin{eqnarray}
&&f(n,t)\Psi(n+1,t,z)-\Psi(n,t,z)=Z^{-1}\hat{Z}^{-n}e^{-iw(z)t\hat{\sigma_3}}\left(Q(n,t)\right)\Psi(n,t,z),
\label{LPS4}\\
&&\Psi_t(n,t,z)=\hat{Z}^{-n}e^{-iw(z)t\hat{\sigma_3}}\left(H(n,t,z)\right)\Psi(n,t,z).
\label{LPT4}
\end{eqnarray}
\label{LPST4}
\end{subequations}

We introduce the notations
\begin{eqnarray}
C(n,t)=\prod^{\infty}_{m=n}f(m,t)=\prod^{\infty}_{m=n}\sqrt{1-q(m,t)p(m,t)},
\quad
C(-\infty)=\lim_{n\rightarrow-\infty}C(n,t).
\end{eqnarray}
Note that the product $C(n,t)$ converges absolutely if $\|q(m,t)\|_1,\|p(m,t)\|_1<\infty$ \cite{AL3}.
One can check directly that $C(-\infty)$ does not depend on $t$. Actually, $C(-\infty)$ is a conserved quantity.

It is straightforward to obtain two particular solutions of equation (\ref{LPS4})
which approach the $2\times 2$ identity matrix $I$ as $n\rightarrow \mp \infty$
by the summation equations:
\begin{eqnarray}
\begin{split}
&\Psi_1(n,t,z)=C(n,t)\left(\frac{1}{C(-\infty)}I
+Z^{-1}\sum_{m=-\infty}^{n-1} \frac{1}{C(m,t)}\hat{Z}^{-m}e^{-iw(z)t\hat{\sigma_3}}(Q(m,t))\Psi_{1}(m,t,z)\right),
\\
&\Psi_2(n,t,z)=C(n,t)\left(I
-Z^{-1}\sum_{m=n}^{\infty} \frac{1}{C(m,t)}\hat{Z}^{-m}e^{-iw(z)t\hat{\sigma_3}}(Q(m,t))\Psi_{2}(m,t,z)\right).
\end{split}
\label{psi}
\end{eqnarray}
Substituting (\ref{psi}) into (\ref{mud}), we
find two particular solutions $\mu_{1}(n,t,z)$ and $\mu_{2}(n,t,z)$ of (\ref{LPS3}) which approach $I$ as $n\rightarrow \mp \infty$:
\begin{subequations}
\begin{eqnarray}
\mu_{1}(n,t,z)&=&C(n,t)\left(\frac{1}{C(-\infty)}I+Z^{-1}\sum_{m=-\infty}^{n-1} \frac{1}{C(m,t)}\hat{Z}^{n-m}(Q(m,t)\mu_{1}(m,t,z))\right),
\label{mua}
 \\
\mu_{2}(n,t,z)&=&C(n,t)\left(I-Z^{-1}\sum_{m=n}^{\infty} \frac{1}{C(m,t)}\hat{Z}^{n-m}(Q(m,t)\mu_{2}(m,t,z))\right).
\label{mub}
\end{eqnarray}
\label{mu}
\end{subequations}
Let $\mu_j^L(n,t,z)$ and $\mu_j^R(n,t,z)$, $j=1,2$, denote the first and second columns of $\mu_j(n,t,z)$ respectively.
It can be shown (see \cite{AL3}) that these columns are analytic in the following domains of the complex $z$-plane:
\begin{eqnarray}
\begin{split}
\mu_1^L(n,t,z),\quad \mu_2^R(n,t,z):  \qquad |z|>1,
\\
\mu_1^R(n,t,z),\quad \mu_2^L(n,t,z):  \qquad |z|<1.
\end{split}
\end{eqnarray}
Moreover, these columns are continuous and bounded on the closure of these domains. The analyticity properties of $\mu_j(n,t,z)$ immediately imply analogous analyticity properties for $\Phi_j(n,t,z)$, namely,
$\Phi_j(n,t,z)$ are analytic in the following domains:
\begin{eqnarray}
\begin{split}
\Phi_1^L(n,t,z),\quad \Phi_2^R(n,t,z):  \qquad |z|>1,
\\
\Phi_1^R(n,t,z),\quad \Phi_2^L(n,t,z):  \qquad |z|<1.
\end{split}
\end{eqnarray}
Using (\ref{mu}), we can derive (see appendix A.1) the asymptotic behaviors of $\mu_j(n,t,z)$, $j=1,2$:
\begin{subequations}
\begin{eqnarray}
\mu_1=\frac{C(n,t)}{C(-\infty)}\left\{I+Q(n-1,t)Z^{-1}+\left( \begin{array}{cc} O(z^{-2},\text{even}) & O(z^{3},\text{odd})  \\
  O(z^{-3},\text{odd}) & O(z^{2},\text{even}) \\ \end{array} \right)\right\},  z\rightarrow (\infty,0),
 \label{aspaivp}
\\
\mu_2=\frac{1}{C(n,t)}\left\{I-Q(n,t)Z+\left( \begin{array}{cc} O(z^{2},\text{even}) & O(z^{-3},\text{odd})  \\
  O(z^{3},\text{odd}) & O(z^{-2},\text{even}) \\ \end{array} \right)\right\},  z\rightarrow (0,\infty),
\label{aspbivp}
\end{eqnarray}
\label{asp}
\end{subequations}
where ``even" indicates that the higher-order terms are even powers of $z$ or $z^{-1}$, while ``odd" indicates that higher-order terms are odd powers,
and the notation $z\rightarrow (\infty,0)$ ($z\rightarrow (0,\infty)$) means $z\rightarrow \infty$ ($z\rightarrow 0$) for the first column of $\mu_j(n,t,z)$ and $z\rightarrow 0$ ($z\rightarrow \infty$) for the second column of $\mu_j(n,t,z)$.

\subsection{The inverse problem}
\subsubsection{Spectral functions}
The $n$-part of the Lax pair (\ref{LPS}) implies the condition $\det\Phi(n+1,t,z)=\det\Phi(n,t,z)$.
This fact together with the limits $\mu_{1}(n,t,z)\rightarrow I$ as $n\rightarrow - \infty$
and $\mu_{2}(n,t,z)\rightarrow I$ as $n\rightarrow \infty$, imply
\begin{eqnarray}
\det\Phi_1(n,t,z)=\det\Phi_2(n,t,z)=1.
\label{detp}
\end{eqnarray}
Since both matrices $\Phi_1(n,t,z)$ and $\Phi_2(n,t,z)$ are fundamental solutions of the same linear spectral problem (\ref{LPS}),
they are related by a matrix $A(z)$, namely,
\begin{eqnarray}
\Phi_1(n,t,z)=\Phi_2(n,t,z)A(z),\quad |z|=1.
\label{s1}
\end{eqnarray}
Thus,
\begin{eqnarray}
\mu_1(n,t,z)=\mu_2(n,t,z)\hat{Z}^ne^{i\omega(z)t\hat{\sigma_3}}A(z),\quad |z|=1.
\label{s2}
\end{eqnarray}
The entries of the $2\times 2$ matrix $A(z)=(a_{jk})_{2\times 2}$, $j,k=1,2$, will be referred to as the spectral functions.
Equation (\ref{detp}) implies
\begin{eqnarray}
\det A(z)=1.
\label{detA}
\end{eqnarray}
\\
{\bf Remark 1.}
In \cite{BH1,AL3}, the determinants of the eigenfunctions and of the matrix $A(z)$ depend on the potential and on the independent variables.
Here, these determinants equal to $1$;
this property is convenient for the formulation of the associated RH problem and for the proof of the vanishing lemma.\\

We note that $A(z)$ is $t$-independent, since $A_t(z)=\lim_{n\rightarrow \infty}\Psi_{1,t}(n,t,z)=0$.

Evaluating (\ref{s2}) at $n\rightarrow \infty$ yields
\begin{eqnarray}
A(z)=\frac{1}{C(-\infty)}I+Z^{-1}\sum_{m=-\infty}^{\infty} \frac{1}{C(m,t)}\hat{Z}^{-m}e^{-i\omega(z)t\hat{\sigma_3}}\left(Q(m,t)\mu_{1}(m,t,z)\right).
\end{eqnarray}
By using (\ref{s1}) we obtain the Wronskian representations for the spectral functions:
\begin{eqnarray}
A(z)=\left( \begin{array}{cc} Wr(\Phi^L_1(n,t,z),\Phi^R_2(n,t,z)) & Wr(\Phi^R_1(n,t,z),\Phi^R_2(n,t,z)) \\
  -Wr(\Phi^L_1(n,t,z),\Phi^L_2(n,t,z)) & -Wr(\Phi^R_1(n,t,z),\Phi^L_2(n,t,z)) \\ \end{array} \right).
\label{Wr}
\end{eqnarray}
The above formula implies that $a_{11}(z)$ and $a_{22}(z)$ have analytic continuation to the domains $|z|>1$ and $|z|<1$ respectively.
Moreover, substituting the $z$ expansions (\ref{asp}) into (\ref{Wr}), we find
\begin{eqnarray}
\begin{split}
a_{11}(z)&=\frac{1}{C(-\infty)}+O(z^{-2},even), \quad z\rightarrow \infty,
\\
a_{22}(z)&=\frac{1}{C(-\infty)}+O(z^{2},even), \quad z\rightarrow 0,
\end{split}
\label{aasp}
\end{eqnarray}
which imply that $a_{11}(z)$ and $a_{22}(z)$ are even functions of $z$.

It is convenient to introduce the reflection coefficients
\begin{eqnarray}
\rho_1(z)=\frac{a_{21}(z)}{a_{11}(z)}, \quad \rho_2(z)=\frac{a_{12}(z)}{a_{22}(z)}.
\label{ro}
\end{eqnarray}

\subsubsection{Symmetries}
Equations (\ref{asp}) and (\ref{Wr}) imply the following symmetry conditions:
\begin{eqnarray}
a_{11}(-z)=a_{11}(z),\quad a_{12}(-z)=-a_{12}(z), \quad a_{21}(-z)=-a_{21}(z),\quad a_{22}(-z)=a_{22}(z),
\label{srivp}
\end{eqnarray}
and therefore
\begin{eqnarray}
\rho_1(-z)=-\rho_1(z),\quad \rho_2(-z)=-\rho_2(z).
\label{sr1ivp}
\end{eqnarray}
For the cases of the DNLS and DMKDV equations, there exist certain relations between $p(n,t)$ and $q(n,t)$,
and these relations imply certain additional symmetry conditions.

\textbf{DNLS equation}
It can be checked directly that for $p(n,t)=\nu q^\ast(n,t)$, furthermore, if $\Phi(n,t,z)$ is a solution of (\ref{LPS}),
then $\hat{\Phi}(n,t)=\sigma_{\nu}\Phi^\ast(n,t,\frac{1}{z^\ast})$ with $\sigma_\nu=\left( \begin{array}{cc} 0 & 1 \\
 \nu &  0 \\ \end{array} \right)$ is also a solution of (\ref{LPS}).
Thus, the eigenfunctions satisfy the following symmetry relations:
\begin{subequations}
\begin{eqnarray}
\Phi_j^L(n,t,z)=\sigma_v\left(\Phi_j^R(n,t,\frac{1}{z^\ast})\right)^*,  \quad j=1,2,
\label{pra}
\\
 \quad \Phi_j^R(n,t,z)=\nu\sigma_v\left(\Phi_j^L(n,t,\frac{1}{z^\ast})\right)^*, \quad j=1,2.
\label{prb}
\end{eqnarray}
\label{pr}
\end{subequations}
These equations imply the following symmetries for the spectral functions:
\begin{eqnarray}
a_{22}(z)=a^{\ast}_{11}(\frac{1}{z^\ast}), \quad a_{21}(z)=\nu a^{\ast}_{12}(\frac{1}{z^\ast}), \quad \rho_2(z)=\nu\rho^{\ast}_1(\frac{1}{z^\ast}).
\label{ror}
\end{eqnarray}
\\
{\bf Remark 2.} In the focusing case ($\nu=-1$),
the symmetry $p(n,t)= -q^\ast(n,t)$ yields $f(n,t)=\sqrt{1+|q(n,t)|^2}$, which implies that $f(n,t)$ is a nonvanishing real function.
In the defocusing case ($\nu=1$), the symmetry $p(n,t)= q^\ast(n,t)$ yields $f(n,t)=\sqrt{1-|q(n,t)|^2}$,
thus it is necessary to require $\mid q(n,t)\mid<1$ in order to guarantee that $f(n,t)$ is a nonvanishing real function such that the symmetries (\ref{pr}) and (\ref{ror}) are valid.
We note that if the potential $q(n,t)$ evolves according to (\ref{nls}), then this condition is time-invariant.
Indeed, by employing equation (\ref{nls}) we find
\begin{eqnarray}
\left(1-|q(n,t)|^2\right)_t=\left(1-|q(n,t)|^2\right)\left\{-2\nu \text{Im} \left[q(n,t)\left(q^*(n-1,t)+q^*(n+1,t)\right)\right]\right\},
\label{remark1}
\end{eqnarray}
where the notation $\text{Im}(\cdot)$ denotes the imaginary part of the complex-valued function $(\cdot)$.
Equation (\ref{remark1}) implies
\begin{eqnarray}
\left(1-|q(n,t)|^2\right)=\left(1-|q(n,0)|^2\right)e^{-2\nu \int_{0}^t\text{Im} \left[q(n,t')\left(q^*(n-1,t')+q^*(n+1,t')\right)\right]dt'}.
\label{remark2}
\end{eqnarray}
Hence if the modulus of initial value $\mid q(n,0)\mid<1$ then $\mid q(n,t)\mid<1$ for all $t>0$.\\

\vspace{0.2cm}

\textbf{DMKDV equation}
For the DMKDV equation, we have similar symmetry relations with those of the DNLS equation.
In fact, under the reduction $p(n,t)=\nu q(n,t)$ with $q(n,t)$ real, if $\Phi(n,t,z)$ is a solution of (\ref{LPS}),
then $\hat{\Phi}(n,t)=\sigma_{\nu}\Phi^\ast(n,t,\frac{1}{z^\ast})$ is also a solution of (\ref{LPS}).
The symmetry relations for the eigenfunctions and spectral functions of the DMKDV equation have the same form as (\ref{pr}) and (\ref{ror}).
\vspace{0.2cm}
\\
{\bf Remark 3.}
The symmetries (\ref{prb}) between the first and second columns involve the transformation $z\rightarrow \frac{1}{z^\ast}$
which maps the point $z=0$ to $z=\infty$ for the second column.
Thus in what follows we will specify the boundary conditions for the associated RH problems as $z\rightarrow(\infty,0)$.
Here, as before, $z\rightarrow(\infty,0)$ denotes $z\rightarrow\infty$ for the first column and $z\rightarrow 0$ for the second column of
a $2\times 2$ matrix function respectively.
This is a significant difference with respect to the case of continuous problems. For example, for the integrable NLS equation,
the relevant transformation is $z\rightarrow z^\ast$, thus one only needs to consider the boundary conditions for both columns of
the associated RH problems  as $z\rightarrow\infty$.

\subsubsection{A Riemann-Hilbert problem}
We define the functions $M_{-}(n,t,z)$ and $M_{+}(n,t,z)$ as follows:
\begin{eqnarray}
\begin{split}
M_{-}(n,t,z)=\frac{1}{C(n,t)}\left(\frac{\mu_1^L(n,t,z)}{a_{11}(z)},\mu_2^R(n,t,z)\right), \quad |z|\geq 1,
 \\
M_{+}(n,t,z)=\frac{1}{C(n,t)}\left(\mu_2^L(n,t,z),\frac{\mu_1^R(n,t,z)}{a_{22}(z)}\right), \quad |z|\leq 1.
\end{split}
\label{M}
\end{eqnarray}
Simple algebraic manipulations imply the following jump condition:
\begin{eqnarray}
M_{-}(n,t,z)=M_{+}(n,t,z)J(n,t,z), \quad |z|=1,
\label{Jc}
\end{eqnarray}
where
\begin{eqnarray}
J(n,t,z)=\left( \begin{array}{cc} 1-\rho_1(z)\rho_2(z) & -z^{2n}e^{2iw(z)t}\rho_2(z) \\
  z^{-2n}e^{-2iw(z)t}\rho_1(z) & 1 \\ \end{array} \right).
\label{Jr}
\end{eqnarray}

Equations (\ref{asp}) and (\ref{aasp}) imply that
\begin{eqnarray}
\begin{split}
M(n,t,z)=I+Q(n-1)Z^{-1}
+\left( \begin{array}{cc} O(z^{-2},\text{even}) & O(z^{3},\text{odd})  \\
  O(z^{-3},\text{odd}) & O(z^{2},\text{even}) \\ \end{array} \right), ~ z\rightarrow (\infty,0).
\end{split}
\label{MA2}
\end{eqnarray}
The solution of the RH problem (\ref{Jc}) can be represented in the form
\begin{eqnarray}
M(n,t,z)=I+\frac{1}{2\pi i}\int_{|z|=1}M_{+}(n,t,\xi)\hat{J}(n,t,\xi)\frac{1}{\xi-z}d\xi,
\label{MS}
\end{eqnarray}
where $\hat{J}(n,t,z)=I-J(n,t,z)$.
Equation (\ref{MS}) implies that as $z\rightarrow 0$,
\begin{eqnarray}
\begin{split}
M(n,t,z)=&I+\frac{1}{2\pi i}\int_{|z|=1}M_{+}(n,t,\xi)\hat{J}(n,t,\xi)\frac{1}{\xi}d\xi+\left(\frac{1}{2\pi i}\int_{|z|=1}M_{+}(n,t,\xi)\hat{J}(n,t,\xi)\frac{1}{\xi^2}d\xi\right)z
\\&+O(z^2).
\end{split}
\label{MS0}
\end{eqnarray}
By comparing the $(1,2)$-entries of equations (\ref{MA2}) and (\ref{MS}),
we arrive at the reconstruction formula for the solutions of the DNLS and DMKDV equations in terms of the solution of the associated RH problem:
\begin{eqnarray}
q(n,t)=\frac{1}{2\pi i}\int_{|z|=1}z^{2n}e^{2iw(z)t}\rho_2(z)\left(M_{+}(n+1,t,z)\right)_{11}dz,
\label{IVPS}
\end{eqnarray}
where, $\omega(z)$ is defined by (\ref{nlsw}) for the DNLS equation, and by (\ref{mkdvw}) for the DMKDV equation, respectively.

\subsubsection{Residue conditions}

\textbf{Residue conditions for the Ablowitz-Ladik lattice (\ref{aleq})}
\\Since $a_{jj}(z)$, $j=1,2$, are even functions, each zero $\xi_k$ of $a_{jj}(z)$ is accompanied by another zero
at $-\xi_k$ .
We make the following assumptions about these zeros:
\begin{itemize}
\item In $|z|>1$, $a_{11}(z)$ has $2\kappa$ simple zeros $\{z_j\}_{1}^{2\kappa}$ such that $z_{j+\kappa}=-z_{j}$, $j=1,\cdots,\kappa$;
\item In $|z|<1$, $a_{22}(z)$ has $2\kappa'$ simple zeros $\{\tilde{z}_j\}_{1}^{2\kappa'}$ such that $\tilde{z}_{j+\kappa'}=-\tilde{z}_{j}$, $j=1,\cdots,\kappa'$.
\end{itemize}

From the Wronskian representations (\ref{Wr}), we conclude that there exist constants $b_j$ and $\tilde{b}_j$ such that
\begin{eqnarray}
\Phi_1^L(n,z_j,t)=b_j\Phi_2^R(n,z_j,t),\quad \Phi_1^R(n,\tilde{z}_j,t)=\tilde{b}_j\Phi_2^L(n,\tilde{z}_j,t).
\label{sc1}
\end{eqnarray}
Hence,
\begin{eqnarray}
\begin{split}
\mu_1^L(n,z_j,t)=b_jz_j^{-2n}e^{-2i\omega(z_j)t}\mu_2^R(n,z_j,t), 
\\
\mu_1^R(n,\tilde{z}_j,t)=\tilde{b}_j\tilde{z}_j^{2n}e^{2i\omega(\tilde{z}_j)t}\mu_2^L(n,\tilde{z}_j,t). 
\end{split}
\label{sc2}
\end{eqnarray}
Using (\ref{sc2}), we find the following residues relations:
\begin{eqnarray}
\begin{split}
{\text{Res}}_{z=z_j}M_{-}^{L}(n,t,z)=c_jM_{-}^{R}(n,t,z_j), 
\\
{\text{Res}}_{z=\tilde{z}_j}M_{+}^{R}(n,t,z)=\tilde{c}_jM_{+}^{L}(n,t,\tilde{z}_j), 
\end{split}
\label{rr1}
\end{eqnarray}
where
\begin{eqnarray}
c_j=\frac{b_jz_j^{-2n}e^{-2i\omega(z_j)t}}{\dot{a}_{11}(z_j)},\quad \tilde{c}_j=\frac{\tilde{b}_j\tilde{z}_j^{2n}e^{2i\omega(\tilde{z}_j)t}}{\dot{a}_{22}(\tilde{z}_j)}.
\label{cj}
\end{eqnarray}

\noindent\textbf{Residue conditions for the DNLS and DMKDV equations}
\\For the DNLS and DMKDV equations, the symmetries (\ref{ror}) imply $\tilde{z_j}=\frac{1}{z_j^{\ast}}$, $\kappa=\kappa'$, $\tilde{b}_j=\nu b_j^{\ast}$, and $\tilde{c}_j=\nu c_j^\ast$.
Thus, the residues relations (\ref{rr1}) become
\begin{eqnarray}
\begin{split}
{Res}_{z=z_j}M_{-}^{L}(n,t,z)=c_jM_{-}^{R}(n,t,z_j), 
\\
{Res}_{z=\frac{1}{z_j^{\ast}}}M_{+}^{R}(n,t,z)=\nu c_j^\ast M_{+}^{L}(n,t,\frac{1}{z_j^{\ast}}), 
\end{split}
\label{rr2}
\end{eqnarray}
with $c_j$ given by (\ref{cj}), where $\omega(z_j)$ is given by equations (\ref{nlsw}) and (\ref{mkdvw}) for the DNLS and DMKDV equations respectively.

\subsubsection{A Vanishing Lemma}

If $a_{11}(z)$ has no zeros, the RH problem (\ref{Jc}) is regular. The unique solvability of this RH
problem is a consequence of the existence of a vanishing lemma (see, for example, \cite{F2} and \cite{Zhou}); i.e., the RH problem obtained from the above RH problem
by replacing the boundary condition $M(n,t,z)\rightarrow I$ as $z\rightarrow (\infty,0)$ with the boundary condition $M(n,t,z)\rightarrow 0$ as $z\rightarrow (\infty,0)$,
has only the trivial solution.
The vanishing lemma will be established in the following section by using the symmetry properties of $J(n,t,z)$.
If $a_{11}(z)$ has zeros,  the RH problem (\ref{Jc}) is singular.
The singular RH problem can be mapped to a regular one coupled with
a system of algebraic equations via the approach developed in \cite{F2}.

We now establish the vanishing lemma for the DNLS and DMKDV equations. Let $J^{\dagger}(n,t,z)$ denote the Hermitian (conjugate transpose) of the matrix $J(n,t,z)$.
Both the DNLS and DMKDV equations possess the symmetry relation $\rho_2(z)=\nu\rho^{\ast}_1(\frac{1}{z^\ast})$.
On the unit circle $|z|=1$, we have $z=\frac{1}{z^\ast}$. Thus, we conclude that $\omega^\ast(z)=\omega(z)$, and $\rho_2(z)=\nu\rho^{\ast}_1(z)$ hold on $|z|=1$
for both the DNLS and DMKDV equations.

\begin{lemma}
For both the DNLS and DMKDV equations, the matrix $J(n,t,z)+J^{\dagger}(n,t,z)$ is positive definite on $|z|=1$.
\end{lemma}
{\bf Proof} \quad
We use the subscript $jk$ to denote the $(j, k)$-element of a matrix.
Direct calculations yield
 \begin{eqnarray}
\left(J(n,t,z)+J^{\dagger}(n,t,z)\right)_{11}=2\left(1-\nu |\rho_1|^2\right)=\frac{2}{|a_{11}(z)|^2}>0,
\label{11}
\end{eqnarray}
where we have used the important fact that $\det A=1$.
Moreover, for $\nu=-1$, we have
\begin{eqnarray}
\det\left(J(n,t,z)+J^{\dagger}(n,t,z)\right)=4>0,
\label{det1}
\end{eqnarray}
and for $\nu=1$, we have
\begin{eqnarray}
\det\left(J(n,t,z)+J^{\dagger}(n,t,z)\right)=\frac{4}{|a_{11}(z)|^2}>0.
\label{det1}
\end{eqnarray}
The above formulas imply that the eigenvalues of $J(n,t,z)+J^{\dagger}(n,t,z)$ are positive. This completes the proof. \QEDB

\begin{lemma}
(Vanishing lemma) The RH problem in (\ref{Jc}) and (\ref{Jr})
with the vanishing boundary condition $M(n,t,z)\rightarrow 0$ as $z\rightarrow (\infty,0)$, has only the zero solution.
\end{lemma}
{\bf Proof} \quad
The relevant symmetries imply that if a function $F(z)$ is analytic inside (outside) the unit circle,
then the function $F^*(\frac{1}{z^\ast})$ is analytic outside (inside) the unit circle.
This implies that we can define analytic functions inside and outside the unit circle by
\begin{eqnarray}
\begin{split}
H_{+}(z)=M_{+}(z)M_{-}^{\dagger}(\frac{1}{z^\ast})+M_{-}(\frac{1}{z^\ast})M_{+}^{\dagger}(z), \quad |z|\leq 1,
\\
H_{-}(z)=M_{-}(z)M_{+}^{\dagger}(\frac{1}{z^\ast})+M_{+}(\frac{1}{z^\ast})M_{-}^{\dagger}(z), \quad |z|\geq 1,
\end{split}
\label{HF}
\end{eqnarray}
where we have suppressed the $n$ and $t$ dependence.
On the contour $|z|=1$, we have
\begin{eqnarray}
\begin{split}
H_{+}(z)=M_{+}(z)\left(J^{\dagger}(z)+J(z)\right)M_{+}^{\dagger}(z), \quad |z|= 1,
\\
H_{-}(z)=M_{+}(z)\left(J(z)+J^{\dagger}(z)\right)M_{+}^{\dagger}(z), \quad |z|= 1,
\end{split}
\label{HF2}
\end{eqnarray}
which implies $H_{+}(z)=H_{-}(z)$ on $|z|=1$.
Thus, we have constructed an entire function $H(z)$ vanishing as as $z\rightarrow (\infty,0)$.
Therefore, $H_{+}(z)$ and $H_{-}(z)$ are identically zero.
Hence, equation (\ref{HF2}) immediately yields
\begin{eqnarray}
M_{+}(z)\left(J(z)+J^{\dagger}(z)\right)M_{+}^{\dagger}(z)=0, \quad |z|= 1.
\label{HF3}
\end{eqnarray}
Since $J(z)+J^{\dagger}(z)$ is a positive definite Hermitian matrix on $|z|=1$, we deduce that
$M_{+}(z)=0$ on $|z|=1$. By analytic continuation, $M_{+}(z)$ and $M_{-}(z)$ vanish identically. This completes the proof. \QEDB

\section{Initial-boundary value problems}

In this section we study IBVPs for the DNLS and DMKDV  equations, namely,
given the initial datum $q_0(n)=q(n,0)$ and an appropriate boundary condition $g_{-1}(t)=q(-1,t)$,
we will solve equations (\ref{nls}) and (\ref{mkdv}) with $n\in\mathbb{N}$ and $0<t<T$, where $\mathbb{N}$ denotes the set of naturals.


In analogy with the integrable PDEs on the half line, we will analyze the DNLS and DMKDV  equations on the naturals by the following three steps.

{\it Step 1. A RH formulation under the assumption of existence.} We assume that there exists a solution $q(n,t)$.
By performing the simultaneous spectral analysis of the Lax pair (\ref{LPST3}), we express $q(n,t)$ in terms of the solution of a $2 \times 2$-matrix RH problem
defined in the complex $z$-plane. This RH problem has explicit $(n,t)$ dependence in the form of
$z^{2n}e^{2i\omega(z)t}$, and it is uniquely defined in terms of the spectral functions $\{a(z),b(z)\}$ and $\{A(z),B(z)\}$.
The spectral functions $\{a(z),b(z)\}$ are expressed in terms of the initial datum $q_0(n)=q(n,0)$,
 while $\{A(z),B(z)\}$ are expressed in terms of the  boundary values $g_{-1}(t)$ and  $g_{0}(t)$, where $g_{-1}(t)=q(-1,0)$ denotes the known boundary datum and
$g_{0}(t)=q(0,t)$ denotes the unknown boundary datum.
 We show that the spectral functions are not independent but satisfy an algebraic relation called global relation.

{\it Step 2. Existence under the assumption that the spectral functions satisfy the global relation.}
We define the spectral functions $\{a(z),b(z)\}$ in terms of the
initial datum $q_0(n)$ and define the spectral functions $\{A(z),B(z)\}$ in terms of the boundary values $g_{-1}(t)$ and $g_{0}(t)$.
We assume that the boundary values are such that the spectral functions satisfy
the global relation. We also define $q(n,t)$ in terms of the solutions of the RH problems
formulated in step 1. We then prove that the formulae for $q(n,t)$ satisfy the DNLS and DMKDV equations,
and furthermore satisfy $q(n,0)=q_0(n)$ and
$q(-1,t)=g_{-1}(t)$, $q(0,t)=g_{0}(t)$.

{\it Step 3. Elimination of the unknown boundary values.}
Given $q_0(n)$ and $g_{-1}(t)$,  by employing the global relation, we characterize the unknown boundary value $g_0(t)$.

In sections 4.1 and 4.2 we implement the first two steps for the DNLS and DMKDV equations.
The third step for these two equations is discussed in section 5.

As mentioned earlier, in \cite{BH1} the authors have implemented the first step for the DNLS equation
(but the second step was not implemented).
Here the implementation of the first step  follows the work of \cite{BH1}.
However, in comparison with the results in \cite{BH1}, the main improvement we have made in the first step is
that the relevant RH problems have explicit $(n, t)$ dependence
and are expressed only in terms of the spectral functions; see section 4.1.2.3.
As a result, our solutions of the IBVPs depend only on the spectral functions
(while the expression of the solution constructed in \cite{BH1} involves the unknown potentials).
This improvement is very important for the implementation of the second step; see section 4.2.


\subsection{ A Riemann-Hilbert formulation under the assumption of existence}

\subsubsection{The direct problem}
We will construct a sectionally holomorphic solution $\mu(n,t,z)$ in terms of $q(n,t)$.
In analogy with the integrable PDEs on the half line (see, for example, \cite{F1}),
we define three different eigenfunctions $\{\mu_j(n,z,t)\}_{1}^3$ as simultaneous solutions
of both parts of the Lax pair (\ref{LPST3}) with normalizations at $(n,t)=(0,0)$, at $(n,t)=(\infty,t)$, and at $(n,t)=(0,T)$, respectively.
By choosing the paths to be parallel to the $n$ and $t$ axes, these eigenfunctions are given by the following formulae:
\begin{eqnarray}
\begin{split}
\mu_{1}(n,t,z)=&\frac{C(n,t)}{C(0,t)}\left(I+\hat{Z}^{n}\int_{0}^{t}e^{iw(z)(t-t')\hat{\sigma}_3}\left(H\mu_{1}(0,t',z)\right)dt'\right)
\\&+C(n,t)Z^{-1}\sum_{m=0}^{n-1} \frac{1}{C(m,t)}\hat{Z}^{n-m}(Q(m,t)\mu_{1}(m,t,z)),
 \\
\mu_{2}(n,t,z)=&C(n,t)\left(I-Z^{-1}\sum_{m=n}^{\infty} \frac{1}{C(m,t)}\hat{Z}^{n-m}(Q(m,t)\mu_{2}(m,t,z))\right),
\\
\mu_{3}(n,t,z)=&\frac{C(n,t)}{C(0,t)}\left(I-\hat{Z}^{n}\int_{t}^{T}e^{iw(z)(t-t')\hat{\sigma}_3}\left(H\mu_{3}(0,t',z)\right)dt'\right)
\\&+C(n,t)Z^{-1}\sum_{m=0}^{n-1} \frac{1}{C(m,t)}\hat{Z}^{n-m}(Q(m,t)\mu_{3}(m,t,z)).
\end{split}
\label{muibv}
\end{eqnarray}

As before, let $\mu_j^L(n,t,z)$ and $\mu_j^R(n,t,z)$, $j=1,2,3$, denote the first and second columns of $\mu_j(n,t,z)$, $j=1,2,3$, respectively.
The eigenfunction $\mu_2(n,t,z)$ coincides with the eigenfunction in section 3 (see (\ref{mub})). It follows that $\mu_2^R(n,t,z)$ is bounded for $|z|\geq 1$.
The second column of $\mu_1(n,t,z)$, $\mu_1^R(n,t,z)$ involves  $\exp[2i\omega(z)(t-t')]$, $t-t'\geq 0$ and $z^{2(n-m)-1}$, $n-m\geq1$.
This exponential term is bounded in the domain of the complex $z$-plane:
$\left\{\text{Im} (\omega(z))\geq 0\right\}$.
As a result, $\mu_1^R(n,t,z)$ is bounded for $\left\{|z|\leq 1 \cap\text{Im} (\omega(z))\geq 0\right\}$.
The second column of $\mu_3(n,t,z)$, $\mu_3^R(n,t,z)$ involves $\exp[2i\omega(z)(t-t')]$, $t-t'\leq 0$ and $z^{2(n-m)-1}$, $n-m\geq1$. This exponential is bounded for
$\left\{\text{Im} (\omega(z))\leq0\right\}$. As a result, $\mu_3^R(n,t,z)$ is bounded for $\left\{|z|\leq 1 \cap\text{Im} (\omega(z))\leq 0\right\}$.
Similar arguments apply for  $\mu_j^L(n,t,z)$, $j=1,2,3$.

We define the domains as follows:
\begin{eqnarray}
\begin{split}
&D_{in}=\left\{z\in\mathbb{C} \Big| |z|< 1 \right\},
~~D_{out}=\left\{z\in\mathbb{C} \Big| |z|> 1 \right\},
\\
&D_{+}=\left\{z\in\mathbb{C} \Big| \text{Im} (\omega(z))>0\right\},
~D_{-}=\left\{z\in\mathbb{C} \Big| \text{Im} (\omega(z))<0\right\},
\\
&D_{+in}=\left\{z\in\mathbb{C}\Big|  |z|< 1 \cap \text{Im} (\omega(z))>0\right\},
\\
&D_{-in}=\left\{z\in\mathbb{C}\Big| |z|< 1 \cap \text{Im} (\omega(z))<0\right\},
\\
&D_{+out}=\left\{z\in\mathbb{C}\Big| |z|> 1 \cap \text{Im} (\omega(z))>0\right\},
\\
&D_{-out}=\left\{z\in\mathbb{C}\Big| |z|> 1 \cap \text{Im} (\omega(z))<0\right\}.
\end{split}
\label{D11}
\end{eqnarray}
Recall that $\omega(z)$ is defined by (\ref{nlsw}) for the DNLS, and by (\ref{mkdvw}) for the DMKDV.
Then, the domains for the DNLS equation become (see Figure 1):
\begin{subequations}
\begin{eqnarray}
D_{+in}&=&\left\{z\in\mathbb{C} \Big|  |z|<1, \arg z\in(\frac{\pi}{2},\pi)\cup (\frac{3\pi}{2},2\pi)\right\},
\label{D1a}
\\
D_{-in}&=&\left\{z\in\mathbb{C} \Big|  |z|<1, \arg z\in(0,\frac{\pi}{2})\cup (\pi,\frac{3\pi}{2})\right\},
\label{D1b}
\\
D_{+out}&=&\left\{z\in\mathbb{C} \Big|  |z|>1, \arg z\in(0,\frac{\pi}{2})\cup (\pi,\frac{3\pi}{2})\right\},
\label{D1c}
\\
D_{-out}&=&\left\{z\in\mathbb{C} \Big|  |z|>1, \arg z\in(\frac{\pi}{2},\pi)\cup (\frac{3\pi}{2},2\pi)\right\},
\label{D1d}
\end{eqnarray}
\label{D1}
\end{subequations}
whereas the domains for the DMKDV equation become (see Figure 3):
\begin{subequations}
\begin{eqnarray}
D_{+in}(z)&=&\left\{z\in\mathbb{C} \Big|  |z|<1, \arg z\in(\frac{\pi}{4},\frac{3\pi}{4})\cup (\frac{5\pi}{4},\frac{7\pi}{4})\right\},
\label{D2a}
\\
D_{-in}(z)&=&\left\{z\in\mathbb{C} \Big|  |z|<1, \arg z\in(-\frac{\pi}{4},\frac{\pi}{4})\cup (\frac{3\pi}{4},\frac{5\pi}{4})\right\},
\label{D2b}
\\
D_{+out}(z)&=&\left\{z\in\mathbb{C} \Big|  |z|>1, \arg z\in(-\frac{\pi}{4},\frac{\pi}{4})\cup (\frac{3\pi}{4},\frac{5\pi}{4})\right\},
\label{D2c}
\\
D_{-out}(z)&=&\left\{z\in\mathbb{C} \Big|  |z|>1, \arg z\in(\frac{\pi}{4},\frac{3\pi}{4})\cup (\frac{5\pi}{4},\frac{7\pi}{4})\right\}.
\label{D2d}
\end{eqnarray}
\label{D2}
\end{subequations}

For the DNLS and DMKDV equations, the above analysis implies the following results:
\begin{itemize}
 \item $\mu_{1}(n,t,z)$ and $\mu_{3}(n,t,z)$  are analytic for $z\in \mathbb{C}\setminus0$;
 \item $\mu^L_{1}(n,t,z)$ is continuous and bounded for $z\in \bar{D}_{-out}$, and $\mu^R_{1}(n,t,z)$ is continuous and bounded for $z\in \bar{D}_{+in}$;
 \item $\mu^L_{3}(n,t,z)$ is continuous and bounded for $z\in \bar{D}_{+out}$, and $\mu^R_{3}(n,t,z)$ is continuous and bounded for $z\in \bar{D}_{-in}$;
  \item $\mu^L_{2}(n,t,z)$ is analytic for $D_{in}$ and it is continuous and bounded for $z\in \bar{D}_{in}$, and $\mu^R_{2}(n,t,z)$ is analytic for $D_{out}$ and it is continuous and bounded for $z\in \bar{D}_{out}$.
\end{itemize}

The asymptotic behavior of $\mu_2(n,t,z)$ as $z\rightarrow (0,\infty)$ is given by equation (\ref{aspbivp}).
Using the Neumann series method, we can derive (see appendix A.2) the asymptotic behavior of $\mu_1(n,t,z)$ and $\mu_3(n,t,z)$:
for $n=0$, we have
\begin{eqnarray}
\begin{split}
\mu_1(0,t,z)=&E_1(t)I+\left(E_1(t)Q(-1,t)-e^{iw(z)t\hat{\sigma}_3}Q(-1,0)\right)Z^{-1}
\\&+\left( \begin{array}{cc} O(z^{-2},\text{even}) & O(z^{3},\text{odd})  \\
  O(z^{-3},\text{odd}) & O(z^{2},\text{even}) \\ \end{array} \right), \quad z\rightarrow (\infty,0),
\\
\mu_3(0,t,z)=&E_2(t)I+\left(E_2(t)Q(-1,t)-e^{iw(z)(t-T)\hat{\sigma}_3}Q(-1,T)\right)Z^{-1}
\\&+\left( \begin{array}{cc} O(z^{-2},\text{even}) & O(z^{3},\text{odd})  \\
  O(z^{-3},\text{odd}) & O(z^{2},\text{even}) \\ \end{array} \right), \quad z\rightarrow (\infty,0),
\end{split}
\label{asp2}
\end{eqnarray}
where
\begin{subequations}
\begin{eqnarray}
&E_1(t)=\exp\left(i\int_{0}^t\left(\alpha q(0,\xi)p(-1,\xi)-\beta q(-1,\xi)p(0,\xi)\right)d\xi\right),
\label{Ea}
\\
&E_2(t)=\exp\left(-i\int_{t}^T\left(\alpha q(0,\xi)p(-1,\xi)-\beta q(-1,\xi)p(0,\xi)\right)d\xi\right).
\label{Eb}
\end{eqnarray}
\label{E}
\end{subequations}
For $n\geq 1$, we find that the following asymptotic behavior:
\begin{eqnarray}
\begin{split}
\mu_1(n,t,z)&=\frac{C(n,t)}{C(0,0)}\left(I+Q(n-1,t)Z^{-1}+\left( \begin{array}{cc} O(z^{-2},\text{even}) & O(z^{3},\text{odd})  \\
  O(z^{-3},\text{odd}) & O(z^{2},\text{even}) \\ \end{array} \right)\right),  z\rightarrow (\infty,0),
\\
\mu_3(n,t,z)&=\frac{C(n,t)}{C(0,0)}E_2(0)\left(I+Q(n-1,t)Z^{-1}+\left( \begin{array}{cc} O(z^{-2},\text{even}) & O(z^{3},\text{odd})  \\
  O(z^{-3},\text{odd}) & O(z^{2},\text{even}) \\ \end{array} \right)\right),  z\rightarrow (\infty,0).
\end{split}
\label{asp1}
\end{eqnarray}

\begin{figure}
\begin{minipage}[t]{0.5\linewidth}
\centering
\includegraphics[width=2.5in]{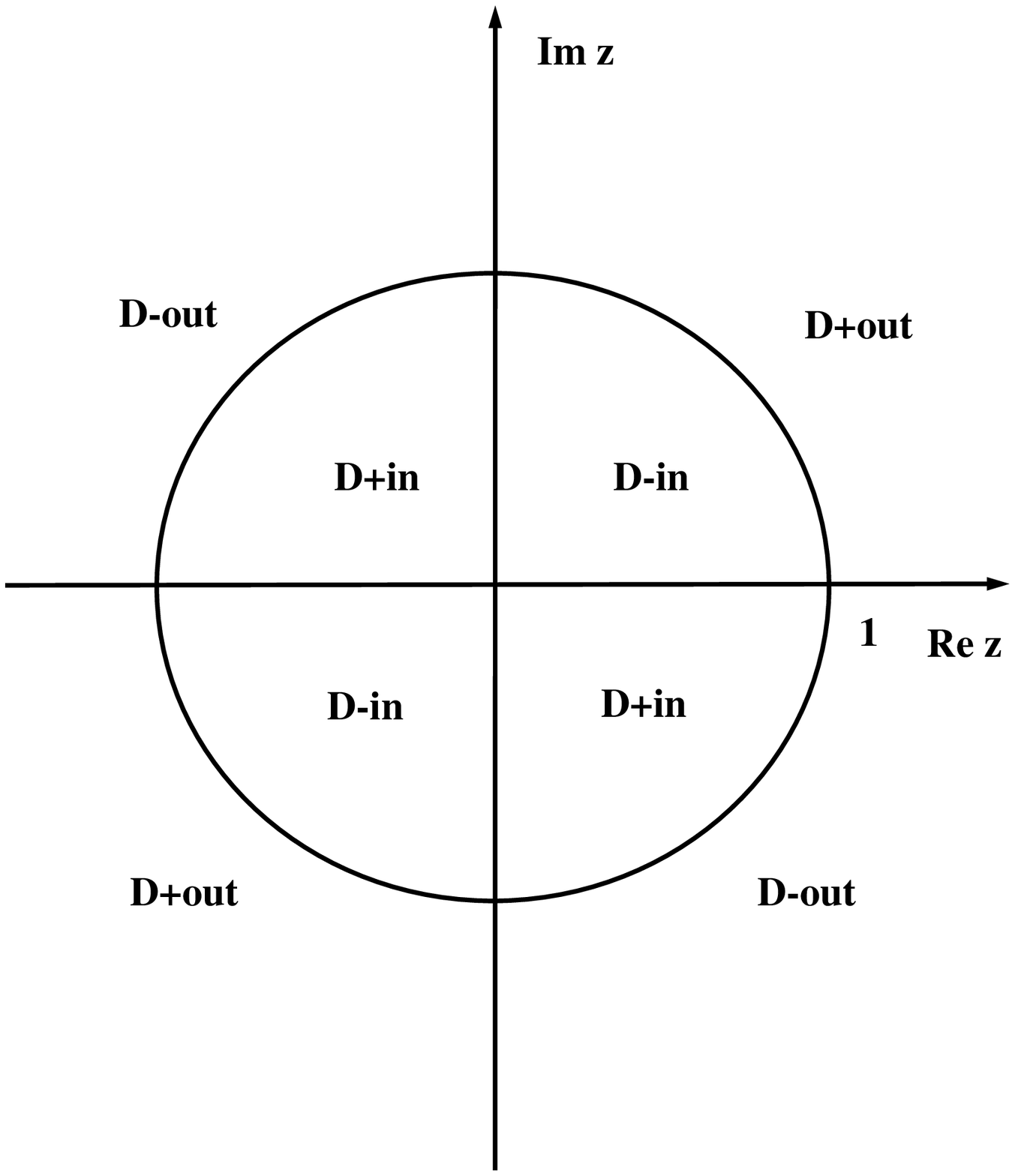}
\caption{\small{ The domains $D_{+}$ and $D_{-}$ of the $z$-plane where $\text{Im}~\omega(z)> 0$ and $\text{Im}~\omega(z)< 0$ for the DNLS equation.}}
\label{F1}
\end{minipage}
\hspace{2.0ex}
\begin{minipage}[t]{0.5\linewidth}
\centering
\includegraphics[width=2.5in]{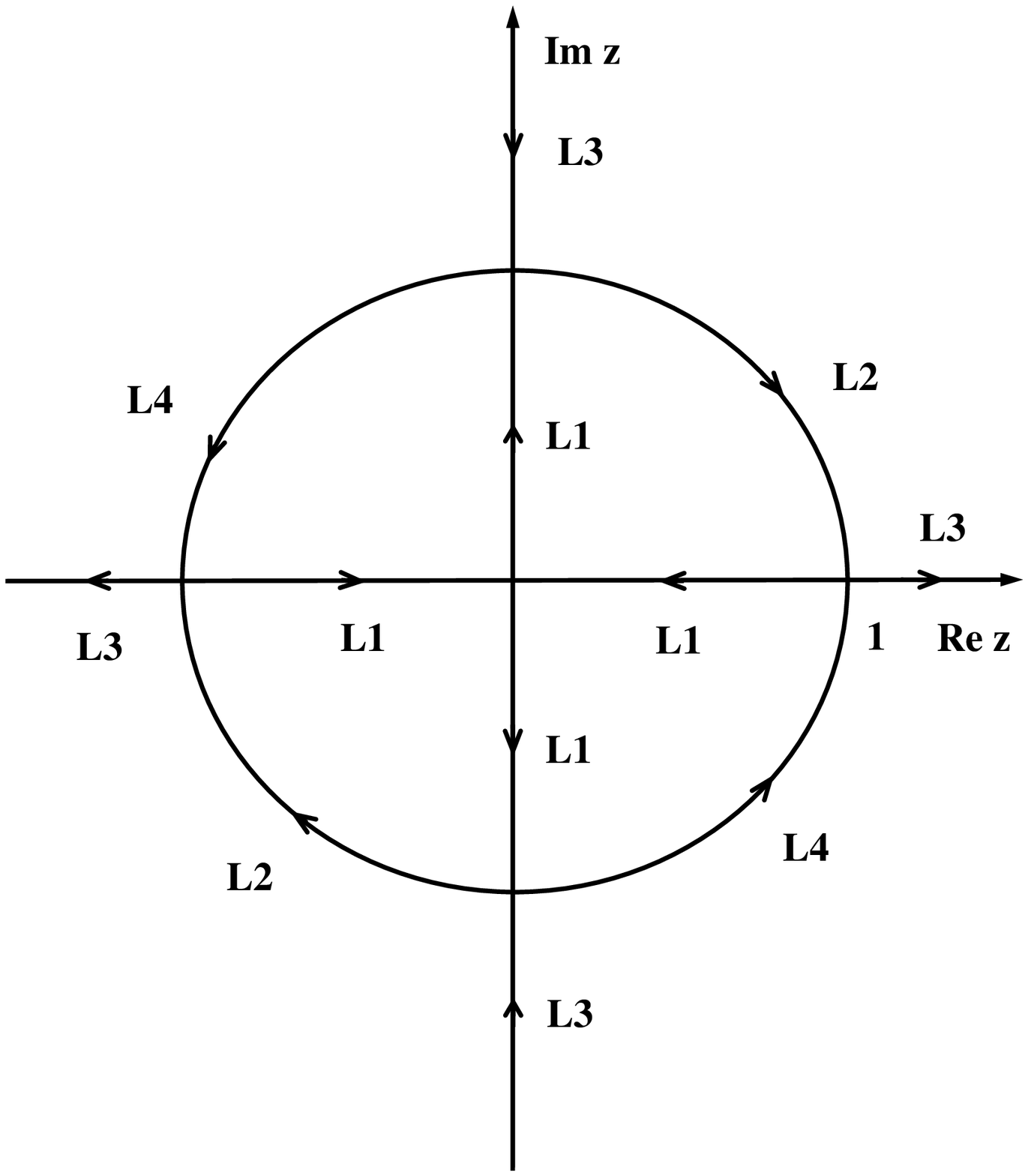}
\caption{\small{ The oriented contours $L_{1}$, $L_{2}$, $L_{3}$, $L_{4}$ that define the
RH problem for the DNLS equation.}}
\label{F2}
\end{minipage}
\end{figure}

\begin{figure}
\begin{minipage}[t]{0.5\linewidth}
\centering
\includegraphics[width=2.5in]{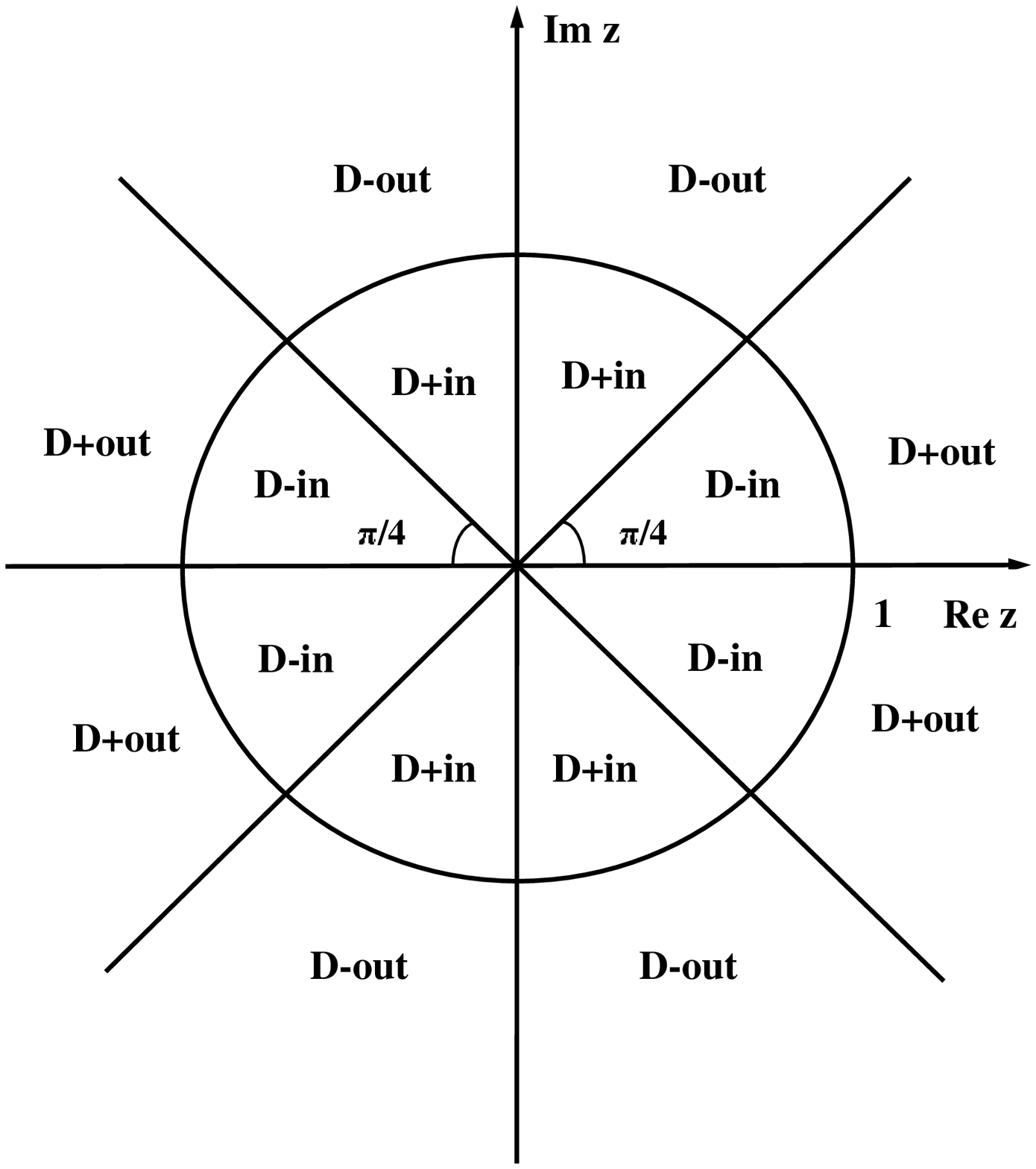}
\caption{\small{ The domains $D_{+}$ and $D_{-}$ of the $z$-plane where $\text{Im}~\omega(z)> 0$ and $\text{Im}~\omega(z)< 0$ for the DMKDV equation.}}
\label{F3}
\end{minipage}
\hspace{2.0ex}
\begin{minipage}[t]{0.5\linewidth}
\centering
\includegraphics[width=2.5in]{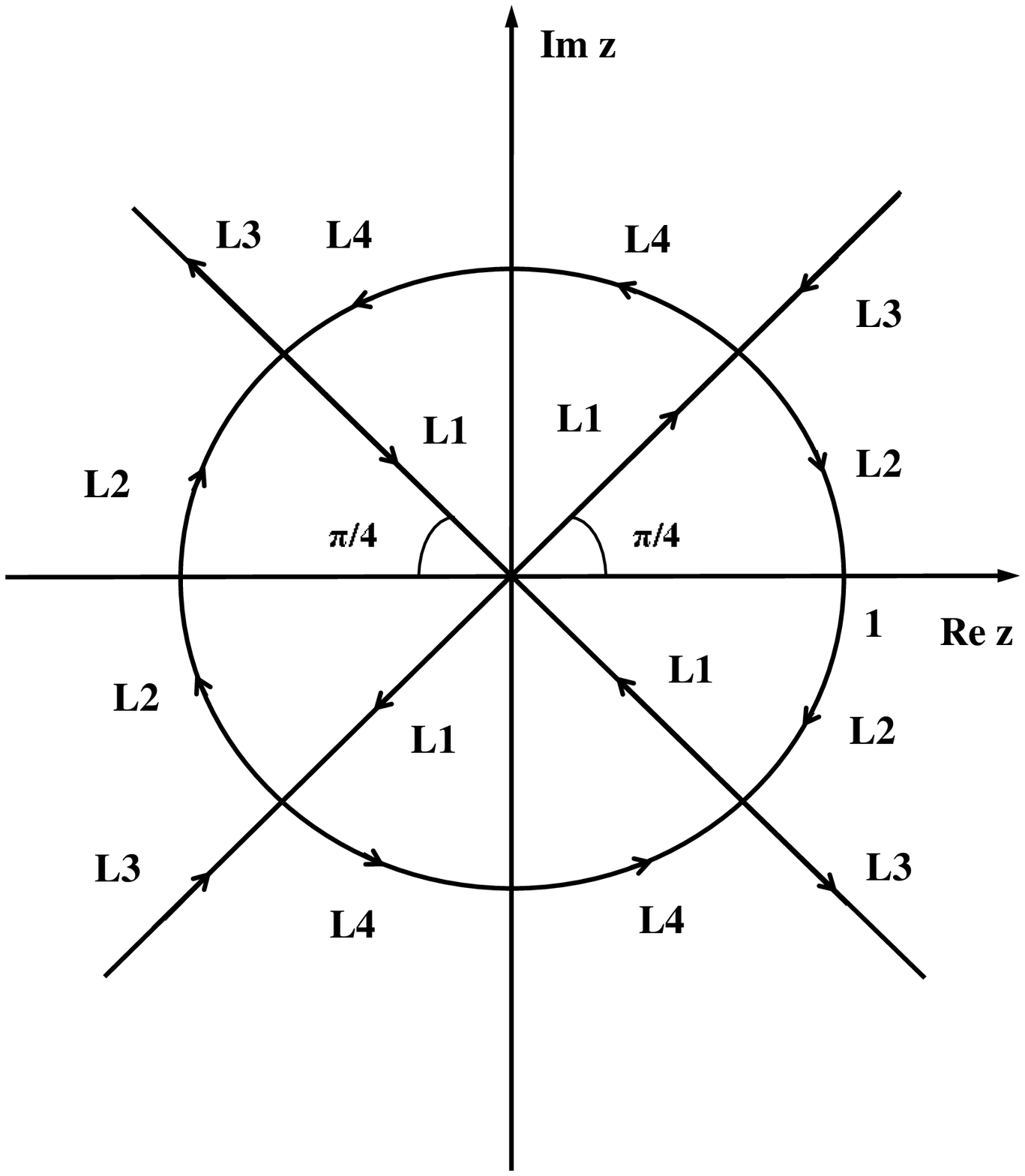}
\caption{\small{ The oriented contours $L_{1}$, $L_{2}$, $L_{3}$, $L_{4}$ that define the
RH problem for the DMKDV equation.}}
\label{F4}
\end{minipage}
\end{figure}

\subsubsection{The inverse problem}
Equation (\ref{muibv}) defines $\mu(n,t,z)$ in terms of $q(n,t)$.
Now we will express $q(n,t)$ in terms of the spectral functions.
This will be achieved by formulating a RH problem.

\subsubsection*{4.1.2.1 The spectral functions}

The $n$-part of Lax pair (\ref{LPS}) implies the equation $\det\Phi(n+1,t,z)=\det\Phi(n,t,z)$.
The matrix $V(n)=i\omega(z)\sigma_3+H(n,t,z)$ in the $t$-part of the Lax pair is traceless.
These two facts together with (\ref{muibv}) imply the important identities:
\begin{eqnarray}
\det\Phi_1(n,t,z)=\det\Phi_2(n,t,z)=\det\Phi_3(n,t,z)=1.
\label{detp1}
\end{eqnarray}

Since the matrices $\{\Phi_j(n,t,z)\}_1^3$ are fundamental solutions of the same Lax pair,
they are related:
\begin{eqnarray}
\Phi_2(n,t,z)=\Phi_1(n,t,z)s(z),\quad \Phi_3(n,t,z)=\Phi_1(n,t,z)S(z),
\label{s3}
\end{eqnarray}
and these equations hold in the domains where $(\Phi_2(n,t,z),\Phi_1(n,t,z))$ and $(\Phi_3(n,t,z),\Phi_1(n,t,z))$ are defined.
Hence, the matrices $\{\mu_j(n,t,z)\}_1^3$ satisfy the equations
\begin{subequations}
\begin{eqnarray}
\mu_2(n,t,z)&=&\mu_1(n,t,z)\hat{Z}^ne^{i\omega(z)t\hat{\sigma_3}}s(z),\label{s4}
\\
 \mu_3(n,t,z)&=&\mu_1(n,t,z)\hat{Z}^ne^{i\omega(z)t\hat{\sigma_3}}S(z).
\label{s5}
\end{eqnarray}
\label{s45}
\end{subequations}

Equation (\ref{detp1}) implies
\begin{eqnarray}
\det s(z)=\det S(z)=1.
\label{det2}
\end{eqnarray}
Evaluating equation (\ref{s4}) at $n=0$, $t=0$, and equation (\ref{s5}) at $n=0$, $t=T$, we find
\begin{subequations}
\begin{eqnarray}
s(z)&=&\mu_2(0,0,z),
\label{sSa}
\\
S(z)&=&\left(e^{-i\omega(z)T\hat{\sigma_3}}\mu_1(0,T,z)\right)^{-1}.
\label{sSb}
\end{eqnarray}
\label{sS}
\end{subequations}
Equations (\ref{muibv}) imply
\begin{subequations}
\begin{eqnarray}
s(z)&=&C(0,0)\left(I-Z^{-1}\sum_{m=0}^{\infty} \frac{1}{C(m,0)}\hat{Z}^{-m}\left(Q(m,0)\mu_{2}(m,0,z)\right)\right),
\label{sS1a}
\\
S^{-1}(z)&=&I+\int_{0}^{T}e^{-iw(z)t\hat{\sigma}_3}\left(H\mu_{1}(0,t,z)\right)dt.
\label{sS1b}
\end{eqnarray}
\label{sS1}
\end{subequations}

Formulas (\ref{sS}) and (\ref{sS1}) imply the following properties for the matrices $s(z)$ and $S(z)$:
\begin{itemize}
\item $s^L(z)$ is analytic for $|z| < 1$ and continuous and bounded for $|z| \leq 1$;
\item $s^R(z)$ is analytic for $|z| > 1$ and continuous and bounded for $|z| \geq 1$;
\item $S(z)$ is analytic in the punctured complex $z$-plane $\mathbb{C}\setminus0$;
\item $S^L(z)$ is continuous and bounded in $\bar{D}_+$, where $\bar{D}_+=\bar{D}_{+in}\cup\bar{D}_{+out}$ is defined by (\ref{D1a}) and (\ref{D1c}) for the DNLS equation,
and by (\ref{D2a}) and (\ref{D2c}) for the DMKDV equation;
\item $S^R(z)$ is continuous and bounded in $\bar{D}_-$, where $\bar{D}_-=\bar{D}_{-in}\cup\bar{D}_{-out}$ is defined by (\ref{D1b}) and (\ref{D1d}) for the DNLS equation,
and by (\ref{D2b}) and (\ref{D2d}) for the DMKDV equation.
\end{itemize}

We will use the following notations:
\begin{eqnarray}
s(z)=\left( \begin{array}{cc} a(z) & \tilde{b}(z) \\ b(z) & \tilde{a}(z) \\ \end{array} \right),
\quad
S(z)=\left( \begin{array}{cc} A(z) & \tilde{B}(z) \\ B(z) & \tilde{A}(z) \\ \end{array} \right).
\label{sm1}
\end{eqnarray}
By employing the symmetries of the spectral functions, we will show below that for the DNLS and DMKDV equations,
$\left\{\tilde{a}(z),\tilde{b}(z)\right\}$ and $\left\{\tilde{A}(z),\tilde{B}(z)\right\}$ can be
expressed respectively in terms of $\left\{a(z),b(z)\right\}$ and $\left\{A(z),B(z)\right\}$.

Equation (\ref{det2}) yields
\begin{eqnarray}
S^{-1}(z)=\left( \begin{array}{cc} \tilde{A}(z) & -\tilde{B}(z) \\ -B(z) & A(z) \\ \end{array} \right).
\label{sm2}
\end{eqnarray}

The asymptotic behavior of the eigenfunctions determines the asymptotic behavior of the spectral functions:
from equations (\ref{aspbivp}) and (\ref{sSa}), we find
\begin{eqnarray}
\begin{split}
a(z)&=\frac{1}{C(0,0)}+O(z^2), ~~b(z)=O(z)~~ z\rightarrow 0 ~ \text{in} ~\bar{D}_{in},
\\
\tilde{a}(z)&=\frac{1}{C(0,0)}+O(\frac{1}{z^2}), ~~\tilde{b}(z)=O(\frac{1}{z}), ~~ z\rightarrow \infty ~ \text{in} ~\bar{D}_{out}.
\end{split}
\label{aspa}
\end{eqnarray}
Furthermore, (\ref{asp2}) and (\ref{s5}) imply
\begin{eqnarray}
\begin{split}
A(z)&=E_2(0)+O(\frac{1}{z^2}),~~ B(z)=O(\frac{1}{z}), \quad z\rightarrow \infty ~ \text{in} ~\bar{D}_{+out},
\\
\tilde{A}(z)&=E_2(0)+O(z^2), ~~ \tilde{B}(z)=O(z), \quad z\rightarrow 0 ~ \text{in} ~\bar{D}_{-in}.
\end{split}
\label{aspA}
\end{eqnarray}

For convenience, we introduce the following notations:
\begin{eqnarray}
\begin{split}
\gamma(z)&=\frac{\tilde{b}(z)}{a(z)}, ~~|z|=1;~~\tilde{\gamma}(z)=\frac{b(z)}{\tilde{a}(z)}, ~~|z|=1;
\\
R(z)&=\frac{B(z)}{A(z)},~~ z\in\mathbb{C}; ~~\tilde{R}(z)=\frac{\tilde{B}(z)}{\tilde{A}(z)},~~z\in\mathbb{C};
\\
d(z)&=a(z)\tilde{A}(z)-b(z)\tilde{B}(z), ~~z\in \bar{D}_{in};
\\
\tilde{d}(z)&=\tilde{a}(z)A(z)-\tilde{b}(z)B(z),~~z\in \bar{D}_{out};
\\
\Gamma(z)&=\frac{B(z)}{\tilde{a}(z)\tilde{d}(z)}, ~~z\in \bar{D}_{out};~~\tilde{\Gamma}(z)=\frac{\tilde{B}(z)}{a(z)d(z)},~~z\in \bar{D}_{in}.
\end{split}
\label{grd}
\end{eqnarray}
Equations (\ref{aspa}) and (\ref{aspA}) imply the following asymptotics:
\begin{eqnarray}
\begin{split}
d(z)&=\frac{E_2(0)}{C(0,0)}+O(z^2), \quad z\rightarrow 0,
\\
\tilde{d}(z)&=\frac{E_2(0)}{C(0,0)}+O(\frac{1}{z^2}), \quad z\rightarrow \infty.
\end{split}
\label{aspd}
\end{eqnarray}

\subsubsection*{4.1.2.2 The symmetries}

Equations (\ref{aspbivp}), (\ref{asp1}) and (\ref{asp2}) imply the following symmetries:
\begin{eqnarray}
\begin{split}
a(-z)&=a(z),\quad b(-z)=-b(z), \quad \tilde{a}(-z)=\tilde{a}(z),\quad \tilde{b}(-z)=-\tilde{b}(z),
\\
A(-z)&=A(z),\quad B(-z)=-B(z), \quad \tilde{A}(-z)=\tilde{A}(z),\quad \tilde{B}(-z)=-\tilde{B}(z).
\end{split}
\label{sr}
\end{eqnarray}
Hence,
\begin{eqnarray}
\begin{split}
d(-z)&=d(z),\quad \tilde{d}(-z)=\tilde{d}(z).
\end{split}
\label{sr1}
\end{eqnarray}
Furthermore, for both the DNLS and DMKDV equations,
following a similar analysis with the analysis presented in section 3.2.2,
we find the following symmetry relations for the eigenfunctions:
\begin{eqnarray}
\Phi_j^L(n,t,z)=\sigma_v\Phi_j^R(n,t,\frac{1}{z^\ast}), \quad \Phi_j^R(n,t,z)=\nu\sigma_v\Phi_j^L(n,t,\frac{1}{z^\ast}),\quad j=1,2,3.
\label{pr1}
\end{eqnarray}
Hence,
\begin{eqnarray}
\begin{split}
\tilde{a}(z)&=a^{\ast}(\frac{1}{z^\ast}), \quad \tilde{b}(z)=\nu b^{\ast}(\frac{1}{z^\ast}), \quad \tilde{A}(z)=A^{\ast}(\frac{1}{z^\ast}), \quad \tilde{B}(z)=\nu B^{\ast}(\frac{1}{z^\ast}),
\\
\tilde{\gamma}&=\nu \gamma^{\ast}(\frac{1}{z^\ast}),\quad \tilde{R}(z)=\nu R^{\ast}(\frac{1}{z^\ast}), \quad \tilde{d}(z)=d^{\ast}(\frac{1}{z^\ast}), \quad \tilde{\Gamma}(z)=\nu \Gamma^{\ast}(\frac{1}{z^\ast}).
\end{split}
\label{ror1}
\end{eqnarray}

We note that as mentioned in Remark 2, for the DNLS equation in the defocusing case ($\nu=1$),
in order to ensure the above symmetries are valid, we should impose the condition $\mid q(n,t)\mid<1$.
Formula (\ref{remark2}) implies that  this condition is time-invariant when the potential evolves according to (\ref{nls});
it imposes a restriction on initial value but not on boundary values.

\subsubsection*{4.1.2.3 The Riemann-Hilbert problem}
We define the matrices $M_{+}(n,t,z)$ and $M_{-}(n,t,z)$ as follows:
\begin{eqnarray}
\begin{split}
M_{+}(n,t,z)&=\left\{\begin{array}{l}\frac{1}{C(n,t)}\left(\mu_2^L(n,t,z),\frac{\mu_1^R(n,t,z)}{a(z)}\right), \quad z\in \bar{D}_{+in},
\\
\frac{1}{C(n,t)}\left(\frac{\mu_3^L(n,t,z)}{\tilde{d}(z)},\mu_2^R(n,t,z)\right), \quad z\in \bar{D}_{+out},
\end{array}\right.
 \\
M_{-}(n,t,z)&=\left\{\begin{array}{l}\frac{1}{C(n,t)}\left(\mu_2^L(n,t,z),\frac{\mu_3^R(n,t,z)}{d(z)}\right), \quad z\in \bar{D}_{-in},
\\
\frac{1}{C(n,t)}\left(\frac{\mu_1^L(n,t,z)}{\tilde{a}(z)},\mu_2^R(n,t,z)\right), \quad z\in \bar{D}_{-out}.
\end{array}\right.
\end{split}
\label{M1}
\end{eqnarray}
It follows from (\ref{asp1}), (\ref{aspa}) and (\ref{aspd}) that
\begin{eqnarray}
\begin{split}
M(n,t,z)=I+Q(n-1,t)Z^{-1}+\left( \begin{array}{cc} O(z^{-2},\text{even}) & O(z^{3},\text{odd})  \\
  O(z^{-3},\text{odd}) & O(z^{2},\text{even}) \\ \end{array} \right), \quad z\rightarrow (\infty,0).
\end{split}
\label{Masp}
\end{eqnarray}

\begin{proposition}
Let $M(n,t,z)$ be defined by equations (\ref{M1}), where $\left\{\mu_{j}(n,t,z)\right\}_{1}^3$
are defined by equations (\ref{muibv}). Then, $M(n,t,z)$ satisfies the jump condition
\begin{eqnarray}
M_{-}(n,t,z)=M_{+}(n,t,z)J(n,t,z), \quad z\in L,
\label{RHP2}
\end{eqnarray}
where the contour $L$ is defined by $L=L_1\cup L_2\cup L_3\cup L_4$, with
\begin{eqnarray}
L_1=\bar{D}_{-in}\cap \bar{D}_{+in},\quad L_2=\bar{D}_{-in}\cap \bar{D}_{+out},
\quad L_3=\bar{D}_{-out}\cap \bar{D}_{+out},\quad L_4=\bar{D}_{-out}\cap \bar{D}_{+in},
\label{L}
\end{eqnarray}
and the jump matrices $J(n,t,z)$ are defined as follows:
\begin{eqnarray}
\begin{split}
J_1(n,t,z)&=\hat{Z}^ne^{i\omega(z)t\hat{\sigma_3}}\left( \begin{array}{cc} 1 & \tilde{\Gamma}(z) \\
  0 & 1 \\ \end{array} \right), \quad z\in L_1,
\\
J_2(n,t,z)&=\hat{Z}^ne^{i\omega(z)t\hat{\sigma_3}}\left( \begin{array}{cc} 1 & (\tilde{\Gamma}(z)-\gamma(z)) \\
 (\tilde{\gamma}(z)-\Gamma(z)) & 1-(\tilde{\gamma}(z)-\Gamma(z))(\gamma(z)-\tilde{\Gamma}(z)) \\ \end{array} \right), \quad z\in L_2,
\\
J_3(n,t,z)&=\hat{Z}^ne^{i\omega(z)t\hat{\sigma_3}}\left( \begin{array}{cc} 1 & 0 \\
  -\Gamma(z) & 1 \\ \end{array} \right), \quad z\in L_3,
\\
J_4(n,t,z)&=\hat{Z}^ne^{i\omega(z)t\hat{\sigma_3}}\left( \begin{array}{cc} 1-\gamma(z)\tilde{\gamma}(z) & \gamma(z) \\
  -\tilde{\gamma}(z) & 1 \\ \end{array} \right), \quad z\in L_4.
\end{split}
\label{JM}
\end{eqnarray}
\end{proposition}
{\bf Proof}
Equations (\ref{s45}) can be written in vector form
\begin{subequations}
\begin{eqnarray}
\mu_2^L&=&a(z)\mu_1^L+z^{-2n}e^{-2i\omega(z)t}b(z)\mu_1^R,
\label{s4va}
\\
\mu_2^R&=&z^{2n}e^{2i\omega(z)t}\tilde{b}(z)\mu_1^L+\tilde{a}(z)\mu_1^R,
\label{s4vb}
\\
 \mu_3^L&=&A(z)\mu_1^L+z^{-2n}e^{-2i\omega(z)t}B(z)\mu_1^R,
\label{s4vc}
\\
\mu_3^R&=&z^{2n}e^{2i\omega(z)t}\tilde{B}(z)\mu_1^L+\tilde{A}(z)\mu_1^R.
\label{s4vd}
\end{eqnarray}
\label{s45v}
\end{subequations}
Here and in what follows, for convenience of notation, we have suppressed the $n$, $t$ and $z$ dependence of $\mu_j(n,t,z)$.
Using (\ref{s4va}) to eliminate $\mu_1^L$ in the right hand side of (\ref{s4vd}) we find the the jump condition across $L_1$.
In order to derive the jump condition across $L_2$, we first eliminate $\mu_1^L$ and $\mu_1^R$ from
equations (\ref{s4vb}) and (\ref{s4vc}):
\begin{subequations}
\begin{eqnarray}
\mu_1^L&=&\frac{\tilde{a}(z)}{\tilde{d}(z)}\mu_3^L-\frac{z^{-2n}e^{-2i\omega(z)t}B(z)}{\tilde{d}(z)}\mu_2^R,
\label{s4vba}
\\
 \mu_1^R&=&-\frac{z^{2n}e^{2i\omega(z)t}\tilde{b}(z)}{\tilde{d}(z)}\mu_3^L+\frac{A(z)}{\tilde{d}(z)}\mu_2^R.
\label{s4vcb}
\end{eqnarray}
\label{s45vab}
\end{subequations}
Then, substituting (\ref{s45vab}) into (\ref{s4va}) and (\ref{s4vd}) we obtain
\begin{subequations}
\begin{eqnarray}
\mu_2^L&=&\frac{\mu_3^L}{\tilde{d}(z)}+z^{-2n}e^{-2i\omega(z)t}(\tilde{\gamma}(z)-\Gamma(z))\mu_2^R,
\label{s4vL2a}
\\
\frac{\mu_3^R}{d(z)}&=&z^{2n}e^{2i\omega(z)t}(-\gamma(z)+\tilde{\Gamma}(z))\frac{\mu_3^L}{\tilde{d}(z)}+
[1-(\tilde{\gamma}(z)-\Gamma(z))(\gamma(z)-\tilde{\Gamma}(z))]\mu_2^R.
\label{s4vL2b}
\end{eqnarray}
\label{s4vL2ab}
\end{subequations}
Equations (\ref{s4vL2ab}) define the jump condition across $L_2$. The jump condition across $L_3$ follows from equation (\ref{s4vba}).
Using (\ref{s4va}) to eliminate $\mu_1^L$ in the right hand side of (\ref{s4vb}) we find
\begin{eqnarray}
\mu_2^R&=&z^{2n}e^{2i\omega(z)t}\gamma(z)\mu_2^L+\frac{\mu_1^R}{a(z)}.
\label{s4vbL4}
\end{eqnarray}
The above equation together with (\ref{s4va}) yield the jump condition across $L_4$. \QEDB

For the DNLS equation the contour of this RH problem is depicted in Figure 2.
For the DMKDV equation the analogous contour is depicted in Figure 4.

We note that $J_2(n,t,z)$ can be expressed in terms of $J_j(n,t,z)$, $j=1,3,4$:
$$J_2(n,t,z)=J_3(n,t,z)(J_4(n,t,z))^{-1}J_1(n,t,z).$$
This fact can be verified  by a direct computation.

The solution of the above RH problem can be represented in the form
\begin{eqnarray}
M(n,t,z)=I+\frac{1}{2\pi i}\int_{L}M_{+}(n,t,\xi)\hat{J}(n,t,\xi)\frac{1}{\xi-z}d\xi,
\label{MS1}
\end{eqnarray}
where $\hat{J}(n,t,z)=I-J(n,t,z)$.
The $z\rightarrow \infty$ limit of equation (\ref{MS1}) yields
\begin{eqnarray}
\begin{split}
M(n,t,z)=&I-\left(\frac{1}{2\pi i}\int_{L}M_{+}(n,t,\xi)\hat{J}(n,t,\xi)d\xi\right)\frac{1}{z}+O(\frac{1}{z^2}).
\end{split}
\label{Masp10}
\end{eqnarray}
The $z\rightarrow 0$ limit of equation (\ref{MS1}) yields
\begin{eqnarray}
\begin{split}
M(n,t,z)=&I+\frac{1}{2\pi i}\int_{L}M_{+}(n,t,\xi)\hat{J}(n,t,\xi)\frac{1}{\xi}d\xi+\left(\frac{1}{2\pi i}\int_{L}M_{+}(n,t,\xi)\hat{J}(n,t,\xi)\frac{1}{\xi^2}d\xi\right)z
\\&+O(z^2).
\end{split}
\label{Masp1}
\end{eqnarray}

By comparing the $(1,2)$-entries of equations (\ref{Masp}) and (\ref{Masp1}), and $(2,1)$-entries of equations (\ref{Masp}) and (\ref{Masp10}), we obtain $q(n,t)$ and $p(n,t)$:
\begin{eqnarray}
q(n,t)=\lim_{z\rightarrow 0}(z^{-1}M(n+1,t,z))^{12}, \quad p(n,t)=\lim_{z\rightarrow \infty}(zM(n+1,t,z))^{21}.
\label{solution1}
\end{eqnarray}
Hence, equations (\ref{Masp1}) and (\ref{solution1}) yield the solutions of the DNLS and DMKDV equations in the following form:
\begin{eqnarray}
\begin{split}
q(n,t)=&\frac{1}{2\pi i}\int_{L_2}[\nu(|\gamma(z)|^2+|\Gamma(z)|^2)-\gamma(z)\Gamma(z)-\gamma^*(z)\Gamma^*(z)]M_{+}^{12}(n+1,t,z)z^{-2}dz
\\
&-\frac{\nu}{2\pi i}\int_{L_1\cup L_2}z^{2n}e^{2iw(z)t}\Gamma^{\ast}(\frac{1}{z^\ast})M_{+}^{11}(n+1,t,z)dz
\\
&-\frac{1}{2\pi i}\int_{|z|=1}z^{2n}e^{2iw(z)t}\gamma(z)M_{+}^{11}(n+1,t,z)dz,
\end{split}
\label{solution2}
\end{eqnarray}
where for the DNLS and DMKDV equations the relevant contours are depicted in Figure 2 and Figure 4, respectively.
\vspace{0.2cm}
\\
{\bf Remark 4.} In \cite{BH1},
the jump matrices, in addition to the spectral functions, they also involve the quantity $C(0,t)$ which depends on the unknown potentials and on the independent variable $t$;
see the equations after (3.35) of \cite{BH1}.
The advantage of our results is that the jump matrices involve only the spectral functions.
The same fact is also valid for the integral representation of the solution $q(n,t)$:
formula (\ref{solution2}) depends only on the spectral functions and the solution of the associated RH problem,
whereas the expression for $q(n,t)$ in \cite{BH1} involves $C(0,t)$ (see equation (3.40) of \cite{BH1}).

\subsubsection*{4.1.2.4 The Residue conditions}
In general, the matrix $M(n,t,z)$ defined by equation (\ref{M1}) is a meromorphic function of $z$.
The possible poles of $M(n,t,z)$ are generated by the zeros of $a(z)$, $d(z)$, $\tilde{a}(z)$ and $\tilde{d}(z)$.
The relations (\ref{sr}) and (\ref{sr1}) imply that $a(z)$, $\tilde{a}(z)$, $d(z)$ and $\tilde{d}(z)$ are even functions of $z$.
Thus, their zeros always appear in opposite pairs. We have the following results.

\begin{proposition}
Assume that
\begin{itemize}
 \item $a(z)$ has $2\mathcal{K}_1$ simple zeros $\{z_j\}_{1}^{2\mathcal{K}_1}$ for $z\in D_{+in}$, such that $z_{j+\mathcal{K}_1}=-z_{j}$, $j=1,\cdots,\mathcal{K}_1$;
\item  $\tilde{a}(z)$ has $2\mathcal{K}_1'$ simple zeros $\{\tilde{z}_j\}_{1}^{2\mathcal{K}_1'}$ for $z\in D_{-out}$, such that $\tilde{z}_{j+\mathcal{K}_1'}=-\tilde{z}_{j}$, $j=1,\cdots,\mathcal{K}_1'$;
\item $d(z)$ has $2\Lambda$ simple zeros $\{\lambda_j\}_{1}^{2\Lambda}$ for $z\in D_{-in}$, such that $\lambda_{j+J}=-\lambda_{j}$, $j=1,\cdots,\Lambda$;
\item $\tilde{d}(z)$ has $2\Lambda'$ simple zeros $\{\tilde{\lambda}_j\}_{1}^{2\Lambda'}$ for $z\in D_{+out}$, such that $\tilde{\lambda}_{j+\Lambda'}=-\tilde{\lambda}_{j}$, $j=1,\cdots,\Lambda'$.
\item  None of the zeros of $d(z)$ for $z\in D_{-in}$ coincides with any of the zeros of $a(z)$.
\item  None of the zeros of $\tilde{d}(z)$ for $z\in D_{+out}$ coincides with any of the zeros of $\tilde{a}(z)$.
\end{itemize}
Then the following residues conditions are valid:
\begin{subequations}
\begin{eqnarray}
&{\text{Res}}_{z=z_j}M_{+}^{R}(n,t,z)=a_jM_{+}^{L}(n,t,z_j), ~~a_j=\frac{z_j^{2n}e^{2i\omega(z_j)t}}{\dot{a}(z_j)b(z_j)},~~j=1,\cdots,2\mathcal{K}_1,
\label{rr3a}
\\
&{\text{Res}}_{z=\tilde{z}_j}M_{-}^{L}(n,t,z)=\tilde{a}_jM_{-}^{R}(n,t,\tilde{z}_j), ~~\tilde{a}_j=\frac{\tilde{z}_j^{-2n}e^{-2i\omega(\tilde{z}_j)t}}{\dot{\tilde{a}}(\tilde{z}_j)\tilde{b}(\tilde{z}_j)},~~j=1,\cdots,2\mathcal{K}_1',
\label{rr3b}
\\
&{\text{Res}}_{z=\lambda_j}M_{-}^{R}(n,t,z)=d_jM_{-}^{L}(n,t,\lambda_j), ~~d_j=\frac{\lambda_j^{2n}e^{2i\omega(\lambda_j)t}\tilde{B}(\lambda_j)}{a(\lambda_j)\dot{d}(\lambda_j)},
~~j=1,\cdots,2\Lambda,
\label{rr3c}
\\
&{\text{Res}}_{z=\tilde{\lambda}_j}M_{+}^{L}(n,t,z)=\tilde{d}_jM_{+}^{R}(n,t,\tilde{\lambda}_j), ~~\tilde{d}_j=\frac{\tilde{\lambda}_j^{-2n}e^{-2i\omega(\tilde{\lambda}_j)t}B(\tilde{\lambda}_j)}{\tilde{a}(\tilde{\lambda}_j)\dot{\tilde{d}}(\tilde{\lambda}_j)},
~~j=1,\cdots,2\Lambda'.
\label{rr3d}
\end{eqnarray}
\label{rr3}
\end{subequations}
\end{proposition}
{\bf Proof}  Evaluating (\ref{s4va}) at $z=z_j$ we find
\begin{eqnarray}
\mu_2^L(n,t,z_j)&=&z_j^{-2n}e^{-2i\omega(z_j)t}b(z_j)\mu_1^R(n,t,z_j),
\label{proofres1}
\end{eqnarray}
which immediately yields (\ref{rr3a}).
Evaluating (\ref{s4vb}) at $z=\tilde{z}_j$ we find
\begin{eqnarray}
\mu_2^R(n,t,\tilde{z}_j)&=&\tilde{z}_j^{2n}e^{2i\omega(\tilde{z}_j)t}\tilde{b}(\tilde{z}_j)\mu_1^L(n,t,\tilde{z}_j),
\label{proofres2}
\end{eqnarray}
which immediately  yields (\ref{rr3b}).
Using (\ref{s4va}) to eliminate $\mu_1^L(n,t,z)$ in (\ref{s4vd}) we find
\begin{eqnarray}
\mu_3^R(n,t,\lambda_j)&=&z^{2n}e^{2i\omega(z)t}\frac{\tilde{B}(z)}{a(z)}\mu_2^L(n,t,z)+\frac{d(z)}{a(z)}\mu_1^R(n,t,z).
\label{proofres3}
\end{eqnarray}
Evaluating the above equation at $z=\lambda_j$ we find (\ref{rr3c}).
Rearranging (\ref{s4vba}) we find
\begin{eqnarray}
\frac{\tilde{d}(z)}{\tilde{a}(z)}\mu_1^L(n,t,z)&=&\mu_3^L(n,t,z)-\frac{z^{-2n}e^{-2i\omega(z)t}B(z)}{\tilde{a}(z)}\mu_2^R(n,t,z).
\label{proofres4}
\end{eqnarray}
The residue formula (\ref{rr3d}) follows by evaluating (\ref{proofres4}) at $z= \tilde{\lambda}_j$.  \QEDB

For the DNLS and DMKDV equations, the symmetry relations (\ref{ror1}) imply
\begin{eqnarray}
\begin{split}
\tilde{z_j}=\frac{1}{z_j^{\ast}}, \quad \tilde{\lambda}_j=\frac{1}{\lambda_j^{\ast}}, \quad \mathcal{K}_1=\mathcal{K}_1', \quad \Lambda=\Lambda',\quad \tilde{a}_j=\nu a_j^{\ast},\quad \tilde{d}_j=\nu d_j^\ast.
\end{split}
\label{DMr}
\end{eqnarray}

The residue conditions (\ref{rr3}) imply the following expression for the solution of the RH problem with the jump condition (\ref{RHP2}):
\begin{eqnarray}
\begin{split}
M(n,t,z)=&I+\frac{1}{2\pi i}\int_{L}M_{+}(n,t,\xi)\hat{J}(n,t,\xi)\frac{1}{\xi-z}d\xi+\sum_{j=1}^{2J}\frac{1}{z-z_j}{\text{Res}}_{z=z_j}M_{+}(n,t,z)
\\
&+\sum_{j=1}^{2J'}\frac{1}{z-\tilde{z}_j}{\text{Res}}_{z=\tilde{z}_j}M_{-}(n,t,z)+\sum_{j=1}^{2\Lambda}\frac{1}{z-\lambda_j}{\text{Res}}_{z=\lambda_j}M_{-}(n,t,z)
\\
&+\sum_{j=1}^{2\Lambda'}\frac{1}{z-\tilde{\lambda}_j}{\text{Res}}_{z=\tilde{\lambda}_j}M_{+}(n,t,z).
\end{split}
\label{MS2}
\end{eqnarray}
Let
\begin{eqnarray}
\alpha_j=\frac{1}{\dot{a}(z_j)b(z_j)}, \quad \beta_j=\frac{\tilde{B}(\lambda_j)}{a(\lambda_j)\dot{d}(\lambda_j)},
\quad
\tilde{\alpha}_j=\frac{1}{\tilde{z}_j^2\dot{\tilde{a}}(\tilde{z}_j)\tilde{b}(\tilde{z}_j)}, \quad \tilde{\beta}_j=\frac{B(\tilde{\lambda}_j)}{\tilde{\lambda}_j^2\tilde{a}(\tilde{\lambda}_j)\dot{\tilde{d}}(\tilde{\lambda}_j)}.
\label{ab}
\end{eqnarray}
Using the asymptotic expansion of (\ref{MS2}) and the residues conditions (\ref{rr3}), we obtain the following expressions for $q(n,t)$ and $p(n,t)$:
\begin{eqnarray}
\begin{split}
q(n,t)=&\frac{1}{2\pi i}\int_{L}\left[M_{+}(n+1,t,z)\hat{J}(n+1,t,z)\right]^{12}\frac{1}{z^2}dz-2\sum_{j=1}^{J}\alpha_jz_j^{2n}e^{2i\omega(z_j)t}M_{+}^{11}(n+1,t,z_j)
\\
&-2\sum_{j=1}^{\Lambda}\beta_j\lambda_j^{2n}e^{2i\omega(\lambda_j)t}M_{-}^{11}(n+1,t,\lambda_j),
\\
p(n,t)=&-\frac{1}{2\pi i}\int_{L}\left[M_{+}(n+1,t,z)\hat{J}(n+1,t,z)\right]^{21}dz+2\sum_{j=1}^{J'}\tilde{\alpha}_j\tilde{z}_j^{-2n}e^{-2i\omega(\tilde{z}_j)t}M_{-}^{22}(n+1,t,\tilde{z}_j)
\\
&+2\sum_{j=1}^{\Lambda'}\tilde{\beta}_j\tilde{\lambda}_j^{-2n}e^{-2i\omega(\tilde{\lambda}_j)t}M_{+}^{22}(n+1,t,\tilde{\lambda}_j),
\end{split}
\label{qs1}
\end{eqnarray}
where we have used the symmetries
\begin{eqnarray}
\begin{split}
&M_{\pm}^{11}(n,t,-z)=M_{\pm}^{11}(n,t,z), ~~M_{\pm}^{12}(n,t,-z)=-M_{\pm}^{12}(n,t,z),
\\
&M_{\pm}^{21}(n,t,-z)=-M_{\pm}^{21}(n,t,z),~~M_{\pm}^{22}(n,t,-z)=M_{\pm}^{22}(n,t,z).
\end{split}
\label{sr2}
\end{eqnarray}

The solitons of the DNLS and DMKDV equations correspond to spectral data  for which $b(z)$, and $B(z)$ vanish identically.
In this case, the jump matrix $J(n,t,z)$ in (\ref{RHP2}) is the identity matrix and thus the first term in the right hand side of (\ref{qs1}) vanishes.
Then the solution becomes
\begin{eqnarray}
\begin{split}
q(n,t)=-2\sum_{j=1}^{J}\alpha_jz_j^{2n}e^{2i\omega(z_j)t}M_{+}^{11}(n+1,t,z_j)
-2\sum_{j=1}^{\Lambda}\beta_j\lambda_j^{2n}e^{2i\omega(\lambda_j)t}M_{-}^{11}(n+1,t,\lambda_j).
\end{split}
\label{qs2}
\end{eqnarray}
The solution (\ref{qs2}) can be accomplished by solving the following algebraic system for $M_{+}^{11}(n,t,z_k)$, $M_{-}^{12}(n,t,\tilde{z}_k)$,
$M_{-}^{11}(n,t,\lambda_k)$ and $M_{+}^{12}(n,t,\tilde{\lambda}_k)$:
\begin{eqnarray}
\begin{split}
M_{+}^{11}(n,t,z_k)&=1+\sum_{j=1}^{J}\tilde{a}_j(\frac{1}{z_k-\tilde{z}_j}-\frac{1}{z_k+\tilde{z}_j})M_{-}^{12}(n,t,\tilde{z}_j)
+\sum_{j=1}^{\Lambda}\tilde{d}_j(\frac{1}{z_k-\tilde{\lambda}_j}-\frac{1}{z_k+\tilde{\lambda}_j})M_{+}^{12}(n,t,\tilde{\lambda}_j),
\\
M_{-}^{12}(n,t,\tilde{z}_k)&=\sum_{j=1}^{J}a_j(\frac{1}{\tilde{z}_k-z_j}+\frac{1}{\tilde{z}_k+z_j})M_{+}^{11}(n,t,z_j)
+\sum_{j=1}^{\Lambda}d_j(\frac{1}{\tilde{z}_k-\lambda_j}+\frac{1}{\tilde{z}_k+\lambda_j})M_{-}^{11}(n,t,\lambda_j),
\\
M_{-}^{11}(n,t,\lambda_k)&=1+\sum_{j=1}^{J}\tilde{a}_j(\frac{1}{\lambda_k-\tilde{z}_j}-\frac{1}{\lambda_k+\tilde{z}_j})M_{-}^{12}(n,t,\tilde{z}_j)
+\sum_{j=1}^{\Lambda}\tilde{d}_j(\frac{1}{\lambda_k-\tilde{\lambda}_j}-\frac{1}{\lambda_k+\tilde{\lambda}_j})M_{+}^{12}(n,t,\tilde{\lambda}_j),
\\
M_{+}^{12}(n,t,\tilde{\lambda}_k)&=\sum_{j=1}^{J}a_j(\frac{1}{\tilde{\lambda}_k-z_j}+\frac{1}{\tilde{\lambda}_k+z_j})M_{+}^{11}(n,t,z_j)
+\sum_{j=1}^{\Lambda}d_j(\frac{1}{\tilde{\lambda}_k-\lambda_j}+\frac{1}{\tilde{\lambda}_k+\lambda_j})M_{-}^{11}(n,t,\lambda_j),
\end{split}
\label{as}
\end{eqnarray}
where $a_j$, $d_j$ are defined by (\ref{rr3a}) and (\ref{rr3b}), and  $\tilde{a}_j$, $\tilde{d}_j$ are given via the symmetry conditions (\ref{DMr}).

Using (\ref{qs2}) and (\ref{as}),
one can easily derive the explicit formulas for the one-soliton solutions of the DNLS and DMKDV equations via direct algebraic computations.
For economy of presentation, we omit the details.

\subsubsection{The Global relation}
\begin{proposition}
The spectral functions $\{a(z), b(z),\tilde{a}(z), \tilde{b}(z)\}$ and $\{A(z), B(z),\tilde{A}(z), \tilde{B}(z)\}$ are
not independent but they satisfy the global relation:
\begin{subequations}
\begin{eqnarray}
&\tilde{A}(z)\tilde{b}(z)-\tilde{B}(z)\tilde{a}(z)=-e^{-2i\omega(z)T}G^{12}(z,T), \quad |z|>1,
\label{gra}
\\
&A(z)b(z)-B(z)a(z)=-e^{2i\omega(z)T}G^{21}(z,T), \quad |z|<1,
\label{grb}
\end{eqnarray}
\label{gr}
\end{subequations}
where
\begin{eqnarray*}
G^{12}(z,t)&=C(0,t)&\sum_{m=0}^{\infty} \frac{1}{C(m,t)}z^{-(2m+1)}q(m,t)\mu_{2}^{22}(m,t,z)),
\\
G^{21}(z,t)&=C(0,t)&\sum_{m=0}^{\infty} \frac{1}{C(m,t)}z^{2m+1}p(m,t)\mu_{2}^{11}(m,t,z)).
\label{gr4}
\end{eqnarray*}
\end{proposition}
{\bf Proof}
Evaluating (\ref{s4}) at $n=0$, $t =T$ and using (\ref{sSb}), we find
\begin{eqnarray}
\mu_2(0,T,z)=e^{i\omega(z)T\hat{\sigma_3}}\left(S^{-1}(z)s(z)\right).
\label{gr1}
\end{eqnarray}
Substituting the second equation of (\ref{muibv}) into (\ref{gr1}), we find
\begin{eqnarray}
S^{-1}(z)s(z)=C(0,T)I-e^{-i\omega(z)T\hat{\sigma_3}}G(z,T),
\label{gr2}
\end{eqnarray}
where
\begin{eqnarray}
G(z,t)=C(0,t)Z^{-1}\sum_{m=0}^{\infty} \frac{1}{C(m,t)}\hat{Z}^{-m}(Q(m,t)\mu_{2}(m,t,z)).
\label{gr3}
\end{eqnarray}
The $(1,2)$ and $(2,1)$-entries of equation (\ref{gr2}) yield the global relation (\ref{gr}). \QEDB

\subsection{Existence under the assumption that the global relation is valid}

The analysis of section 4.1 motivates the following definitions and results for the spectral
functions for the DNLS and DMKDV equations.
\begin{definition} (the spectral functions $a(z)$ and $b(z)$).
Given $q_0(n)\in l^1(\mathbb{N})$, define the vector $\phi(n, z) = (\phi_1(n,z), \phi_2(n,z))^\mathcal{T}$
(here and in what follows the symbol $\mathcal{T}$ denotes the transpose of a vector) as the unique solution of
\begin{eqnarray}
\begin{split}
&f_0(n)\phi_1(n+1,z)-\phi_1(n,z)=z^{-1}q_0(n)\phi_{2}(n,z),
\\
&f_0(n)\phi_2(n+1,z)-z^{-2}\phi_2(n,z)=z^{-1}p_0(n)\phi_{1}(n,z),
\\
&\lim_{n\rightarrow \infty}\phi(n, z)=(1,0)^\mathcal{T},
\end{split}
\label{def1phi}
\end{eqnarray}
where $f_0(n)=\sqrt{1-q_0(n)p_0(n)}$, with $p_0(n)=\nu q^*_0(n)$ for the DNLS equation, and $p_0(n)=\nu q_0(n)$ for the DMKDV equation, respectively.
Given $\phi(n, z)$  define the spectral functions $a(z)$ and $b(z)$ by
\begin{eqnarray}
a(z)=\phi_1(0,z),~~b(z)=\phi_2(0,z), ~~|z|\leq 1.
\label{dab}
\end{eqnarray}
\end{definition}

\begin{proposition} (properties of $a(z)$ and $b(z)$).
The spectral functions $a(z)$ and $b(z)$
defined by (\ref{dab}) have the following properties:
\\(i) $a(z)$ and $b(z)$ are analytic for $|z| < 1$ and continuous and bounded for $|z| \leq 1$.
\\(ii) $a(-z)=a(z)$, $b(-z)=-b(z)$.
\\(iii) $|a(z)|^2 -\nu |b(z)|^2=1,~~ |z|=1.$
\\(iv) $a(z)=\frac{1}{C(0,0)}+O(z^2,\text{even})$, $b(z)=O(z,\text{odd})$, $z\rightarrow 0$.
\\(v) $q_0(n)$ can be reconstructed in terms of $a(z)$ and $b(z)$ by
\begin{eqnarray}
q_0(n)=\lim_{z\rightarrow 0}(z^{-1}M^{(n)}(n+1,z))^{12},
\label{abis}
\end{eqnarray}
where $M^{(n)}(n,z)$ is the unique solution of the following RH problem:
\begin{itemize}
\item
\begin{eqnarray}
M^{(n)}(n, z)=\left\{ \begin{array}{cc} M_{-}^{(n)}(n, z), ~~|z|\geq 1,   \\
  M_{+}^{(n)}(n, z), ~~|z|\leq 1,  \\ \end{array} \right.
\label{abm}
\end{eqnarray}
is a sectionally meromorphic function.
\item
\begin{eqnarray}
M_{-}^{(n)}(n,z)=M_{+}^{(n)}(n,z)J^{(n)}(n,z), \quad |z|=1,
\label{abJc}
\end{eqnarray}
where
\begin{eqnarray}
J^{(n)}(n,z)=\hat{Z}^{n}\left( \begin{array}{cc} \frac{1}{|a(z)|^2} & \nu \frac{b^*(z)}{a(z)} \\
  - \frac{b(z)}{a^*(z)} & 1 \\ \end{array} \right).
\label{abJr}
\end{eqnarray}
\item
\begin{eqnarray}
M^{(n)}(n,z)=I+\left( \begin{array}{cc} O(z^{-2},\text{even}) & O(z,\text{odd})  \\
  O(z^{-1},\text{odd}) & O(z^2,\text{even}) \\ \end{array} \right), \quad z\rightarrow (\infty,0).
\label{abMA}
\end{eqnarray}
\item
Suppose that $a(z)$ has at most $2\mathcal{K}$ simple zeros $\{z_j\}_{1}^{2\mathcal{K}}$, $\mathcal{K}=\mathcal{K}_1+\mathcal{K}_2$, where $z_j\in D_{+in}$, $j=1,\cdots,2\mathcal{K}_1$, such that $z_{j+\mathcal{K}_1}=-z_{j}$, $j=1,\cdots,\mathcal{K}_1$; $z_j\in D_{-in}$,
$j=2\mathcal{K}_1+1,\cdots,2\mathcal{K}_1+2\mathcal{K}_2$, such that $z_{j+\mathcal{K}_2}=-z_{j}$, $j=2\mathcal{K}_1+1,\cdots,2\mathcal{K}_1+\mathcal{K}_2$.
  The first column of $M_{-}^{(n)}(n,z)$
 can have simple poles at $\left\{\frac{1}{z^*_j}\right\}_1^{2\mathcal{K}}$, and the second column of $M_{+}^{(n)}(n,z)$
can have simple poles at $\{z_j\}_1^{2\mathcal{K}}$. The
associated residues are given by
\begin{subequations}
\begin{eqnarray}
&&{{\rm Res}}_{z=z_j}[M^{(n)}(n, z)]^R=\frac{z_j^{2n}}{\dot{a}(z_j)b(z_j)}[M^{(n)}(n, z_j)]^L, ~~j=1,\cdots,2\mathcal{K}, \label{abrr1a}
\\
&&{{\rm Res}}_{z=\frac{1}{z^*_j}}[M^{(n)}(n, z)]^L=\frac{\nu (z^*_j)^{2n}}{\left(\dot{a}(z_j)b(z_j)\right)^*}[M^{(n)}(n, \frac{1}{z^*_j})]^R, ~~j=1,\cdots,2\mathcal{K}. \label{abrr1b}
\end{eqnarray}
\label{abrr1}
\end{subequations}
\end{itemize}
\end{proposition}
{\bf Proof} \quad Properties (i)-(iv) follow from Definition 1.
Property (v) can be derived by performing the spectral analysis of the $n$-part of Lax pair (\ref{LPS3}) evaluated at $t=0$; see appendix B.
The unique solvability of the above RH problem is a consequence of a vanishing lemma for the above RH problem with the vanishing condition as
$z\rightarrow (\infty,0)$. The vanishing lemma can be established in the same manner as used in section 3.2.5. \QEDB

\begin{definition}
Denote by $H_0(t, z)$ the matrix $H(0,t, z)$, in which $q(-1, t)$ and  $q(0, t)$ are
replaced by $g_{-1}(t)$ and  $g_{0}(t)$.
Given the smooth functions $g_{-1}(t)$ and  $g_{0}(t)$ define the vector $\varphi(t, z) = (\varphi_1(t,z), \varphi_2(t,z))^\mathcal{T}$
by the unique solution of
\begin{eqnarray}
\begin{split}
&(\varphi_1)_t=H_0^{11}\varphi_1+H_0^{12}\varphi_2,
\\
&(\varphi_2)_t+2i\omega(z)\varphi_2=H_0^{21}\varphi_1+H_0^{22}\varphi_2,
\\
&\varphi(0, z) =(1,0)^\mathcal{T},
\end{split}
\label{ABphidef}
\end{eqnarray}
where $\omega(z)$ is defined by (\ref{nlsw}) for the DNLS equation, and by (\ref{mkdvw})  for the DMKDV equation respectively.
Given $\varphi(t, z)$ define the spectral functions $A(z)$ and $B(z)$ by
\begin{eqnarray}
A(z)=\varphi_1^*(T,\frac{1}{z^*}),~~B(z)=- e^{2iw(z)T}\varphi_2(T,z), ~~z\in \mathbb{C}\setminus 0.
\label{dAB1}
\end{eqnarray}
\end{definition}

\begin{proposition} (properties of $A(z)$ and $B(z)$).
The spectral functions $A(z)$ and $B(z)$
defined by (\ref{dAB1}) have the following properties:
\\(i) $A(z)$ and $B(z)$ are analytic for $z\in \mathbb{C}\setminus0$ and continuous and bounded for $z\in\bar{D}_+$.
\\(ii) $A(-z)=A(z)$, $B(-z)=-B(z)$.
\\(iii) $A(z)A^*(\frac{1}{z^*})-\nu B(z)B^*(\frac{1}{z^*})=1,~~ z\in \mathbb{C}\setminus0$.
\\(iv)  $A(z)=E_2(0)+O(z^{-2},\text{even})$, $B(z)=O(z^{-1},\text{odd})$, $z\rightarrow \infty$.
\\(v) $g_{-1}(t)$ and  $g_{0}(t)$ can be reconstructed as follows:
\begin{eqnarray}
\begin{split}
&\text{DNLS}:  ~g_{-1}(t)=\lim_{z\rightarrow 0}\left(z^{-1}M^{(t)}\right)^{12},
\\
&~~~~~g_{0}(t)=(\nu|g_{-1}(t)|^2-1)^{-1}\lim_{z\rightarrow 0}\left(i(g_{-1}(t))_t+2g_{-1}(t)-z^{-2}g_{-1}(t)(M^{(t)})^{22}
+z^{-3}(M^{(t)})^{12}\right);
\\
&\text{DMKDV}:  ~g_{-1}(t)=\lim_{z\rightarrow 0}\left(z^{-1}M^{(t)}\right)^{12},
\\
&~~~~~~~~~~~g_{0}(t)=(\nu (g_{-1}(t))^2-1)^{-1}\lim_{z\rightarrow 0}\left((g_{-1}(t))_t+z^{-2}g_{-1}(t)(M^{(t)})^{22}
-z^{-3}(M^{(t)})^{12}\right),
\end{split}
\label{bvs}
\end{eqnarray}
where $M^{(t)}=M^{(t)}(t,z)$ is the unique solution of the following RH problem:
\begin{itemize}
\item
\begin{eqnarray}
M^{(t)}(t, z)=\left\{ \begin{array}{cc} M_{-}^{(t)}(t, z), ~~z\in \bar{D}_{-}=\left\{z\Big| z\in\mathbb{C}, \text{Im} (\omega(z))\leq0\right\},   \\
  M_{+}^{(t)}(t, z), ~~z\in \bar{D}_+=\left\{z\Big| z\in\mathbb{C}, \text{Im} (\omega(z))\geq 0\right\},  \\ \end{array} \right.
\label{ABm}
\end{eqnarray}
is a sectionally meromorphic function.
\item
\begin{eqnarray}
M_{-}^{(t)}(t,z)=M_{+}^{(t)}(t,z)J^{(t)}(t,z), \quad z\in L,
\label{ABJc}
\end{eqnarray}
where the contours $L=L_1\cup L_2\cup L_3\cup L_4$ are defined by (\ref{L}) (see figures 2 and 4 for DNLS and DMKDV equations respectively), and
\begin{eqnarray}
J^{(t)}(t,z)=e^{i\omega(z)t\hat{\sigma_3}}\left( \begin{array}{cc} 1 & \nu \frac{B^*(\frac{1}{z^*})}{A^*(\frac{1}{z^*})} \\
  - \frac{B(z)}{A(z)} & \frac{1}{A(z)A^*(\frac{1}{z^*})} \\ \end{array} \right).
\label{AB1Jr}
\end{eqnarray}
\item
\begin{eqnarray}
M^{(t)}(t,z)=I+\left( \begin{array}{cc} O(z^{-2},\text{even}) & O(z,\text{odd})  \\
  O(z^{-1},\text{odd}) & O(z^2,\text{even}) \\ \end{array} \right), \quad z\rightarrow (\infty,0).
\label{ABMA}
\end{eqnarray}
\item
The first column of $M_{+}^{(t)}(t,z)$
 can have simple poles at $\left\{k_j\right\}_1^{2K}$, and the second column of $M_{-}^{(t)}(t,z)$
can have simple poles at $\frac{1}{k^*_j}$, where $\left\{k_j\right\}_1^{2K}$, $k_{j+K}=-k_{j}$, $j=1,\cdots,K$, are the simple zeros of $A(z)$, $z\in D_{+}$. The
associated residues are given by
\begin{subequations}
\begin{eqnarray}
&&{{\rm Res}}_{z=k_j}[M^{(t)}(t, z)]^L=\frac{e^{-2i\omega(k_j)t}B(k_j)}{\dot{A}(k_j)}[M^{(t)}(t, k_j)]^R, \label{ABrr1a}
\\
&&{{\rm Res}}_{z=\frac{1}{k^*_j}}[M^{(t)}(t, z)]^R=\frac{\nu e^{2i\omega^*(k_j)t}B^*(k_j)}{(\dot{A}(k_j))^*}[M^{(t)}(t, \frac{1}{k^*_j})]^L. \label{ABrr1b}
\end{eqnarray}
\label{ABrr1}
\end{subequations}
\end{itemize}
\end{proposition}
{\bf Proof} \quad Properties (i)-(iv) follow from Definition 2.
Property (v) can be derived by performing the spectral analysis of the $t$-part of Lax pair (\ref{LPT3}) evaluated at $n=0$; see appendix C.
The unique solvability of this RH problem is a consequence of a vanishing lemma which can be established by using a similar manner as in section 3.2.5. \QEDB

The main result is the following:
\begin{theorem}
Given $q_0(n)\in l^1(\mathbb{N})$, define the spectral functions $\left\{a(k),b(k)\right\}$ by Definition 1.
Suppose that there exist smooth functions $g_{-1}(t)$ and $g_{0}(t)$ such that the spectral functions $\left\{A(k),B(k)\right\}$  defined by Definition 2, satisfy the global condition (\ref{gr}). Assume that
\begin{itemize}
 \item
   The spectral function $a(z)$ has at most $2\mathcal{K}$ simple zeros $\{z_j\}_{1}^{2\mathcal{K}}$, $\mathcal{K}=\mathcal{K}_1+\mathcal{K}_2$, where $z_j\in D_{+in}$, $j=1,\cdots,2\mathcal{K}_1$, such that $z_{j+\mathcal{K}_1}=-z_{j}$, $j=1,\cdots,\mathcal{K}_1$; $z_j\in D_{-in}$,
$j=2\mathcal{K}_1+1,\cdots,2\mathcal{K}_1+2\mathcal{K}_2$, such that $z_{j+\mathcal{K}_2}=-z_{j}$, $j=2\mathcal{K}_1+1,\cdots,2\mathcal{K}_1+\mathcal{K}_2$.
 Here $D_{\pm in}$ are defined by (\ref{D1a}), (\ref{D1b}) and (\ref{D2a}), (\ref{D2b}) for the DNLS and DMKDV equations respectively.
 \item  The function $d(z)=a(z)A^{\ast}(\frac{1}{z^\ast})-\nu b(z)B^{\ast}(\frac{1}{z^\ast})$ has at most $2\Lambda$ simple zeros $\{\lambda_j\}_{1}^{2\Lambda}$ with $\lambda_{j+\Lambda}=-\lambda_{j}$, $j=1,\cdots,\Lambda$, for $z \in D_{-in}$, where $D_{-in}$ are defined by (\ref{D1b}) and (\ref{D2b}) for the DNLS and DMKDV equations respectively.
\item None of the zeros of $a(z)$ for $z\in D_{-in}$ coincides with a zero of $d(z)$.
\end{itemize}
Define $M(n, t, z)$ as the solution of the following $2 \times 2$ matrix RH problem:
\begin{itemize}
\item $M$ is sectionally meromorphic in $z\in \mathbb{C}\backslash L$, where $L=L_1\cup L_2\cup L_3\cup L_4$ are defined by (\ref{L}); see figures 2 and 4 for these contours for the DNLS and DMKDV equations respectively.
\item \begin{eqnarray}
M_{-}(n,t,z)=M_{+}(n,t,z)J(n,t,z), \quad z\in L,
\label{RHP3}
\end{eqnarray}
where $M(n,t,z)$ is $M_{-}(n,t,z)$ for $ k\in \bar{D}_{-in}\cup \bar{D}_{-out}$,
$M(n,t,z)$ is $M_{+}(n,t,z)$ for $ k\in \bar{D}_{+in}\cup \bar{D}_{+out}$, and
$J(n,t,z)$ is defined in terms of $\left\{a(z), b(z), A(z), B(z)\right\}$ by equations (\ref{JM}); see figures 1-4 for the domains and contours for DNLS and DMKDV equations respectively.
\item \begin{eqnarray}
\begin{split}
&M(n,t,z)=I+\left( \begin{array}{cc} O(z^{-2},\text{even}) & O(z,\text{odd})  \\
  O(z^{-1},\text{odd}) & O(z^2,\text{even}) \\ \end{array} \right), \quad z\rightarrow (\infty,0).
\end{split}
\label{Masp1Th}
\end{eqnarray}
\item M satisfies the residue conditions (\ref{rr3}).
\end{itemize}

Then $M(n,t, z)$ exists and is unique.

Define $q(n,t)$ in terms of $M(n,t, z)$ by
\begin{eqnarray}
q(n,t)=\lim_{z\rightarrow 0}\left(z^{-1}M(n+1,t,z)\right)^{12}.
\label{stheorem}
\end{eqnarray}
Then $q(n,t)$ solves the DNLS and DMKDV equations with
\begin{eqnarray}
q(n,0)=q_0(n), ~q(-1,t)=g_{-1}(t), ~q(0,t)=g_0(t).
\label{ibv}
\end{eqnarray}

\end{theorem}
{\bf Proof} \quad The proof follows the analogous steps as in the case of integrable PDEs
on the half-line; see \cite{F3,F4}. The main steps are as follows.

If $a(z)$ and $d(z)$ have no zeros for $D_{+in}$ and $D_{-in}$ respectively,
then the function $M(n, t, z)$ satisfies a regular RH problem. The unique solvability of this RH problem is
a consequence of the existence of a vanishing lemma; see \cite{F2, Zhou}.
In case that $a(z)$ and $d(z)$ have a finite number of zeros, the above RH problem is singular.
The singular RH problem can be mapped to a regular RH problem supplemented by a system of algebraic equations; see \cite{F2}.

{\it Proof that $q(n,t)$ solves the DNLS and DMKDV equations.}
It can be verified directly that if $M(n, t, z)$ solves the above RH problem and if $q(n,t)$ is defined by (\ref{stheorem}), then $q(n,t)$ solves the DNLS and DMKDV equations;
see appendix D.

{\it Proof that $q(n,0)=q_0(n)$.} The proof that $q(n,t)$ satisfies the initial condition $q(n,0)=q_0(n)$ follows from the fact
that it is possible to map the RH problem for $M(n,0,z)$ to that for $M^{(n)}(n, z)$.
In fact, we define $M^{(n)}(n, z)$ by
\begin{subequations}
\begin{eqnarray}
&M^{(n)}(n, z)=M(n,0,z),~~z\in \bar{D}_{+in}\cup \bar{D}_{-out},\label{Mxta}
\\&M^{(n)}(n, z)=M(n,0,z)\left(J_1(n,0,z)\right)^{-1},~~z\in \bar{D}_{-in},\label{Mxtb}
\\&M^{(n)}(n, z)=M(n,0,z)\left(J_3(n,0,z)\right)^{-1},~~z\in \bar{D}_{+out}.\label{Mxc}
\end{eqnarray}
\label{Mxt}
\end{subequations}
Then $M^{(n)}(n, z)$ has no jumps across $\mathbb{R}\cup i\mathbb{R}$ and it has a jump on $|z|=1$ with the jump matrix precisely given by (\ref{abJr}).
Furthermore, by a straightforward calculation one can verify that the transformation (\ref{Mxt})
replaces poles at $\{z_j,\frac{1}{z_j^*}\}_{1}^{2\mathcal{K}_1}$, $\{\lambda_j,\frac{1}{\lambda_j^*}\}_{1}^{2\Lambda}$ by poles at $\{z_j,\frac{1}{z_j^*}\}_1^{2\mathcal{K}}$,
with the residue conditions (\ref{rr3}), replaced by the residue conditions (\ref{abrr1}).
In fact, since $M^{(n)}(n, z)=M(n,0,z)$, $z\in \bar{D}_{+in}$, from (\ref{rr3a}) we immediately find $M^{(n)}(n, z)$ has poles at
$\{z_j\}_{1}^{2\mathcal{K}_1}$ with the residue relation (\ref{abrr1a}) for $j=1,\cdots,2\mathcal{K}_1$.
Moreover, equation (\ref{Mxtb}) can be written as
\begin{eqnarray}
\left(\left(M^{(n)}(n, z)\right)^L,\left(M^{(n)}(n, z)\right)^R\right)=\left(M^L(n,0,z),M^R(n,0,z)-\frac{z^{2n}\tilde{B}(z)}{a(z)d(z)}M^L(n,0,z)\right).\label{Mxtb1}
\end{eqnarray}
The residue condition (\ref{rr3c}) yields
${{\rm Res}}_{z=z_j}\left(M^{(n)}(n, z)\right)^R=0,$
 which implies that $M^{(n)}(n, z)$ has no poles at $\left\{\lambda_j\right\}_{1}^{2\Lambda}$.
 In addition,  (\ref{Mxtb1}) yields
$$
{\rm \mathop{Res}\limits_{z=z_j}}\left(M^{(n)}(n, z)\right)^R=-\frac{z_j^{2n}\tilde{B}(z_j)}{\dot{a}(z_j)d(z_j)}M^L(n,0,z_j)
 =\frac{z_j^{2n}}{\dot{a}(z_j)b(z_j)}\left(M^{(n)}(n, z_j)\right)^L, ~j=2\mathcal{K}_1+1,\cdots,2\mathcal{K},
$$
which is the expected residue condition (\ref{abrr1a}) at $\{z_j\}_{2\mathcal{K}_1+1}^{2\mathcal{K}}$.
The case of $\frac{1}{z_j^*}$ and $\frac{1}{\lambda_j^*}$ can be discussed in the same manner.
Thus, $M^{(n)}(n, z)$ satisfies the same RH problem as the RH problem in the Proposition 4.
Comparing equation (\ref{abis}) with equation (\ref{stheorem}) evaluated at $t=0$, we arrive at $q(n, 0)=q_0(n)$.

{\it Proof that $q(-1,t)=g_{-1}(t)$ and $q(0,t)=g_0(t)$.} The proof that $q(n,t)$ satisfies the boundary conditions $q(-1,t)=g_{-1}(t)$ and $q(0,t)=g_0(t)$ follows from the fact
that it is possible to map the RH problem for $M(0,t,z)$ to that for $M^{(t)}(t, z)$.  In this respect, we introduce the transformation matrix $G(t,z)$ defined as follows:
\begin{eqnarray}
\begin{split}
&G^{(1)}(t,z)=C(0,0)\left( \begin{array}{cc} \frac{1}{a(z)} & 0 \\
 \frac{e^{2i\omega(z)(T-t)}G^{21}(z,T)}{A(z)} & a(z) \\ \end{array} \right), ~~z\in \bar{D}_{+in},
 \\
&G^{(2)}(t,z)=C(0,0)\left( \begin{array}{cc} \frac{A^*(\frac{1}{z^*})}{d(z)} & 0 \\
 -b(z)e^{-2i\omega(z)t} & \frac{d(z)}{A^*(\frac{1}{z^*})} \\ \end{array} \right), ~~z\in \bar{D}_{-in},
 \\
 &G^{(3)}(t,z)=C(0,0)\left( \begin{array}{cc} \frac{d^*(\frac{1}{z^*})}{A(z)} &  -\nu b^*(\frac{1}{z^*})e^{2i\omega(z)t} \\
 0 & \frac{A(z)}{d^*(\frac{1}{z^*})} \\ \end{array} \right), ~~z\in \bar{D}_{+out},
 \\
&G^{(4)}(t,z)=C(0,0)\left( \begin{array}{cc} a^*(\frac{1}{z^*}) & \frac{e^{2i\omega(z)(t-T)}G^{12}(z,T)}{A^*(\frac{1}{z^*})}  \\
 0 & \frac{1}{a^*(\frac{1}{z^*})} \\ \end{array} \right), ~~z\in \bar{D}_{-out}.
\end{split}
\label{Gy}
\end{eqnarray}
Using the global relation (\ref{gr}) we can verify directly that the constructed
$G(t,z)$ satisfies
\begin{eqnarray}
\begin{split}
&J_1(0,t,z)G^{(2)}(t,z)=G^{(1)}(t,z)J^{(t)}(t,z), ~~z\in L_1,
 \\
&J_2(0,t,z)G^{(2)}(t,z)=G^{(3)}(t,z)J^{(t)}(t,z), ~~z\in L_2,
 \\
 &J_3(0,t,z)G^{(4)}(t,z)=G^{(3)}(t,z)J^{(t)}(t,z), ~~z\in L_3,
 \\
&J_4(0,t,z)G^{(4)}(t,z)=G^{(1)}(t,z)J^{(t)}(t,z), ~~z\in L_4.
\end{split}
\label{gtr}
\end{eqnarray}
Moreover, equations (\ref{Gy}) imply
\begin{eqnarray}
G(t,z)\rightarrow I,~~z\rightarrow (\infty,0).
\label{tmasp}
\end{eqnarray}

Define $M^{(t)}(t,z)$ by
\begin{eqnarray}
M^{(t)}(t,z)=M(0,t,z)G(t,z),
\label{Myt}
\end{eqnarray}
where $G(t,z)$ is given by $G^{(1)}(t,z)$, $\cdots$, $G^{(4)}(t,z)$ for
$z\in \bar{D}_{+in},\bar{D}_{-in},\bar{D}_{+out}, \bar{D}_{-out}$.
Equation (\ref{Myt}) together with (\ref{gtr}) and (\ref{tmasp}) imply that $M^{(t)}(t,z)$ satisfies the jump condition (\ref{ABJc}) with the jump matrix
defined by (\ref{AB1Jr}) and the estimate (\ref{ABMA}).
In analogy with the proof used for the equation $q(n,0)=q_0(n)$, it can be verified that the the transformation (\ref{Myt}) replaces the residue conditions (\ref{rr3})
by the residue conditions (\ref{ABrr1}).
Thus, the transformation (\ref{Myt}) maps the RH problem of Theorem 1  evaluated at $n=0$ to the RH problem of Proposition 5. \QEDB

\section {Elimination of the unknown boundary values}
\subsection {Linearizable boundary conditions}
The spectral functions $\{A(z),B(z)\}$ are expressed in terms of both the known boundary value $q(-1,t)$ and the unknown boundary datum $q(0,t)$.
In general it is not possible to obtain directly the spectral functions $A(z)$, $B(z)$ only in terms of the known initial and boundary data.
However, for a particular class of boundary conditions, called linearizable, the unknown boundary value can be eliminated by employing the associated global relations.

By direct calculations we arrive at the following result.
\begin{proposition}
Let $z\mapsto f(z)$ be the transformation in the complex $z$-plane which leaves
$\omega(z)$ invariant, i.e., $\omega(f(z))=\omega(z)$, $f(z)\neq z$.
Let $V(t,z)=i\omega(z)\sigma_3+H_0(t,z)$.
If there exists a nonsingular matrix $N(z)$ such that
\begin{eqnarray}
V(t,f(z))N(z)=N(z)V(t,z),
\label{NV}
\end{eqnarray}
then the spectral functions ${A(z), B(z), \tilde{A}(z), \tilde{B}(z)}$ possess the following symmetry
properties:
\begin{eqnarray}
\begin{split}
A(f(z))\det N(z)&=N^{22}(z)\left(N^{11}(z)A(z)+N^{12}(z)B(z)\right)
\\&~~~~-N^{21}(z)e^{2i\omega(z)T}\left(N^{12}(z)\tilde{A}(z)+N^{11}(z)\tilde{B}(z)\right),
\\
B(f(z))\det N(z)&=N^{22}(z)\left(N^{21}(z)A(z)+N^{22}(z)B(z)\right)
\\&~~~~-N^{21}(z)e^{2i\omega(z)T}\left(N^{22}(z)\tilde{A}(z)+N^{21}(z)\tilde{B}(z)\right),
\\
\tilde{A}(f(z))\det N(z)&=N^{11}(z)\left(N^{22}(z)\tilde{A}(z)+N^{21}(z)\tilde{B}(z)\right)
\\&~~~~-N^{12}(z)e^{-2i\omega(z)T}\left(N^{21}(z)A(z)+N^{22}(z)B(z)\right),
\\
\tilde{B}(f(z))\det N(z)&=N^{11}(z)\left(N^{12}(z)\tilde{A}(z)+N^{11}(z)\tilde{B}(z)\right)
\\&~~~~-N^{12}(z)e^{-2i\omega(z)T}\left(N^{11}(z)A(z)+N^{12}(z)B(z)\right).
\end{split}
\label{AB}
\end{eqnarray}
\end{proposition}

Equation (\ref{NV}) implies that a necessary condition for
the existence of linearizable boundary conditions is that
$\det V(t,f(z))=\det V(t,z)$, i.e.,
\begin{eqnarray}
\left(\frac{1}{z^{2}}-\frac{1}{[f(z)]^{2}}\right)q(0,t)p(-1,t)+\left(z^{2}-[f(z)]^{2}\right)p(0,t)q(-1,t)=0.
\label{LC}
\end{eqnarray}
In what follows we discuss in detail linearizable boundary conditions for the DNLS equation and for the DMKDV equation, respectively;
for convenience, we assume that $T = \infty$.

We note that as $T= \infty$, for $z\in \bar{D}_{-out}$ and for $z\in \bar{D}_{+in}$, the global relation (\ref{gr}) becomes
\begin{eqnarray}
\begin{split}
\tilde{A}(z)\tilde{b}(z)-\tilde{B}(z)\tilde{a}(z)=0, \quad z\in \bar{D}_{-out},
\\
A(z)b(z)-B(z)a(z)=0, \quad z\in \bar{D}_{+in}.
\end{split}
\label{gr5}
\end{eqnarray}
Using the symmetry relations (\ref{ror1}), the global relation (\ref{gr5}) becomes
\begin{eqnarray}
A^\ast(\frac{1}{z^\ast})b^\ast(\frac{1}{z^\ast})-B^\ast(\frac{1}{z^\ast})a^\ast(\frac{1}{z^\ast})=0, \quad z\in \bar{D}_{-out}.
\label{gr7}
\end{eqnarray}

\subsubsection{ Linearizable boundary conditions for the DNLS equation }
In the case of the DNLS equation, $\omega(z)$ is defined via (\ref{nlsw}). The invariance of $\omega(z)$ yields $f(z)=\frac{1}{z}$.
Equation (\ref{LC}) becomes
\begin{eqnarray}
q(0,t)q^\ast(-1,t)-q^\ast(0,t)q(-1,t)=0.
\label{LCDNLS}
\end{eqnarray}
We consider three particular solutions of equation (\ref{LCDNLS}).

\textbf{Case 1 } $q(0,t)=0$. In this case, in order to satisfy equation (\ref{NV}) we take $N^{12}(z)=N^{21}(z)=0$ and $N^{11}(z)=z^2N^{22}(z)$.
Then (\ref{AB}) yields
\begin{eqnarray}
A^\ast(z^\ast)=A^\ast(\frac{1}{z^\ast}),\quad B^\ast(z^\ast)=z^2B^\ast(\frac{1}{z^\ast}).
\label{LCDNLS1}
\end{eqnarray}
Letting $z\mapsto \frac{1}{z}$ in the global relation (\ref{gr5}) and using the symmetries (\ref{LCDNLS1}), we find
\begin{eqnarray}
A^\ast(\frac{1}{z^\ast})b^\ast(z^\ast)-z^2B^\ast(\frac{1}{z^\ast})a^\ast(z^\ast)=0, \quad z\in \bar{D}_{-in}.
\label{gr8}
\end{eqnarray}
This equation and the definition of $d(z)$ (see (\ref{grd})) are two algebraic equations for $A(k)$ and $B(k)$,
which yield the following expressions:
\begin{eqnarray}
A^\ast(\frac{1}{z^\ast})=\frac{z^2a^\ast(z^\ast)d(z)}{\Delta_{\infty}(z)}, \quad B^\ast(\frac{1}{z^\ast})=\frac{ b^\ast(z^\ast)d(z)}{\Delta_{\infty}(z)}, \quad z\in \bar{D}_{-in},
\label{sgr1}
\end{eqnarray}
where
\begin{eqnarray*}
\Delta_{\infty}(z)=z^2a(z)a^\ast(z^\ast)-\nu b(z)b^\ast(z^\ast).
\end{eqnarray*}
Using (\ref{sgr1}), we can express the ratio $R^\ast(\frac{1}{z^\ast})$ in the form
\begin{eqnarray*}
R^\ast(\frac{1}{z^\ast})=\frac{B^\ast(\frac{1}{z^\ast})}{A^\ast(\frac{1}{z^\ast})}=\frac{b^\ast(z^\ast)}{z^2a^\ast(z^\ast)}, \quad z\in \bar{D}_{-in}.
\end{eqnarray*}
Finally we find the following expression for $\Gamma^\ast(\frac{1}{z^\ast})$:
\begin{eqnarray}
\Gamma^\ast(\frac{1}{z^\ast})=\frac{B^\ast(\frac{1}{z^\ast})}{a(z)d(z)}
=\frac{R^\ast(\frac{1}{z^\ast})}{a(z)(a(z)-\nu b(z)R^\ast(\frac{1}{z^\ast}))}
=\frac{b^\ast(z^\ast)}{a(z)\Delta_{\infty}(z)}.
\label{G1}
\end{eqnarray}
Hence, we have expressed the jump matrices only in terms of known spectral functions.

\textbf{Case 2 } $q(-1,t)=0$. In this case we take $N^{12}(z)=N^{21}(z)=0$ and $N^{22}(z)=z^2N^{11}(z)$.
Then (\ref{AB}) yields
\begin{eqnarray}
A^\ast(z^\ast)=A^\ast(\frac{1}{z^\ast}),\quad B^\ast(z^\ast)=\frac{1}{z^2}B^\ast(\frac{1}{z^\ast}).
\label{LCDNLS2}
\end{eqnarray}
Proceeding as in Case 1, we find
\begin{eqnarray}
A^\ast(\frac{1}{z^\ast})=\frac{a^\ast(z^\ast)d(z)}{z^2\Delta_0(z)}, \quad B^\ast(\frac{1}{z^\ast})=\frac{ b^\ast(z^\ast)d(z)}{\Delta_0(z)}, \quad z\in \bar{D}_{-in},
\label{sgr2}
\end{eqnarray}
where
\begin{eqnarray*}
\Delta_0(z)=z^{-2}a(z)a^\ast(z^\ast)-\nu b(z)b^\ast(z^\ast).
\end{eqnarray*}
The expression for $\Gamma^\ast(\frac{1}{z^\ast})$ becomes
\begin{eqnarray}
\Gamma^\ast(\frac{1}{z^\ast})=\frac{b^\ast(z^\ast)}{a(z)\Delta_0(z)}.
\label{G2}
\end{eqnarray}

\textbf{Case 3 } $q(-1,t)=\chi q(0,t)$, where $\chi$ is an arbitrary real constant.
In this case we take $N^{12}(z)=N^{21}(z)=0$ and $N^{11}(z)=\frac{1-z^2\chi}{z^2-\chi}N^{22}(z)$.
Then, (\ref{AB}) yields
\begin{eqnarray}
A^\ast(z^\ast)=A^\ast(\frac{1}{z^\ast}),\quad B^\ast(z^\ast)=\frac{1-z^2\chi}{z^2-\chi}B^\ast(\frac{1}{z^\ast}).
\label{LCDNLS3}
\end{eqnarray}
Following the same steps as in Case 1, we find
\begin{eqnarray}
A^\ast(\frac{1}{z^\ast})=\frac{(1-z^2\chi) a^\ast(z^\ast)d(z)}{(z^2-\chi)\Delta_{\chi}(z)},
\quad B^\ast(\frac{1}{z^\ast})=\frac{ b^\ast(z^\ast)d(z)}{\Delta_{\chi}(z)}, \quad z\in \bar{D}_{-in},
\label{sgr3}
\end{eqnarray}
where
\begin{eqnarray*}
\Delta_{\chi}(z)=\frac{1-z^2\chi}{z^2-\chi}a(z)a^\ast(z^\ast)-\nu b(z)b^\ast(z^\ast).
\end{eqnarray*}
Then the expression for $\Gamma^\ast(\frac{1}{z^\ast})$ becomes
\begin{eqnarray}
\Gamma^\ast(\frac{1}{z^\ast})=\frac{b^\ast(z^\ast)}{a(z)\Delta_{\chi}(z)}.
\label{G3}
\end{eqnarray}
\\
{\bf Remark 5.}
The linearizable boundary conditions of case 3 is the discrete analogue of
the homogeneous Robin boundary conditions of the continuous problem.
For $\chi=0$, these  boundary conditions are reduced to the boundary condition of case 2, while for $\chi\rightarrow \infty$,
they are reduced to the boundary condition of case 1.
We note that these linearizable boundary conditions have been identified earlier via the unified transform method by the authors in \cite{BH1}.
We also note that these linearizable boundary conditions have been found in \cite{IB1,IB2} via an algebraic method based on
B\"{a}cklund transformations.

\subsubsection{ Linearizable boundary conditions for the DMKDV equation }
In the case of the DMKDV equation, $\omega(z)$ is defined via (\ref{mkdvw}). The invariance of $\omega(z)$ yields $f(z)=\frac{i}{z}$.
Then, equation (\ref{LC}) yield
\begin{eqnarray}
q(0,t)q(-1,t)=0.
\label{LCDDMKDV}
\end{eqnarray}

\textbf{Case 1 } $q(0,t)=0$. In this case we take $N^{12}(z)=N^{21}(z)=0$ and $N^{11}(z)=-iz^2N^{22}(z)$.
Then, (\ref{AB}) yields
\begin{eqnarray}
A^\ast(iz^\ast)=A^\ast(\frac{1}{z^\ast}),\quad B^\ast(iz^\ast)=-iz^2B^\ast(\frac{1}{z^\ast}).
\label{LCDMKDV1}
\end{eqnarray}
Following the same arguments used in the case of the DNLS equation we find
\begin{eqnarray}
A^\ast(\frac{1}{z^\ast})=\frac{iz^2a^\ast(iz^\ast)d(z)}{\Delta_0(z)}, \quad B^\ast(\frac{1}{z^\ast})=-\frac{ b^\ast(iz^\ast)d(z)}{\Delta_0(z)}, \quad z\in \bar{D}_{-in},
\label{sgr4}
\end{eqnarray}
where
\begin{eqnarray*}
\Delta_0(z)=iz^2a(z)a^\ast(iz^\ast)+\nu b(z)b^\ast(iz^\ast).
\end{eqnarray*}
Then, the expression for $\Gamma^\ast(\frac{1}{z^\ast})$ is given by
\begin{eqnarray}
\Gamma^\ast(\frac{1}{z^\ast})
=-\frac{b^\ast(iz^\ast)}{a(z)\Delta_0(z)}.
\label{G4}
\end{eqnarray}

\textbf{Case 2 } $q(-1,t)=0$. In this case we have $N^{12}(z)=N^{21}(z)=0$ and $N^{11}(z)=iz^{-2}N^{22}(z)$.
Equations (\ref{AB}) imply
\begin{eqnarray}
A^\ast(iz^\ast)=A^\ast(\frac{1}{z^\ast}),\quad B^\ast(iz^\ast)=iz^{-2}B^\ast(\frac{1}{z^\ast}).
\label{LCDMKDV2}
\end{eqnarray}
Then,
\begin{eqnarray}
A^\ast(\frac{1}{z^\ast})=\frac{iz^{-2}a^\ast(iz^\ast)d(z)}{\Delta_1(z)}, \quad B^\ast(\frac{1}{z^\ast})=\frac{ b^\ast(iz^\ast)d(z)}{\Delta_1(z)}, \quad z\in \bar{D}_{-in},
\label{sgr5}
\end{eqnarray}
where
\begin{eqnarray*}
\Delta_1(z)=iz^{-2}a(z)a^\ast(iz^\ast)-\nu b(z)b^\ast(iz^\ast).
\end{eqnarray*}
The expression for $\Gamma^\ast(\frac{1}{z^\ast})$ becomes
\begin{eqnarray}
\Gamma^\ast(\frac{1}{z^\ast})
=\frac{b^\ast(iz^\ast)}{a(z)\Delta_1(z)}.
\label{G5}
\end{eqnarray}

To the best of our knowledge, the above  linearizable boundary conditions for the DMKDV equation have not been reported in the literature.
We note that for the DNLS equation the homogeneous Robin boundary conditions are linearizable boundary conditions,
but this is not the case for the DMKDV equation.

\subsection {General non-linearizable boundary conditions}
In the continuous case, there exist two effective approaches for determining the unknown boundary values for the general case of non-linearizable
boundary conditions. The first approach employs the so-called Gelfand-Levitan-Marchenko (GLM) representations \cite{F4,MFS}.
The second approach uses certain asymptotic considerations \cite{FL1}-\cite{FL3}.
In this section by extending the second approach to the discrete case,
we will show how to characterise the unknown boundary values $q(0,t)$ for both the DNLS and DMKDV equations.

We write $\mu_1(0,t,z)$ and $\mu_2(n,0,z)$ in the following form:
\begin{eqnarray}
\mu_1(0,t,z)&=&\left( \begin{array}{cc} \varphi_1(t,z) & \tilde{\varphi}_2(t,z) \\ \varphi_2(t,z) & \tilde{\varphi}_1(t,z) \\ \end{array} \right),
\label{mu0}
\\
\mu_2(n,0,z)&=&\left( \begin{array}{cc} \phi_1(n,z) & \tilde{\phi}_2(n,z) \\  \phi_2(n,z)  & \tilde{\phi}_1(n,z) \\ \end{array} \right).
\label{mu2}
\end{eqnarray}
Then $\varphi_1(t,z)$ and $\varphi_2(t,z)$ satisfy the following system of nonlinear integral equations:
\begin{eqnarray}
\begin{split}
\varphi_1(t,z)=&1+i\int_0^t[2\left(\alpha zq(0,t')-\beta z^{-1}q(-1,t')\right)\varphi_2 (t',z)
\\&\qquad-\left(\alpha q(0,t')p(-1,t')+\beta q(-1,t')p(0,t')\right)\varphi_1 (t',z)]dt',
\\
\varphi_2(t,z)=&i\int_0^te^{-2i\omega(z)(t-t')}[2\left(\alpha zp(-1,t')-\beta z^{-1}p(0,t')\right)\varphi_1 (t',z)
\\&\qquad+\left(\alpha q(0,t')p(-1,t')+\beta q(-1,t')p(0,t')\right)\varphi_2 (t',z)]dt';
\end{split}
\label{vphi}
\end{eqnarray}
whereas $\tilde{\phi}_1(n,z)$, $\tilde{\phi}_2(n,z)$ satisfy the summation equations
\begin{eqnarray}
\begin{split}
\tilde{\phi}_1(n,z)=&C(n,0)\left(1-\sum_{m=n}^{\infty} \frac{z}{C(m,0)}p(m,0)\tilde{\phi}_{2}(m,z)\right),
\\
\tilde{\phi}_2(n,z)=&-C(n,0)\sum_{m=n}^{\infty} \frac{z^{2(n-m)-1}}{C(m,0)}q(m,0)\tilde{\phi}_{1}(m,z).
\end{split}
\label{phi}
\end{eqnarray}

Equation (\ref{sS}) implies
\begin{eqnarray}
\tilde{A}(z)=\varphi_1(T,z),\quad \tilde{B}(z)=-e^{-2i\omega(z)T}\tilde{\varphi}_2(T,z),
\label{ABvp}
\end{eqnarray}
and
\begin{eqnarray}
\tilde{a}(z)=\tilde{\phi}_1(0,z),\quad \tilde{b}(z)=\tilde{\phi}_2(0,z).
\label{abp}
\end{eqnarray}
Following (\ref{asp2}) we find
\begin{eqnarray}
\begin{split}
\varphi_1(t,z)&=E_1(t)+O(z^{-2}), \quad z\rightarrow \infty,
\\
\widetilde{\varphi}_2(t,z)&=[E_1(t)q(-1,t)-e^{2i\omega(z)t}q(-1,0)]z+O(z^3), \quad z\rightarrow 0.
\end{split}
\label{aspvp}
\end{eqnarray}
Letting
\begin{eqnarray}
\widehat{G}(z,t)=\frac{G^{12}(z,t)}{\widetilde{a}(z)},
\label{g}
\end{eqnarray}
it follows that the global relation (\ref{gra}) can be written in the form
\begin{eqnarray}
\tilde{\varphi}_2(z,t)+e^{2i\omega(z)t}\frac{\tilde{b}(z)}{\tilde{a}(z)}\varphi_1(t,z)=-\widehat{G}(z,t), \quad |z|>1.
\label{grvp}
\end{eqnarray}
Using (\ref{aspbivp}) we find the following asymptotic behavior for $\widehat{G}(z,t)$:
\begin{eqnarray}
\widehat{G}(z,t)=\frac{C(0,0)}{C(0,t)}q(0,t)z^{-1}+O(z^{-3}), \quad z\rightarrow \infty.
\label{gasp}
\end{eqnarray}
The quantity $\frac{C(0,0)}{C(0,t)}$ appeared in (\ref{gasp}), can be expressed in terms of the boundary values $q(-1,t)$ and $q(0,t)$.
In fact, by employing the formula (\ref{cts}) of appendix A.2, we find
\begin{eqnarray}
\frac{C(0,0)}{C(0,t)}=\frac{1}{E_1(t)}.
\label{cts2}
\end{eqnarray}
Thus, equation (\ref{gasp}) becomes
\begin{eqnarray}
\widehat{G}(z,t)=\frac{q(0,t)}{E_1(t)}z^{-1}+O(z^{-3}), \quad z\rightarrow \infty.
\label{gaspnew}
\end{eqnarray}

\subsubsection{Elimination of the unknown boundary values for the DNLS equation}
For the DNLS equation (\ref{nls}), the unknown boundary value $q(0,t)$ can be characterized as follows.
\begin{proposition}
Define $\omega(z)$ via (\ref{nlsw}). Let $\partial D_{\pm in}$ and $\partial D_{\pm out}$ denote the boundaries of the domains $D_{\pm in}$ and $D_{\pm out}$ defined in (\ref{D1}), oriented
so that $D_{\pm in}$ and $D_{\pm out}$ lie in the left of $\partial D_{\pm in}$ and $\partial D_{\pm out}$.
Let $\varphi_1(t,z)$ and $\varphi_2(t,z)$ be defined by (\ref{vphi}) and $\widetilde{\varphi}_1(t,z)$, $\widetilde{\varphi}_2(t,z)$ be defined by the symmetries $\widetilde{\varphi}_1(t,z)= \varphi^*_1(t,\frac{1}{z^*})$, $\widetilde{\varphi}_2(t,z)=\nu \varphi^*_2(t,\frac{1}{z^*})$.
The unknown boundary value $q(0,t)$ associated with the DNLS equation (\ref{nls}) is given by
\begin{eqnarray}
q(0,t)=-q(-1,t)+\frac{1}{\pi i}\left[\int_{\partial D_{+in}}\psi_2 (t,z)dz+E_1(t)\int_{\partial D_{+out}}
e^{2i\omega(z)t}\frac{\tilde{b}(z)}{\tilde{a}(z)}\varphi_1 (t,z)dz\right],
\label{DN1}
\end{eqnarray}
where
\begin{eqnarray}
\psi_2 (t,z)=z^{-2}\left(\frac{1}{E_1(t)}\tilde{\varphi}_2(t,z)-E_1(t)\tilde{\varphi}_2(t,z^{-1})\right).
\label{psinls}
\end{eqnarray}
\end{proposition}
{\bf Proof} \quad Using (\ref{psinls}) we find
\begin{eqnarray}
\int_{\partial D_{+in}}\psi_2 (t,z)dz=\frac{1}{E_1(t)}\int_{\partial D_{+in}}z^{-2}\tilde{\varphi}_2(t,z)dz
-E_1(t)\int_{\partial D_{+in}}z^{-2}\tilde{\varphi}_2(t,z^{-1})dz.
\label{psi2dnlsin}
\end{eqnarray}
Using the asymptotics of $\tilde{\varphi}_2(t,z)$ (the second of equations (\ref{aspvp})),
the first integral in the right hand side of (\ref{psi2dnlsin}) can be evaluated explicitly:
\begin{eqnarray}
\frac{1}{E_1(t)}\int_{\partial D_{+in}}z^{-2}\tilde{\varphi}_2(t,z)dz=q(-1,t)\pi i,
\label{psi2dnlsin1}
\end{eqnarray}
where we have used the fact that the term involving $q(-1,0)$ in (\ref{aspvp}) vanishes in (\ref{psi2dnlsin1})
due to the term $e^{2i\omega(z)t}$ which is bounded and decays as $z\rightarrow 0$ in $D_{+in}$.
Replacing in the second integral in the right hand side of (\ref{psi2dnlsin}) $z^{-1}$ by $z$, we find
\begin{eqnarray}
 -E_1(t)\int_{\partial D_{+in}}z^{-2}\tilde{\varphi}_2(t,z^{-1})dz=E_1(t)\int_{\partial D_{+out}}\tilde{\varphi}_2(t,z)dz.
\label{psi2dnlsin2}
\end{eqnarray}
Employing the global relation (\ref{grvp}) and the asymptotic behavior (\ref{gaspnew}), the integral (\ref{psi2dnlsin2}) can be evaluated as follows:
\begin{eqnarray}
E_1(t)\int_{\partial D_{+out}}\tilde{\varphi}_2(t,z)dz=q(0,t)\pi i-E_1(t)\int_{\partial D_{+out}}
e^{2i\omega(z)t}\frac{\tilde{b}(z)}{\tilde{a}(z)}\varphi_1 (t,z)dz.
\label{psi2dnlsin3}
\end{eqnarray}
Equations (\ref{psi2dnlsin})-(\ref{psi2dnlsin3}) imply the formula (\ref{DN1}). \QEDB

In what follows we describe an effective characterization of the unknown boundary value $q(0,t)$ by employing a suitable perturbation expansion.
We expand $\varphi_j(t,z)$, $\tilde{\phi}_j(n,z)$, $q(-1,t)$, $q(0,t)$, $q(n,0)$ in the following forms:
\begin{eqnarray}
\begin{split}
\varphi_j(t,z)&=\varphi_{j,0}(t,z)+\varphi_{j,1}(t,z)\epsilon+\varphi_{j,2}(t,z)\epsilon^2+\cdots, \quad j=1,2,
\\
\tilde{\phi}_j(n,z)&=C(n,0)[\tilde{\phi}_{j,0}(n,z)+\tilde{\phi}_{j,1}(n,z)\epsilon+\tilde{\phi}_{j,2}(n,z)\epsilon^2+\cdots], \quad j=1,2,
\\
q(j,t)&=q_1(j,t)\epsilon+q_2(j,t)\epsilon^2+\cdots, \quad j=-1,0,
\\
q(n,0)&=q_1(n,0)\epsilon+q_2(n,0)\epsilon^2+\cdots, \quad n\in \mathbb{N},
\end{split}
\label{expvp}
\end{eqnarray}
where $\epsilon$ is a small perturbation parameter. Substituting the above expansions into (\ref{vphi}) we find
\begin{eqnarray}
\begin{split}
\varphi_{1,0}(t,z)=&1,\quad \varphi_{2,0}(t,z)=0,
\\
\varphi_{1,1}(t,z)=&0,\quad \varphi_{2,1}(t,z)=\nu i\int_0^te^{-2i\omega(z)(t-t')}\left( zq^*_1(-1,t')- z^{-1}q^*_1(0,t')\right)dt',
\\
\varphi_{1,2}(t,z)=&i\int_0^t[\left( zq_1(0,t')- z^{-1}q_1(-1,t')\right)\varphi_{2,1}(t',z)
\\&~~~~~~~ -\frac{\nu}{2}( q_1(0,t')q^*_1(-1,t')+ q_1(-1,t')q^*_1(0,t'))]dt',
 \\
 \varphi_{2,2}(t,z)=&\nu i\int_0^te^{-2i\omega(z)(t-t')}\left( zq^*_2(-1,t')- z^{-1}q^*_2(0,t')\right)dt',
 \end{split}
\label{expvp1}
\end{eqnarray}
as well as the following recursive formulas for $k\geq 3$:
\begin{eqnarray}
\begin{split}
 \varphi_{1,k}(t,z)=&i\int_0^t[2\sum_{l=1}^{k-1}\left(\alpha zq_l(0,t')-\beta z^{-1}q_l(-1,t')\right)\varphi_{2,k-l}(t',z)
\\& -\frac{\nu}{2}\sum_{m=2}^{k}\sum_{r=1}^{m-1}(q_r(0,t')q^*_{m-r}(-1,t')+ q_r(-1,t')q^*_{m-r}(0,t'))\varphi_{1,k-m}(t',z)]dt',
 \\
 \varphi_{2,k}(t,z)=& \nu i\int_0^te^{-2i\omega(z)(t-t')}[\sum_{l=1}^{k}\left( zq^*_l(-1,t')- z^{-1}q^*_l(0,t')\right)\varphi_{1,k-l}(t',z)
 \\& +\frac{1}{2}\sum_{m=2}^{k-1}\sum_{r=1}^{m-1}( q_r(0,t')q^*_{m-r}(-1,t')+ q_r(-1,t')q^*_{m-r}(0,t'))\varphi_{2,k-m}(t',z)]dt'.
\end{split}
\label{expvp2}
\end{eqnarray}
Substituting (\ref{expvp}) into (\ref{phi}) we find
\begin{eqnarray}
\begin{split}
\tilde{\phi}_{1,0}(n,z)=&1,\quad \tilde{\phi}_{2,0}(n,z)=0,
\\
\tilde{\phi}_{1,1}(n,z)=&0,\quad \tilde{\phi}_{2,1}(n,z)=-\sum_{m=n}^\infty z^{2(n-m)-1}q_1(m,0),
\\
\tilde{\phi}_{1,2}(n,z)=&\nu\sum_{m=n}^\infty \sum_{n'=m}^\infty z^{2(m-n')}q^*_1(m,0)q_1(n',0),
 \\
\tilde{\phi}_{2,2}(n,z)=&-\sum_{m=n}^\infty z^{2(n-m)-1}q_2(m,0),
 \end{split}
\label{expp1}
\end{eqnarray}
as well as the following recursive formulas for $k\geq 3$:
\begin{eqnarray}
\begin{split}
 \tilde{\phi}_{1,k}(n,z)=& -\nu\sum_{m=n}^{\infty}\sum_{l=1}^{k-1}z q^*_{l}(m,0)\tilde{\phi}_{2,k-l}(m,z),
 \\
  \tilde{\phi}_{2,k}(n,z)=& -\sum_{m=n}^{\infty}\sum_{l=1}^{k}z^{2(n-m)-1} q_{l}(m,0)\tilde{\phi}_{1,k-l}(m,z).
\end{split}
\label{expp2}
\end{eqnarray}
Equation (\ref{abp}) yields the expansion
\begin{eqnarray}
\frac{\tilde{b}(z)}{\tilde{a}(z)}&=\gamma_0(z)+\gamma_1(z)\epsilon+\gamma_2(z)\epsilon^2+\cdots+\gamma_k(z)\epsilon^k+\cdots,
\label{expba}
\end{eqnarray}
where
\begin{eqnarray}
\begin{split}
\gamma_0(z)&=0, \quad \gamma_1(z)=-\sum_{m=0}^{\infty}z^{-2m-1}q_1(m,0), \quad \gamma_2(z)=-\sum_{m=0}^{\infty}z^{-2m-1}q_2(m,0),
\\
\gamma_k(z)&=\tilde{\phi}_{2,k}(0,z)-\sum_{l=1}^{k-1}\gamma_l(z)\tilde{\phi}_{1,k-l}(0,z),\quad k\geq3.
\end{split}
\label{expbac}
\end{eqnarray}
Recall that for the DNLS equation, the quantity $E_1(t)$ is defined by
\begin{eqnarray}
E_1(t)=\exp\left(\frac{\nu i}{2}\int_0^t\left(q(0,t')q^*(-1,t')- q^*(0,t')q(-1,t')\right)dt'\right).
\label{ct1}
\end{eqnarray}
The above formula implies that the quantity $E_1(t)$ has the expansion
\begin{eqnarray}
E_1(t)=c_0(t)+c_1(t)\epsilon+c_2(t)\epsilon^2+c_3(t)\epsilon^3+O(\epsilon^4),
\label{ct1exp}
\end{eqnarray}
where
\begin{eqnarray}
\begin{split}
c_0(t)&=1, \quad c_1(t)=0,
\\
c_2(t)&=\frac{i\nu}{2} \int_0^t\left(q_1(0,t')q^*_1(-1,t')-q^*_1(0,t')q_1(-1,t')\right)dt',
\\
c_3(t)&=\frac{i\nu}{2} \int_0^t\left(q_1(0,t')q^*_2(-1,t')+q_2(0,t')q^*_1(-1,t')-q^*_1(0,t')q_2(-1,t')-q^*_2(0,t')q_1(-1,t')\right)dt'.
\label{ct1expc}
\end{split}
\end{eqnarray}
Using equations (\ref{psinls}) and (\ref{ct1exp}) we find
\begin{eqnarray}
\psi_2 (t,z)=\psi_{2,1} (t,z)\epsilon+\psi_{2,2} (t,z)\epsilon^2+\psi_{2,3} (t,z)\epsilon^3+O(\epsilon^4),
\label{psinlsexp}
\end{eqnarray}
where
\begin{eqnarray}
\begin{split}
\psi_{2,1} (t,z)=&i(z^{-1}-z^{-3}) \int_0^te^{2i\omega(z)(t-t')}\left(q_1(-1,t')+q_1(0,t')\right)dt',
\\
\psi_{2,2} (t,z)=&i(z^{-1}-z^{-3}) \int_0^te^{2i\omega(z)(t-t')}\left(q_2(-1,t')+q_2(0,t')\right)dt',
\\
\psi_{2,3} (t,z)=&\nu z^{-2} \left(\varphi_{2,3}^* (t,\frac{1}{z^*})-\varphi_{2,3}^* (t,z^*)\right)
\\&-\frac{i}{2}z^{-2}\left(\varphi_{2,1}^* (t,\frac{1}{z^*})+\varphi_{2,1}^* (t,z^*)\right) \int_0^t\left(q_1(0,t')q^*_1(-1,t')-q^*_1(0,t')q_1(-1,t')\right)dt'.
\label{psinlsexpc}
\end{split}
\end{eqnarray}
Substituting expansions (\ref{expvp}), (\ref{expba}) and (\ref{psinlsexp}) into (\ref{DN1}),
we can obtain the expansion for $q(0,t)$ to all orders of $\epsilon$.
For example, the first few terms are given by the following formulae:
\begin{eqnarray}
\begin{split}
q_1(0,t)=&-q_1(-1,t)+\frac{1}{\pi i}\int_{\partial D_{+in}}\psi_{2,1} (t,z)dz+\frac{1}{\pi i}\int_{\partial D_{+out}}
e^{2i\omega(z)t}\gamma_1(z)dz,
\\
q_2(0,t)=&-q_2(-1,t)+\frac{1}{\pi i}\int_{\partial D_{+in}}\psi_{2,2} (t,z)dz+\frac{1}{\pi i}\int_{\partial D_{+out}}
e^{2i\omega(z)t}\gamma_2(z)dz,
\\
q_3(0,t)=&-q_3(-1,t)+\frac{1}{\pi i}\int_{\partial D_{+in}}\psi_{2,3} (t,z)dz+\frac{1}{\pi i}\int_{\partial D_{+out}}
e^{2i\omega(z)t}(\gamma_1(z)\varphi_{1,2}(t,z)+\gamma_3(z))dz
\\&
+\frac{c_2(t)}{\pi i}\int_{\partial D_{+out}}
e^{2i\omega(z)t}\gamma_1(z)dz.
\label{DN1exp}
\end{split}
\end{eqnarray}

\subsubsection{Elimination of the unknown boundary values for the DMKDV equation}

Using similar calculations as in section 5.2.1, we can characterize the unknown boundary value $q(0,t)$ for the DMKDV eqaution (\ref{mkdv}).

\begin{proposition}
Define $\omega(z)$ via (\ref{mkdvw}).
Let $\partial D_{\pm in}$ and $\partial D_{\pm out}$ denote the boundaries of the domains $D_{\pm in}$ and $D_{\pm out}$ defined in (\ref{D2}),
oriented so that $D_{\pm in}$ and $D_{\pm out}$ lie in the left of $\partial D_{\pm in}$ and $\partial D_{\pm out}$.
Let $\varphi_1(t,z)$ and $\varphi_2(t,z)$ be defined by (\ref{vphi}) and $\widetilde{\varphi}_1(t,z)$, $\widetilde{\varphi}_2(t,z)$ be defined by the symmetries $\widetilde{\varphi}_1(t,z)= \varphi^*_1(t,\frac{1}{z^*})$, $\widetilde{\varphi}_2(t,z)=\nu \varphi^*_2(t,\frac{1}{z^*})$.
The unknown boundary value $q(0,t)$ associated with the DMKDV equation (\ref{mkdv}) is given by
\begin{eqnarray}
q(0,t)=-q(-1,t)+\frac{1}{\pi i}\int_{\partial D_{+in}}\psi_2 (t,z)dz+\frac{E_1(t)}{\pi i}\int_{\partial D_{+out}}
e^{2i\omega(z)t}\frac{\tilde{b}(z)}{\tilde{a}(z)}\varphi_1 (t,z)dz,
\label{DN2}
\end{eqnarray}
where
\begin{eqnarray}
\psi_2 (t,z)=-z^{-2}\left(\frac{1}{E_1(t)}\tilde{\varphi}_2(t,z)+iE_1(t)\tilde{\varphi}_2(t,iz^{-1})\right).
\label{psimkdv}
\end{eqnarray}
\end{proposition}

Proceeding as with the DNLS equation, we can derive an effective characterization of the unknown boundary value $q(0,t)$ for the DMKDV equation.
Indeed, assuming $\varphi_j(t,z)$, $\tilde{\phi}_j(n,z)$, $q(-1,t)$, $q(0,t)$, $q(n,0)$ in the same forms as in (\ref{expvp}), we find the expansions
\begin{eqnarray}
\begin{split}
\varphi_1(t,z)&=1+\varphi_{1,2}(t,z)\epsilon^2+\cdots+\varphi_{1,k}(t,z)\epsilon^k+\cdots,
\\
\varphi_2(t,z)&=\varphi_{2,1}(t,z)\epsilon+\varphi_{2,2}(t,z)\epsilon^2+\cdots+\varphi_{2,k}(t,z)\epsilon^k+\cdots,
\end{split}
\label{expmkdvvp}
\end{eqnarray}
where
\begin{eqnarray}
\begin{split}
\varphi_{1,2}(t,z)=&-\int_0^t[\left( zq_1(0,t')+z^{-1}q_1(-1,t')\right)\varphi_{2,1}(t',z)dt',
 \\
 \varphi_{2,1}(t,z)=&-\nu \int_0^te^{-2i\omega(z)(t-t')}\left( zq_1(-1,t')+ z^{-1}q_1(0,t')\right)dt',
\\
 \varphi_{2,2}(t,z)=&-\nu \int_0^te^{-2i\omega(z)(t-t')}\left( zq_2(-1,t')+ z^{-1}q_2(0,t')\right)dt',
 \\
 \varphi_{1,k}(t,z)=&-\int_0^t[2\sum_{l=1}^{k-1}\left( zq_l(0,t')+ z^{-1}q_l(-1,t')\right)\varphi_{2,k-l}(t',z)
\\& -\frac{\nu}{2}\sum_{m=2}^{k}\sum_{r=1}^{m-1}(q_r(0,t')q_{m-r}(-1,t')- q_r(-1,t')q_{m-r}(0,t'))\varphi_{1,k-m}(t',z)]dt', ~~k\geq 3,
 \\
 \varphi_{2,k}(t,z)=& -\nu \int_0^te^{-2i\omega(z)(t-t')}[\sum_{l=1}^{k}\left( zq_l(-1,t')+ z^{-1}q_l(0,t')\right)\varphi_{1,k-l}(t',z)
 \\& +\frac{1}{2}\sum_{m=2}^{k-1}\sum_{r=1}^{m-1}( q_r(0,t')q_{m-r}(-1,t')-q_r(-1,t')q_{m-r}(0,t'))\varphi_{2,k-m}(t',z)]dt', ~~k\geq 3.
\end{split}
\label{expvp2mkdv}
\end{eqnarray}
Furthermore,
\begin{eqnarray}
\begin{split}
\tilde{\phi}_1(n,z)&=C(n,0)[1+\tilde{\phi}_{1,2}(n,z)\epsilon^2+\cdots+\tilde{\phi}_{1,k}(n,z)\epsilon^k+\cdots],
\\
\tilde{\phi}_2(n,z)&=C(n,0)[\tilde{\phi}_{2,1}(n,z)\epsilon+\tilde{\phi}_{2,2}(n,z)\epsilon^2+\cdots+\tilde{\phi}_{2,k}(n,z)\epsilon^k+\cdots],
\end{split}
\label{expmkdvp}
\end{eqnarray}
where
\begin{eqnarray}
\begin{split}
\tilde{\phi}_{1,2}(n,z)=&\nu\sum_{m=n}^\infty \sum_{n'=m}^\infty z^{2(m-n')}q_1(m,0)q_1(n',0),
\\
 \tilde{\phi}_{2,1}(n,z)=&-\sum_{m=n}^\infty z^{2(n-m)-1}q_1(m,0),
~~
\tilde{\phi}_{2,2}(n,z)=-\sum_{m=n}^\infty z^{2(n-m)-1}q_2(m,0),
\\
 \tilde{\phi}_{1,k}(n,z)=& -\nu\sum_{m=n}^{\infty}\sum_{l=1}^{k-1}z q_{l}(m,0)\tilde{\phi}_{2,k-l}(m,z),~~k\geq 3,
 \\
  \tilde{\phi}_{2,k}(n,z)=& -\sum_{m=n}^{\infty}\sum_{l=1}^{k}z^{2(n-m)-1} q_{l}(m,0)\tilde{\phi}_{1,k-l}(m,z),~~k\geq 3.
\end{split}
\label{expp2mkdv}
\end{eqnarray}
The expansion of $\frac{\tilde{b}(z)}{\tilde{a}(z)}$ is given by equations (\ref{expba}) and (\ref{expbac}),
where $\tilde{\phi}_{1,j}(n,z)$ and $\tilde{\phi}_{2,j}(n,z)$ are replaced by (\ref{expp2mkdv}).
Recall that for the DMKDV equation, the quantity $E_1(t)$ is defined by
\begin{eqnarray}
E_1(t)=\exp\left(-\nu \int_0^tq(0,t')q(-1,t')dt'\right).
\label{ct2}
\end{eqnarray}
The above formula implies the expansion
\begin{eqnarray}
E_1(t)=1+c_2(t)\epsilon^2+c_3(t)\epsilon^3+O(\epsilon^4),
\label{ct1mkdvexp}
\end{eqnarray}
where
\begin{eqnarray}
\begin{split}
c_2(t)&=-\nu\int_0^tq_1(0,t')q_1(-1,t')dt',
\\
c_3(t)&=-\nu\int_0^t\left(q_1(0,t')q_2(-1,t')+q_2(0,t')q_1(-1,t')\right)dt'.
\label{ct1mkdvexpc}
\end{split}
\end{eqnarray}
Using equation (\ref{psimkdv}) we find the expansion
$$\psi_2 (t,z)=\psi_{2,1} (t,z)\epsilon+\psi_{2,2} (t,z)\epsilon^2+\psi_{2,3} (t,z)\epsilon^3+O(\epsilon^4),$$
where
\begin{eqnarray}
\begin{split}
\psi_{2,1} (t,z)=&z^{-2} \int_0^te^{2i\omega(z)(t-t')}\left((z+z^{-1})q_1(-1,t')+(z-z^{-1})q_1(0,t')\right)dt',
\\
\psi_{2,2} (t,z)=&z^{-2} \int_0^te^{2i\omega(z)(t-t')}\left((z+z^{-1})q_2(-1,t')+(z-z^{-1})q_2(0,t')\right)dt',
\\
\psi_{2,3} (t,z)=&-\nu z^{-2} \left(\varphi_{2,3}^* (t,\frac{1}{z^*})+i\varphi_{2,3}^* (t,iz^*)\right)
\\&-z^{-2}\left(\varphi_{2,1}^* (t,\frac{1}{z^*})-i\varphi_{2,1}^* (t,iz^*)\right) \int_0^t q_1(0,t')q_1(-1,t')dt'.
\label{psimkdvexpc}
\end{split}
\end{eqnarray}
Substituting expansions (\ref{expmkdvvp}), (\ref{ct1mkdvexp}) and (\ref{psimkdvexpc}) into (\ref{DN2}),
we can obtain an expansion of $q(0,t)$ for the DMKDV equation.
The first few terms are listed in the formulas (\ref{DN1exp}), where the domains $D_{+ in}$ and $D_{+ out}$ are replaced by (\ref{D2}),
and the expressions for $\varphi_{k,j}(t,z)$, $c_j(t)$ and $\psi_{2,j}(t,z)$  are replaced by (\ref{expvp2mkdv}), (\ref{ct1mkdvexpc}) and (\ref{psimkdvexpc}).

\section {Conclusions}
By implementing the unified transform (Fokas method),
we have expressed the solutions of certain IBVPs for the DNLS equation (\ref{nls}) and DMKDV  equation (\ref{mkdv})
in terms of solutions of appropriate RH problems formulated in the complex $z$-plane.
We have also studied the elimination of the unknown boundary values for both the DNLS and DMKDV equations.
As with the case of the unified method in the continuous case,
the main advantage of this method in the discrete case is that the associated RH problem involves a jump matrix with explicit
exponential $(n, t)$ dependence,
and thus it is possible to study the long-time asymptotics for the associated integrable nonlinear differential-difference equations.

For integrable PDEs there exists a well developed formalism for the rigorous implementation of the Fokas method.
For example, for the nonlinear Schr\"{o}dinger equation this formalism involves the following steps:
(i) assuming that there exists a solution $q(x,t)$ with sufficient smoothness and decay,
express this solution in terms of the solution of a $2\times 2$ matrix RH problem which is uniquely characterized in terms of
certain spectral functions. These functions are uniquely defined in terms of the initial datum $q_0(x)=q(x,0)$ and the boundary values $g_0(t)=q(0,t)$
and $g_1(t)=q_x(0,t)$. Furthermore, they satisfy an algebraic relation called the global relation.
The analogous step for the nonlinear differential-difference equations analysed here is implemented in sections 4.1.
(ii) Prove that the above RH problem has a unique solution and show that the inverse problem solved the direct problem, namely,
show that if there exist functions $g_0(t)$ and $g_1(t)$ which satisfy the global relation, then the function $q(x,t)$ which is defined in terms of
the above RH problem in terms of $\left\{q_0(x), g_0(t), g_1(t)\right\}$, satisfies the nonlinear Schr\"{o}dinger equation, and furthermore,
satisfies $q(x,0)=q_0(x)$, as well as $q(0,t)=g_0(t)$, $q_x(0,t)=g_1(t)$.
The analogous step here is implemented in section 4.2.
(iii) Characterize the Dirichlet to Neumann map, namely, given $g_0(t)$ show that the global relation uniquely determines $g_1(t)$.
The formal implementation of the analogous step here is presented in section 5. However, the rigorous implementation of this step remains open
(for the nonlinear Schr\"{o}dinger equation the rigorous implementation was carried out only recently in the important work of Antonopoulou and Kamvissis \cite{AK}).

The approach presented here can be generalized to other integrable discrete nonlinear evolution equations, such as the Volterra lattice \cite{V1,V2}, the Toda lattice \cite{T1} and the four-potential Ablowitz-Ladik lattice \cite{AL1,Geng2,ZC}.

\section*{ACKNOWLEDGMENTS}

B. Xia was supported by the National Natural Science Foundation of China (Grant No. 11301229)
and by the Jiangsu Government Scholarship for Overseas Studies. A.S. Fokas was supported by EPSRC in the form of a senior fellowship.

\begin{appendices}
\section{Asymptotic behavior of the eigenfunctions}
\subsection{Asymptotic behavior of the eigenfunctions of the initial value problems}
Let us first fix some notations: we will denote the diagonal and off-diagonal parts of a matrix $M$ by $M_D$ and $M_O$ respectively.
Since $\mu_1^L(n,t,z)$ are analytic for $|z| > 1$, they have a
convergent Laurent series expansion about the point $z =\infty$, while $\mu_1^R(n,t,z)$ are analytic for $|z| < 1$, they have a convergent power series expansion about $z=0$.
This implies that we can expand $\mu_1(n,t,z)$ in the following form
\begin{eqnarray}
\mu_1(n,t,z)=\frac{C(n,t)}{C(-\infty)}\left(M^{(0)}(n,t)+M^{(1)}(n,t)Z^{-1}+M^{(2)}(n,t)Z^{-2}+\cdots\right).
\label{A11}
\end{eqnarray}
Substituting this expansion into  (\ref{mua}) and matching the powers of $Z^{-1}$ we find
\begin{eqnarray}
M^{(0)}(n,t)=I,~~M^{(1)}(n,t)=Q(n-1,t),~~M^{(2)}(n,t)=\sum_{m=-\infty}^{n-1}Q(m,t)Q(m-1,t).
\label{A12}
\end{eqnarray}
 Moreover, by induction we find that
\begin{eqnarray}
M_O^{(2j)}(n,t)=0,~~M_D^{(2j+1)}(n,t)=0, ~~j\geq 1.
\label{A13}
\end{eqnarray}
Equations (\ref{A11})-(\ref{A13}) imply the asymptotic behavior (\ref{aspaivp}).
The asymptotic behavior (\ref{aspbivp}) can be derived in a similar procedure but with the following adjustment:
 instead of deriving the expansion of $\mu_2(n,t,z)$ directly, we derive the expansion of $\hat{\mu}_2(n,t,z)=C(n,t)\mu_2(n,t,z)$.
Since $\hat{\mu}_2^L(n,t,z)$ are analytic for $|z| < 1$, they have a convergent power series expansion at $z=0$,
while $\hat{\mu}_2^R(n,t,z)$  are analytic for $|z| < 1$, they have a convergent Laurent series expansion at $z =\infty$.
Thus we suppose the expression of $\hat{\mu}_2(n,t,z)$ as the form
\begin{eqnarray}
\hat{\mu}_2(n,t,z)=\hat{M}^{(0)}(n,t)+\hat{M}^{(1)}(n,t)Z+\hat{M}^{(2)}(n,t)Z^{2}+\cdots.
\label{A14}
\end{eqnarray}
Proceeding as earlier, we find
\begin{eqnarray}
\hat{\mu}_2(n,t,z)=I-Q(n,t)Z+\left( \begin{array}{cc} O(z^{2},\text{even}) & O(z^{-3},\text{odd})  \\
  O(z^{3},\text{odd}) & O(z^{-2},\text{even}) \\ \end{array} \right),  ~z\rightarrow (0,\infty).
\label{A15}
\end{eqnarray}
The above formula immediately implies the asymptotic behavior (\ref{aspbivp}).

\subsection{Asymptotic behavior of the eigenfunctions of the IBVPs}

By splitting the first equation of (\ref{muibv}) into its diagonal and off-diagonal parts, we can construct the Neumann series
\begin{eqnarray}
\mu_{1}(n,t,z)=\frac{C(n,t)}{C(0,t)}\sum_{j=0}^{\infty} \mu^j_{1}(n,t,z),
\label{NS}
\end{eqnarray}
where
\begin{eqnarray}
\begin{split}
\mu^0_{1,D}(n,t,z)=&I, ~~\mu^0_{1,O}(n,t,z)=0,
\\
\mu^{j+1}_{1,D}(n,t,z)=&\int_{0}^{t}\left(H_D\mu^j_{1,D}(0,t',z)+H_O\mu^{j+1}_{1,O}(0,t',z)\right)dt'
+\sum_{m=0}^{n-1} Q(m,t)\mu^{j+1}_{1,O}(m,t,z)Z^{-1},
\\
\mu^{j+1}_{1,O}(n,t,z)=&\int_{0}^{t}e^{iw(z)(t-t')\hat{\sigma}_3}\left(H_O\mu^j_{1,D}(0,t',z)+H_D\mu^j_{1,O}(0,t',z)\right)Z^{-2n}dt'
\\
&+\sum_{m=0}^{n-1} Q(m,t)\mu^j_{1,D}(m,t,z)Z^{-2(n-m)+1}.
\end{split}
\label{NSE}
\end{eqnarray}
For the case $n=0$, equations (\ref{NSE}) become
\begin{subequations}
\begin{eqnarray}
&&\mu^0_{1,D}(0,t,z)=I, ~~\mu^0_{1,O}(0,t,z)=0,
\label{0NSEa}
\\
&&\mu^{j+1}_{1,D}(0,t,z)=\int_{0}^{t}\left(H_D\mu^j_{1,D}(0,t',z)+H_O\mu^{j+1}_{1,O}(0,t',z)\right)dt',
\label{0NSEb}
\\
&&\mu^{j+1}_{1,O}(0,t,z)=\int_{0}^{t}e^{iw(z)(t-t')\hat{\sigma}_3}\left(H_O\mu^j_{1,D}(0,t',z)+H_D\mu^j_{1,O}(0,t',z)\right)dt'.
\label{0NSEc}
\end{eqnarray}
\label{0NSE}
\end{subequations}
Note that
\begin{eqnarray}
\frac{1}{2\omega(z)}I=\left( \begin{array}{cc} \frac{1}{2\alpha} & 0 \\
 0 &  \frac{1}{2\beta} \\ \end{array} \right)Z^{-2}+O(Z^{-6}),~~z\rightarrow(\infty,0).
\label{omep}
\end{eqnarray}
Integration by parts, implies as $z\rightarrow(\infty,0)$ that (\ref{0NSEc}) has the expansion
\begin{eqnarray}
\begin{split}
\mu^{j+1}_{1,O}(0,t,z)=&[Q(-1,t)\mu^{j}_{1,D}(0,t,z)-e^{i\omega(z)t\hat{\sigma}_3}Q(-1,0)\mu^{j}_{1,D}(0,0,z)]Z^{-1}
\\&+\frac{1}{2}[(D_1Q(0,t)Q(-1,t)+Q(-1,t)Q(0,t))\mu^{j}_{1,O}(0,t,z)
\\&-e^{i\omega(z)t\hat{\sigma}_3}(D_1Q(0,0)Q(-1,0)+Q(-1,0)Q(0,0))\mu^{j}_{1,O}(0,0,z)]Z^{-2}
\\&+O( Z^{-3}),
\end{split}
\label{mOrl2}
\end{eqnarray}
where $$D_1=\left( \begin{array}{cc} \frac{\alpha}{\beta} & 0 \\
 0 &  \frac{\beta}{\alpha} \\ \end{array} \right).$$
Substituting the above equation into (\ref{0NSEb}) we find
\begin{eqnarray}
\begin{split}
\mu^{j+1}_{1,D}(0,t,z)=&i\int_{0}^{t}\left(\Gamma_1Q(0,t')Q(-1,t')-\Gamma_2Q(-1,t')Q(0,t')\right)\mu^j_{1,D}(0,t',z)dt'
\\
&-2i\int_{0}^{t}\Gamma_1Q(0,t')e^{i\omega(z)t'\hat{\sigma}_3}\left(Q(-1,0)\mu^{j}_{1,D}(0,0,z)\right)dt'
\\
&+\{i\int_{0}^{t}\Gamma_1Q(0,t')[(D_1Q(0,t')Q(-1,t')+Q(-1,t')Q(0,t'))\mu^{j}_{1,O}(0,t',z)
\\&-e^{i\omega(z)t'\hat{\sigma}_3}(D_1Q(0,0)Q(-1,0)+Q(-1,0)Q(0,0))\mu^{j}_{1,O}(0,0,z)]dt'\}Z^{-1}
\\&+O( Z^{-2}),
\end{split}
\label{mDrl2}
\end{eqnarray}
where
$$\Gamma_1=\left( \begin{array}{cc} \alpha & 0 \\
 0 &  -\beta \\ \end{array} \right),~~\Gamma_2=\left( \begin{array}{cc} \beta & 0 \\
 0 &  -\alpha \\ \end{array} \right).$$
As $z\rightarrow(\infty,0)$, the first term in the right hand side of (\ref{mDrl2}) contributes to the leading term of $\mu^{j+1}_{1,D}(0,t,z)$.
More precisely, as $z\rightarrow(\infty,0)$, we obtain
\begin{eqnarray}
\begin{split}
&\mu^0_{1,D}(0,t,z)=I,
\\
&\mu^{1}_{1,D}(0,t,z)=\left(i \int_{0}^{t}(\alpha q(0,t')p(-1,t')-\beta q(-1,t')p(0,t'))dt'\right)I+O(Z^{-2}),
\\
&\mu^{2}_{1,D}(0,t,z)=\frac{1}{2}\left(i \int_{0}^{t}(\alpha q(0,t')p(-1,t')-\beta q(-1,t')p(0,t'))dt'\right)^2I+O(Z^{-2}),
\\
&\mu^{j}_{1,D}(0,t,z)=\frac{1}{j!}\left(i \int_{0}^{t}(\alpha q(0,t')p(-1,t')-\beta q(-1,t')p(0,t'))dt'\right)^{j}I+O(Z^{-2}), ~j\geq2.
\end{split}
\label{leadtermmjD}
\end{eqnarray}
Thus, we have
\begin{eqnarray}
\begin{split}
&\mu_{1,D}(0,t,z)=E_1(t)I+O(Z^{-2}),
\end{split}
\label{leadtermmD}
\end{eqnarray}
where
$$E_1(t)=\exp\left(i \int_{0}^{t}\left(\alpha q(0,t')p(-1,t')-\beta q(-1,t')p(0,t')\right)dt'\right).$$
Using (\ref{mOrl2}) and (\ref{leadtermmjD}), we find
\begin{eqnarray}
\begin{split}
&\mu_{1,O}(0,t,z)=\left(E_1(t)Q(-1,t)-e^{i\omega(z)t\hat{\sigma}_3}Q(-1,0)\right)Z^{-1}+O(Z^{-3}).
\end{split}
\label{leadtermmO}
\end{eqnarray}
Moreover, from (\ref{0NSE}) it can be shown by induction that the expansion of $\mu_{1,D}(0,t,z)$ only involves the even powers of $Z^{-1}$, whereas the expansion of $\mu_{1,O}(0,t,z)$ only involves the odd powers of $Z^{-1}$. This fact together with (\ref{leadtermmD}) and (\ref{leadtermmO}) yield that the asymptotic behavior of $\mu_{1}(0,t,z)$  are given by the first of (\ref{asp2}). The asymptotic behavior of $\mu_{3}(0,t,z)$ can be derived in a similar manner.

For the case $n\geq 1$, instead of (\ref{mOrl2}) and (\ref{mDrl2}) we find the following formulae:
\begin{eqnarray}
\begin{split}
\mu^{j+1}_{1,O}(n,t,z)=&\sum_{m=0}^{n-1}Q(m,t)\mu^{j}_{1,D}(m,t,z)Z^{-2(n-m)+1}
\\&+[Q(-1,t)\mu^{j}_{1,D}(0,t,z)-e^{i\omega(z)t\hat{\sigma}_3}Q(-1,0)\mu^{j}_{1,D}(0,0,z)]Z^{-(2n+1)}
\\&+\frac{1}{2}[(D_1Q(0,t)Q(-1,t)+Q(-1,t)Q(0,t))\mu^{j}_{1,O}(0,t,z)
\\&-e^{i\omega(z)t\hat{\sigma}_3}(D_1Q(0,0)Q(-1,0)+Q(-1,0)Q(0,0))\mu^{j}_{1,O}(0,0,z)]Z^{-2(n+1)}
\\&+O( Z^{-(2n+3)}),
\end{split}
\label{mOrl3}
\end{eqnarray}
\begin{eqnarray}
\begin{split}
\mu^{j+1}_{1,D}(n,t,z)=&\sum_{m=0}^{n-1}Q(m,t)\mu^{j+1}_{1,O}(m,t,z)Z^{-1}
\\&+i\int_{0}^{t}\left(\Gamma_1Q(0,t')Q(-1,t')-\Gamma_2Q(-1,t')Q(0,t')\right)\mu^j_{1,D}(0,t',z)dt'
\\
&-2i\int_{0}^{t}\Gamma_1Q(0,t')e^{i\omega(z)t'\hat{\sigma}_3}\left(Q(-1,0)\mu^{j}_{1,D}(0,0,z)\right)dt'
\\
&+\{i\int_{0}^{t}\Gamma_1Q(0,t')[(D_1Q(0,t')Q(-1,t')+Q(-1,t')Q(0,t'))\mu^{j}_{1,O}(0,t',z)
\\&-e^{i\omega(z)t'\hat{\sigma}_3}(D_1Q(0,0)Q(-1,0)+Q(-1,0)Q(0,0))\mu^{j}_{1,O}(0,0,z)]dt'\}Z^{-1}
\\&+O( Z^{-2}).
\end{split}
\label{mDrl3}
\end{eqnarray}
Proceeding as in the case $n=0$, we find
\begin{eqnarray}
\begin{split}
\mu_1(n,t,z)&=\frac{C(n,t)}{C(0,t)}E_1(t)\left(I+Q(n-1,t)Z^{-1}+\left( \begin{array}{cc} O(z^{-2},\text{even}) & O(z^{3},\text{odd})  \\
  O(z^{-3},\text{odd}) & O(z^{2},\text{even}) \\ \end{array} \right)\right), z\rightarrow (\infty,0),
\\
\mu_3(n,t,z)&=\frac{C(n,t)}{C(0,t)}E_2(t)\left(I+Q(n-1,t)Z^{-1}+\left( \begin{array}{cc} O(z^{-2},\text{even}) & O(z^{3},\text{odd})  \\
  O(z^{-3},\text{odd}) & O(z^{2},\text{even}) \\ \end{array} \right)\right),  z\rightarrow (\infty,0).
\end{split}
\label{asp10}
\end{eqnarray}
By employing equation (\ref{aleq}) we can compute the $t$-derivative of $C(0,t)$:
\begin{eqnarray}
C_t(0,t)=i\left(\alpha q(0,t)p(-1,t)-\beta p(0,t)q(-1,t)\right)C(0,t).
\label{ct}
\end{eqnarray}
Hence,
\begin{eqnarray*}
C(0,t)=C(0,0)\exp\left(i\int_0^t\left(\alpha q(0,t')p(-1,t')-\beta p(0,t')q(-1,t')\right)dt'\right),
\label{cts00}
\end{eqnarray*}
which is
\begin{eqnarray}
C(0,t)=C(0,0)E_1(t).
\label{cts}
\end{eqnarray}
Using (\ref{cts}) in equations (\ref{asp10}) we find the asymptotic behavior (\ref{asp1}).

\section{ Proof of property (v) in Proposition 4}
In order to derive property (v) in Proposition 4 we define the vector function  $\psi(n, z) = (\psi_1(n,z),\psi_2(n,z))^\mathcal{T}$ as the unique solution of
\begin{eqnarray}
\begin{split}
&f_0(n)\psi_1(n+1,z)-\psi_1(n,z)=z^{-1}q_0(n)\psi_{2}(n,z),
\\
&f_0(n)\psi_2(n+1,z)-z^{-2}\psi_2(n,z)=z^{-1}p_0(n)\psi_{1}(n,z),
\\
&\psi(0, z)=(1,0)^\mathcal{T},
\end{split}
\label{Proofdef1phi}
\end{eqnarray}
where as before, $f_0(n)=\sqrt{1-q_0(n)p_0(n)}$, with $p_0(n)=\nu q^*_0(n)$ for the DNLS equation, and $p_0(n)=\nu q_0(n)$ for the DMKDV equation, respectively.
Note that equations (\ref{def1phi}) are equivalent to the summation equations
\begin{eqnarray}
\begin{split}
\phi_1(n,z)=&C(n,0)\left(1-z^{-1}\sum_{m=n}^{\infty} \frac{1}{C(m,0)}q_0(m)\phi_{2}(m,z)\right),
\\
\phi_2(n,z)=&-C(n,0)\sum_{m=n}^{\infty} \frac{z^{-2(n-m)+1}}{C(m,0)}p_0(m)\phi_{1}(m,z).
\end{split}
\label{def11phi}
\end{eqnarray}
Similarly, equations (\ref{Proofdef1phi}) are equivalent to the summation equations
\begin{eqnarray}
\begin{split}
\psi_1(n,z)=&C(n,0)\left(\frac{1}{C(0,0)}+z^{-1}\sum_{m=0}^{n-1} \frac{1}{C(m,0)}q_0(m)\psi_{2}(m,z)\right),
\\
\psi_2(n,z)=&C(n,0)\sum_{m=0}^{n-1} \frac{z^{-2(n-m)+1}}{C(m,0)}p_0(m)\psi_{1}(m,z).
\end{split}
\label{Proofdef11phi}
\end{eqnarray}
Because $\phi_1(n,z)$ and $\phi_2(n,z)$ are analytic for $|z| < 1$, they have a convergent power series expansion about $z=0$.
Substituting this expansion into the summation equations (\ref{def11phi}) and matching the powers of $z$, we find:
\begin{eqnarray}
\begin{split}
\phi(n,z)=\frac{1}{C(n,0)}\left( \begin{array}{cc} 1+O(z^{2},\text{even})  \\
 -p_0(n)z+ O(z^{3},\text{odd}) \\ \end{array} \right), \quad z\rightarrow 0.
\end{split}
\label{aspdef1proofphi}
\end{eqnarray}
Similarly, because $\psi_1(n,z)$ and $\psi_2(n,z)$ are analytic for $|z| > 1$, they have a convergent Laurent series expansion about $z=\infty$.
Substituting this expansion into the summation equations (\ref{Proofdef11phi}) and matching the powers of $z^{-1}$, we find:
\begin{eqnarray}
\begin{split}
\psi(n,z)=\frac{C(n,0)}{C(0,0)}\left( \begin{array}{cc} 1+O(z^{-2},\text{even})  \\
 p_0(n-1)z^{-1}+ O(z^{-3},\text{odd}) \\ \end{array} \right), \quad z\rightarrow \infty.
\end{split}
\label{aspdef1proofpsi}
\end{eqnarray}

We introduce the notations
\begin{eqnarray}
\phi^*(n,z)=\left(\nu\phi_2^*(n,\frac{1}{z^*}), \phi_1^*(n,\frac{1}{z^*})\right)^T, \quad \psi^*(n,z)=\left(\nu \psi_2^*(n,\frac{1}{z^*}),\psi_1^*(n,\frac{1}{z^*})\right)^T.
\label{proofpr}
\end{eqnarray}
Let the matrices $\mu_1(n,z)$ and $\mu_2(n,z)$ be defined by
\begin{eqnarray}
\mu_1(n,z)=\left(\psi(n,z),\psi^*(n,z)\right), \quad \mu_2(n,z)=\left(\phi(n,z),\phi^*(n,z)\right).
\label{proofmu}
\end{eqnarray}
Then $\mu_1(n,z)$ and $\mu_2(n,z)$ satisfy the $n$-part of Lax pair (\ref{LPS3}) evaluated at $t=0$, namely,
\begin{eqnarray}
&&f_0(n)\mu(n+1,z)-\hat{Z}\mu(n,z)=Q_0(n)\mu(n,z)Z^{-1},
\label{LPS3proof}
\end{eqnarray}
where
$$Q_0(n)=\left( \begin{array}{cc} 0 & q_0(n) \\ p_0(n) &  0 \\ \end{array} \right).$$
 This in turn implies that $\mu_1(n,z)$ and $\mu_2(n,z)$ are related by:
\begin{eqnarray}
\mu_2(n,z)=\mu_1(n,z)\hat{Z}^ns(z), ~~|z|=1.\label{proofs}
\end{eqnarray}
We define
\begin{eqnarray}
\begin{split}
&M_{-}^{(n)}(n,z)=\frac{1}{C(n,0)}\left(\frac{\psi(n,z)}{a^*(\frac{1}{z^*})},\phi^*(n,z)\right), \quad |z|\geq 1,
 \\
&M_{+}^{(n)}(n,t,z)=\frac{1}{C(n,0)}\left(\phi(n,z),\frac{\psi^*(n,z)}{a(z)}\right), \quad |z|\leq 1.
\end{split}
\label{proofM1}
\end{eqnarray}
It can be checked directly that the function $M^{(n)}(n,z)$ defined in (B.6) satisfies the
RH problem (\ref{abJc}) with the jump matrix defined by (\ref{abJr}).
Moreover, equations (\ref{proofM1}) together with (\ref{aspdef1proofphi}) and (\ref{aspdef1proofpsi})
yield the asymptotic expansion (\ref{abMA}).

We now  prove the residue conditions at the
possible simple zeros $\{z_j\}_{1}^{2\mathcal{K}}$ of $a(z)$. It follows from equation (\ref{proofs}) that
\begin{eqnarray}
\phi(n,z)=a(z)\psi(n,z)+z^{-2n}b(z)\psi^*(n,z).\label{proofs1}
\end{eqnarray}
Evaluating (\ref{proofs1}) at $z=z_j$, we find
\begin{eqnarray}
\phi(n,z_j)=z_j^{-2n}b(z_j)\psi^*(n,z_j).\label{proofs2}
\end{eqnarray}
Equations (\ref{proofM1}) and (\ref{proofs2}) gives rise to the residue conditions (\ref{abrr1a}).
The residue conditions (\ref{abrr1b}) can be derived similarly.

Substituting the asymptotic expansion
\begin{eqnarray*}
C(n,0)M^{(n)}(n,z)=C(n,0)\left(I+M_1(n)Z^{-1}+\left( \begin{array}{cc} O(z^{-2},\text{even}) & O(z^{3},\text{odd})  \\
  O(z^{-3},\text{odd}) & O(z^{2},\text{even}) \\ \end{array} \right)\right), ~ z\rightarrow (\infty,0),
\label{abMAproof}
\end{eqnarray*}
into (\ref{LPS3proof}), we find
\begin{eqnarray}
q_0(n)=\left(M_1(n)\right)^{12}=\lim_{z\rightarrow 0}(z^{-1}M^{(n)}(n+1,z))^{12},
\label{abisproof}
\end{eqnarray}
which is the formula (\ref{abis}). This completes the proof of the property (v) in Proposition 4.
\vspace{0.2cm}
\\
{\bf Remark 6.} Definition 1 gives rise to the map
$\mathbb{S} : \left\{q_0(n)\right\}\rightarrow\left\{a(z), b(z)\right\}$.
The inverse of this map $\mathbb{Q} : \left\{a(z), b(z)\right\}\rightarrow\left\{q_0(n)\right\}$ is defined by (\ref{abis}).
Following the same lines as with the case of integrable nonlinear Schr\"{o}dinger equation (see, for example, \cite{F4}),
it can be shown that $\mathbb{S}^{-1}=\mathbb{Q}$.

\section{Proof of property (v) in Proposition 5}
In order to derive property (v) in Proposition 5 we define the vector function  $\eta(t, z) = (\eta_1(t,z),\eta_2(t,z))^\mathcal{T}$ by
\begin{eqnarray}
\begin{split}
&(\eta_1)_t=H_0^{11}\eta_1+H_0^{12}\eta_2,
\\
&(\eta_2)_t+2i\omega(z)\eta_2=H_0^{21}\eta_1+H_0^{22}\eta_2,
\\
&\eta(T, z) =(1,0)^\mathcal{T},
\end{split}
\label{ABphidefproof}
\end{eqnarray}
where $\omega(z)$ is defined by (\ref{nlsw}) for the DNLS equation, and by (\ref{mkdvw})  for the DMKDV equation respectively.
Note that equations (\ref{ABphidef}) are equivalent to the linear Volterra integral equations
\begin{eqnarray}
\begin{split}
&\varphi_1(t,z)=1+\int_0^{t}\left(H_0^{11}\varphi_1+H_0^{12}\varphi_2\right)(t',z)dt',
\\
&\varphi_2(t,z)=\int_0^{t}e^{-2i\omega(z)(t-t')}\left(H_0^{21}\varphi_1+H_0^{22}\varphi_2\right)(t',z)dt'.
\end{split}
\label{ABphidef1}
\end{eqnarray}
Similarly, equations (\ref{ABphidefproof}) are equivalent to the linear Volterra integral equations
\begin{eqnarray}
\begin{split}
&\eta_1(t,z)=1-\int_t^{T}(H_0^{11}\eta_1+H_0^{12}\eta_2)(t',z)dt',
\\
&\eta_2(t,z)=-\int_t^{T}e^{-2i\omega(z)(t-t')}(H_0^{21}\eta_1+H_0^{22}\eta_2)(t',z)dt'.
\end{split}
\label{ABphidefproof1}
\end{eqnarray}
Integration by parts implies that, as $z\rightarrow \infty$ with $z\in\bar{D}_{-}$, we have
\begin{eqnarray}
\begin{split}
\varphi(t,z)&=\left( \begin{array}{cc} E_1(t)+O(z^{-2},\text{even})  \\
 O(z^{-1},\text{odd}) \\ \end{array} \right), \quad z\rightarrow \infty,
\end{split}
\label{aspdef21proof}
\end{eqnarray}
where $E_1(t)$ are defined by (\ref{Ea}).
As $z\rightarrow \infty$ with $z\in\bar{D}_{+}$, we have
\begin{eqnarray}
\begin{split}
\eta(t,z)&=\left( \begin{array}{cc} E_2(t)+O(z^{-2},\text{even})  \\
 O(z^{-1},\text{odd}) \\ \end{array} \right), \quad z\rightarrow \infty,
\end{split}
\label{aspdef22proof}
\end{eqnarray}
where $E_2(t)$ are defined by (\ref{Eb}).

We introduce the notations
\begin{eqnarray}
\varphi^*(t,z)=\left(\nu \varphi_2^*(t,\frac{1}{z^*}), \varphi_1^*(t,\frac{1}{z^*})\right)^T, \quad \eta^*(t,z)=\left(\nu\eta_2^*(t,\frac{1}{z^*}), \eta_1^*(t,\frac{1}{z^*})\right)^T.
\label{proofpr}
\end{eqnarray}
Let
\begin{eqnarray}
\mu_1(t,z)=\left(\varphi(t,z),\varphi^*(t,z)\right), \quad \mu_3(t,z)=\left(\eta(t,z),\eta^*(t,z)\right).
\label{proofmu}
\end{eqnarray}
These functions satisfy the $t$-part of Lax pair (\ref{LPT3}) evaluated at $n=0$, namely,
\begin{eqnarray}
&&\mu_t-i\omega(z)[\sigma_3,\mu]=
H_0(t,z)\mu.
\label{LPT3proof}
\end{eqnarray}
This implies that they are related by:
\begin{eqnarray}
\mu_3(t,z)=\mu_1(t,z)e^{i\omega(z)t\hat{\sigma}}S(z), ~~z\in\left\{z|~ z\in \mathbb{C}, ~\text{Im} (\omega(z))=0\right\}.\label{proofS}
\end{eqnarray}
Let
\begin{eqnarray}
\begin{split}
&M_{-}^{(t)}(t,z)=\frac{1}{E_1(t)}\left(\varphi(t,z),\frac{\eta^*(t,z)}{A^*(\frac{1}{z^*})}\right), \quad z\in \bar{D}_{-}=\left\{z\Big| z\in\mathbb{C}, \text{Im} (\omega(z))\leq0\right\},
 \\
&M_{+}^{(t)}(t,z)=\frac{1}{E_1(t)}\left(\frac{\eta(t,z)}{A(z)},\varphi^*(t,z)\right), \quad z\in \bar{D}_{+}=\left\{z\Big| z\in\mathbb{C}, \text{Im} (\omega(z))\geq0\right\}.
\end{split}
\label{proofM2}
\end{eqnarray}
It can be verified directly that the function  $M^{(t)}(t,z)$ defined in (C.6) can be rewritten as (\ref{ABJc}) with the jump matrix defined by (\ref{AB1Jr}).
Moreover, equations (\ref{aspdef21proof}),  (\ref{aspdef22proof}) together with (\ref{proofM2}) yield the estimate (\ref{ABMA}).

The proof of the residue
conditions (\ref{ABrr1a}) and  (\ref{ABrr1b}) at the possible simple zeros $\left\{k_j\right\}_1^{2K}$ of $A(k)$ follows the same lines as in the case
of the function $M^{(n)}(n,z)$. The formulae (\ref{bvs}) can be derived by substituting the asymptotic expansion
$E_1(t)M^{(t)}(t,z)$ into (\ref{LPT3proof}).
\vspace{0.2cm}
\\
{\bf Remark 7.} Definition 2 gives rise to the map
$\hat{\mathbb{S}} : \left\{g_{-1}(t),g_{0}(t)\right\}\rightarrow\left\{A(z), B(z)\right\}$.
The inverse of this map $\hat{\mathbb{Q}} : \left\{A(z), B(z)\right\}\rightarrow\left\{g_{-1}(t),g_{0}(t)\right\}$ is defined by (\ref{bvs}).
In analogy with the integrable nonlinear Schr\"{o}dinger equation in the continuous case (see, for example, \cite{F4}),
it can be shown that $\hat{\mathbb{S}}^{-1}=\hat{\mathbb{Q}}$.

\section{ Proof that the formulae for $q(n,t)$ satisfy the DNLS and DMKDV equations}
Following (\ref{Masp1Th}) we can write the asymptotic expansion of $M(n, t, z)$ in the form
\begin{eqnarray}
\begin{split}
M=I+\left( \begin{array}{cc} 0 & M^{12}_1(n,t) \\
 M^{21}_1(n,t) &  0 \\ \end{array} \right)Z^{-1}+\left( \begin{array}{cc} O(z^{-2},\text{even}) & O(z^{3},\text{odd})  \\
  O(z^{-3},\text{odd}) & O(z^{2},\text{even}) \\ \end{array} \right),  z\rightarrow (\infty,0).
  \end{split}
\label{appasym}
\end{eqnarray}

We define a pair of linear operators $\mathcal{L}$ and $\mathcal{P}$ by
\begin{eqnarray}
\begin{split}
&\mathcal{L}M(n, t, z)=M(n+1,t,z)Z-ZM(n, t, z)-Q(n,t)M(n, t, z),
\\
&\mathcal{P}M(n, t, z)=\left(M(n, t, z)\right)_t-i\omega(z)[\sigma_3,M(n, t, z)]-H(n,t,z)M(n, t, z),
\end{split}
\label{LPA}
\end{eqnarray}
with
\begin{eqnarray}
\begin{split}
&Q(n,t)=\left( \begin{array}{cc} 0 & M^{12}_1(n+1,t) \\
 M^{21}_1(n+1,t) &  0 \\ \end{array} \right),
 \\
&H(n,t,z)=2i\left(D_1Q(n-1)Z+D_2Q(n)Z^{-1}-D_2Q(n)Q(n-1)\right),
\end{split}
\label{LPQH}
\end{eqnarray}
where $D_1$ and $D_2$ are the diagonal matrices
\begin{eqnarray*}
D_1=\left( \begin{array}{cc} -\beta & 0 \\
0 & \alpha\\ \end{array} \right), ~~D_2=\left( \begin{array}{cc} \alpha & 0 \\
0 & -\beta \\ \end{array} \right).
\end{eqnarray*}
It can be verified directly that $\mathcal{L}M$ and $\mathcal{P}M$ satisfy  the same jump condition as $M$,  i.e.,
\begin{eqnarray}
\begin{split}
\mathcal{L}M_{-}(n,t,z)=\left(\mathcal{L}M_{+}(n,t,z)\right)J(n,t,z),
\\
\mathcal{P}M_{-}(n,t,z)=\left(\mathcal{P}M_{+}(n,t,z)\right)J(n,t,z).
\end{split}
\label{RHP3AppJ}
\end{eqnarray}
Furthermore, the asymptotic expansion  (\ref{appasym}) implies  that
\begin{eqnarray}
\begin{split}
\mathcal{L}M(n,t,z)=\left( \begin{array}{cc} O(\frac{1}{z}) & O(z)  \\
  O(\frac{1}{z}) & O(z) \\ \end{array} \right),  ~~z\rightarrow (\infty,0),
\\
\mathcal{P}M(n,t,z)=\left( \begin{array}{cc} O(\frac{1}{z}) & O(z)  \\
  O(\frac{1}{z}) & O(z) \\ \end{array} \right),  ~~z\rightarrow (\infty,0).
\end{split}
\label{RHP3AppAS}
\end{eqnarray}
Equations (\ref{RHP3AppJ}) and (\ref{RHP3AppAS}) imply that $\mathcal{L}M$ and $\mathcal{P}M$ satisfy a RH problem with the same jump condition as $M$, but with a vanishing boundary condition. Hence, $\mathcal{L}M=0$ and $\mathcal{P}M=0$, namely,
\begin{eqnarray}
\begin{split}
&M(n+1,t,z)Z-ZM(n, t, z)-Q(n,t)M(n, t, z)=0,
\\
&\left(M(n, t, z)\right)_t-i\omega(z)[\sigma_3,M(n, t, z)]-H(n,t,z)M(n, t, z)=0.
\end{split}
\label{LPA2}
\end{eqnarray}
The compatibility condition of (\ref{LPA2}) yields
\begin{eqnarray}
\begin{split}
\frac{d Q(n,t)}{d t}=2i\{&c\sigma_3Q(n,t)-D_1Q(n-1,t)+D_2Q(n+1,t)
\\&+D_1\left(Q(n,t)\right)^2Q(n-1,t)-D_2Q(n+1,t)\left(Q(n,t)\right)^2\}.
\end{split}
\label{ALApp}
\end{eqnarray}
The reductions $\alpha=\beta=\frac{1}{2}$, $c=-1$ and $M_1^{21}(n,t)=\nu \left(M_1^{12}(n,t)\right)^*$ reduce (\ref{ALApp}) into the DNLS equation,
whereas the reductions $\alpha=-\beta=\frac{i}{2}$, $c=0$ and $M_1^{21}(n,t)=\nu M_1^{12}(n,t)$ reduce (\ref{ALApp}) into the DMKDV equation.

\end{appendices}

\vspace{1cm}
\small{

}

\begin{thebibliography}{99}

\bibitem{F1} A.S. Fokas, Proc. R. Soc. London, Ser. A, 53 (1997) 1411.
\bibitem{F2} A.S. Fokas and A.R. Its, SIAM J. Math. Anal., 27 (1996) 738-764. 
\bibitem{F3} A.S. Fokas, Commun. Math. Phys., 230 (2002) 1-39. 
\bibitem{F4} A.S. Fokas, A.R. Its and L.Y. Sung, Nonlinearity, 18 (2005) 1771-1822. 
\bibitem{F5} A.S. Fokas, Commun. Pure Appl. Math., 58 (2005) 639-670. 
\bibitem{F6} A.S. Fokas, A Unified Approach to Boundary Value Problems, Society for Industrial and Applied Mathematics, Philadelphia, 2008, 27(2).

\bibitem{L1} J. Lenells, Physica D, 241 (2012) 857-875. 
\bibitem{L2} J. Lenells, Nonlinear Analysis, 76 (2013) 122-139. 
\bibitem{XF} J. Xu and E. Fan, Proc. R. Soc. A., 469 (2013) 20130068. 
\bibitem{XF2} J. Xu and E. Fan, Stud. Appl. Math., 136 (2016) 321. 
\bibitem{GLZ} X. Geng, H. Liu and J. Zhu, Stud. Appl. Math., 135 (2015) 310-346.


\bibitem{BH1} G. Biondini and G. Hwang, Inverse Problems, 24 (2008) 065011.
\bibitem{BH2} G. Biondini and D. Wang, IMA J. Appl. Math., 75 (2010) 968-997. 
\bibitem{BH3} G. Biondini and A. Bui, J. Phys. A: Math. Theor., 48 (2015) 375202.
\bibitem{AL1} M.J. Ablowitz and J.F. Ladik, J. Math. Phys., 16 (1975) 598. 
\bibitem{AL2} M.J. Ablowitz and J.F. Ladik, J. Math. Phys., 17 (1976) 1011-1018. 
\bibitem{AL3} M.J. Ablowitz, B. Prinari and A.D. Trubatch, Discrete and Continuous Nonlinear Schr¡§odinger Systems, London Mathematical Society Lecture Note Series vol 302, Cambridge: Cambridge University Press, 2003.
\bibitem{Zhou}  X. Zhou, SIAM J. Math. Anal. 20 (1989) 966-986. 


\bibitem{IB1} I. Habibullin, Phys. Lett. A, 207 (1995) 263-268. 
\bibitem{IB2} B. G\"{u}rel, M. G\"{u}rses and I. Habibullin, J. Math. Phys., 36 (1995) 6809-6821.

\bibitem{MFS} A.B. Monvel, A.S. Fokas, and D. Shepelsky, Lett. Math. Phys., 65 (2003) 199-212. 
\bibitem{FL1} A.S. Fokas and J. Lenells, J. Phys. A: Math. Theor., 45 (2012) 195201. 
\bibitem{FL2} J. Lenells and A.S. Fokas, Proc. R. Soc. A, 471 (2015) 20140925. 
\bibitem{FL3} J. Lenells and A.S. Fokas, Proc. R. Soc. A, 471 (2015) 20140926. 
\bibitem{AK} D.C. Antonopoulou and S. Kamvissis, Nonlinearity, 28 (2015) 3073-3099. 

\bibitem{Geng1} X. Geng, H.H. Dai and J. Zhu, Stud. Appl. Math., 118 (2007) 281-312. 
\bibitem{Geng2} X. Geng and H.H. Dai, J. Math. Anal. Appl., 327 (2007) 829-853. 
\bibitem{ZC} D. Zhang and S. Chen, Stud. Appl. Math., 125 (2010) 393-418. 


\bibitem{V1} V. Volterra, Lecons sur la theorie Mathematique da la Lutte pour la Vie, Gauthier-Villars, Paris 1931 (in French).
\bibitem{V2} M.J. Ablowitz, P.A. Clarkson, Solitons, nonlinear evolution equations and inverse scattering, London Math. Soc. Lecture Note Ser.,
149, Cambridge University Press, Cambridge, 1991.
\bibitem{T1} M. Toda, Theory of Nonlinear Lattices, Berlin: Springer, 1981.
\bibitem{Tu} G.Z. Tu, J. Phys. A: Math. Gen., 23 (1990) 3903-3922. 


\end{thebibliography}
\end{document}